\documentclass[aps,prd,superscriptaddress,groupedaddress,nofootinbib]{article}
\usepackage{dcolumn}
\usepackage{bm}
\usepackage{lscape}
\usepackage{amsmath,amssymb,graphicx}
\usepackage{epsfig}
\usepackage{pstricks}
\usepackage{multirow}
\usepackage[bookmarks,linktocpage]{hyperref}
\usepackage{booktabs}
\usepackage{graphicx}
\usepackage{axodraw4j}
\usepackage{subfigure}
\usepackage{cite}
\usepackage{rotating}
\usepackage{float}
\usepackage{soul}
\usepackage[a4paper, left=2.5cm, right=2.5cm, top=2.5cm, bottom=2.5cm]{geometry}

\usepackage{array}
\newcolumntype{L}[1]{>{\raggedright\let\newline\\\arraybackslash\hspace{0pt}}m{#1}}
\newcolumntype{C}[1]{>{\centering\let\newline\\\arraybackslash\hspace{0pt}}m{#1}}
\newcolumntype{R}[1]{>{\raggedleft\let\newline\\\arraybackslash\hspace{0pt}}m{#1}}

\RequirePackage{lineno}

\begin{document}

\setlength{\arraycolsep}{1.5pt}

\begin{center}
{\Large \textbf{Production of extra quarks at the Large Hadron Collider\\ 
beyond the Narrow Width Approximation}}

\vskip.1cm
\end{center}

\vskip0.2cm

\begin{center}
\textbf{{Stefano Moretti$^{1,2,3}$, Dermot O'Brien$^{1,2}$, Luca Panizzi$^{4,3,1,2}$ and Hugo Prager$^{1,2}$} \vskip 8pt }

{\small
$^1$\textit{School of Physics and Astronomy, University of Southampton, Highfield, Southampton SO17 1BJ, UK}\\[0pt]
\vspace*{0.1cm} $^2$\textit{Particle Physics Department, Rutherford Appleton Laboratory, Chilton, Didcot, Oxon OX11 0QX, UK}\\[0pt]
\vspace*{0.1cm} $^3$\textit{Physics Department, CERN, CH-1211, Geneva 23, Switzerland}\\[0pt]
\vspace*{0.1cm} $^4$\textit{Universit\`a di Genova and INFN Genova, Via Dodecaneso 33, 16146, Genova, Italy}\\[0pt]
}
\end{center}

\begin{abstract} 
\noindent
This paper explores the effects of both finite width and interference (with background) in the pair production and decay of extra heavy quarks with charge 2/3 at the Large Hadron Collider (LHC). This dynamics is normally ignored in standard experimental searches and we assess herein the regions of validity of current approaches, also evaluating the performances of a set of current experimental analyses at 8 and 13 TeV for the deterimination of the excluded regions in the $(M_{\rm VLQ},\Gamma_{\rm VLQ})$ plane, $M_{\rm VLQ}$ being the mass of the VLQ and $\Gamma_{\rm VLQ}$ its width. Further, we discuss the configurations of masses, widths and couplings where the latter breaks down.
\end{abstract}


\tableofcontents

\section{Introduction}

Following the discovery of a Higgs boson \cite{Aad:2012tfa,Chatrchyan:2012xdj} with essentially a Standard Model (SM) nature \cite{Djouadi:2012ae,Eberhardt:2012gv}, the existence of a fourth generation of chiral quarks (i.e., with SM-like $V-A$ structure in gauge boson charged currents) has been excluded \cite{Eberhardt:2012gv}\footnote{It has to be specified here that new chiral quarks have been excluded in the context of a minimal extension of the SM where an extra quark would be the only new particle. If the Higgs sector is also enlarged to contain new states \cite{Alves:2013dga,Banerjee:2013hxa,Holdom:2014bla}, or if more than one quark multiplet is introduced \cite{Bizot:2015zaa}, new chiral quarks can indeed be accomodated. We will not discuss such non-minimal extensions in this context.}. However, the same Large Hadron Collider (LHC) data constrain Vector-Like Quarks (VLQs) significantly less. These hypothised states of matter are heavy spin 1/2 particles that transform as triplets under colour but, unlike SM quarks, their left- and right-handed couplings have the same Electro-Weak (EW) quantum numbers. These objects are predicted by various theoretical scenarios (composite Higgs models \cite{Dobrescu:1997nm,Chivukula:1998wd,He:2001fz,Hill:2002ap,Agashe:2004rs,Contino:2006qr,Barbieri:2007bh,Anastasiou:2009rv}, models with extra dimensions, little Higgs models \cite{ArkaniHamed:2002qy,Schmaltz:2005ky}, models with gauging of the flavour group \cite{Davidson:1987tr,Babu:1989rb,Grinstein:2010ve,Guadagnoli:2011id},  non-minimal
supersymmetric  extensions of the Standard Model (SM) \cite{Moroi:1991mg,Moroi:1992zk,Babu:2008ge,Martin:2009bg,Graham:2009gy,Martin:2010dc}, Grand Unified Theories  \cite{Rosner:1985hx,Robinett:1985dz}) and can be observed in a large number of final states, depending on how they interact with SM particles (see for example \cite{AguilarSaavedra:2009es,Okada:2012gy,DeSimone:2012fs,Buchkremer:2013bha,Aguilar-Saavedra:2013qpa} for general reviews). 
 
In order to be as model independent as possible,  experimental searches for VLQs exploit an economical approach, assuming that only one new VLQ is present beyond the SM, consider QCD processes alone  and parametrise the production and decay dynamics using the Narrow Width Approximation (NWA). While this procedure is clearly very appropriate for several parameter configurations of new physics models containing VLQs, there are others where the aforementioned assumptions would not be correct. For example, most VLQ models predict in general the existence of a new {quark sector}, which implies the presence of more than just one new coloured state, so that, especially when such states are (nearly) degenerate in mass, 
significant interference effects may occur. This has been pointed out in Ref.~\cite{Barducci:2013zaa} where it was also shown how to account for them in a model-independent way, at least in the case of two VLQs being present and with moderate intrinsic width (say, less than 10\% of the mass). As for the reliance on QCD pair production only, this approach has been  {recently superseeded too}, as EW processes have also been explored \cite{Aad:2013rna}), using a parameterisation which largely maintains a model independent approach.

An aspect that has received less attention so far is the adoption of the NWA and its limitations. It is well known that, in the case of the top quark, effects induced onto the inclusive cross-section by its finite width are of ${\cal O}(\Gamma_t/m_t)^2$, hence generally negligible, as $m_t\approx173$  GeV and $\Gamma_t\approx1.5$  GeV. A study of finite width effects in final states corresponding to top pair production has been performed  in Ref.\cite{Kauer:2001sp}. One would naively expect that similar effects in the case of VLQs would be of the same size, i.e., of ${\cal O}(\Gamma_{\rm VLQ}/M_{\rm VLQ})^2$. However, it should be noted that, as
$M_{\rm VLQ}$ is unknown, also $\Gamma_{\rm VLQ}$  is, so that the aforementioned corrections may not be negligible,
if $\Gamma_{\rm VLQ}/M_{\rm VLQ}$ is not very small. In fact, also  differences between the case of the top quark and a VLQ  due to the different structure of their couplings in the charged decay currents would play a role\footnote{Notice that  {VLQs may also decay through flavour changing neutral currents}, involving both the Higgs and $Z$ bosons.}. In this connection, one should recall that, in taking the NWA, as generally done in most Monte Carlo (MC) programs used {in phenomenological and experimental analyses}, one neglects off-diagonal spin effects which stem from the quark (top or vector-like) being massive and whose size is intimately related to the vector/axial (or left/right) composition of the fermionic state entering the charged decay currents and, of course, to the value of the ratio $\Gamma_{\rm VLQ}/M_{\rm VLQ}$. Furthermore, these 
very same two aspects also enter the interfering terms between the heavy quark (top or vector-like) signal (whichever way this is defined in terms of Feynman diagrams) and the background (which would then be
represented by all the other graphs leading to the same final state). Needless to say, one should then not assume
that what is valid for the treatment of off-shellness effects of the top quark (and consequent interferences) remains so for VLQs as well. 

Very recently experimental searches for VLQs have started to explore the large width regime, considering single production of top and bottom VLQ partners~\cite{CMS:2017oef,CMS:2017zhw}. However, to our knowledge, no experimental limit has been set for topologies compatible with the pair production channels.
It is the purpose of this paper to assess the regions of validity of the NWA for final states compatible with pair production and decay of a VLQ with charge 2/3 but where, due to its finite width, the VLQ is produced, via both QCD and EW interactions, in pairs or even singly. Interference effects of various nature will also be considered. We will do so under the assumption that all the decay products of the heavy quark are visible SM states.
The plan of the paper is as follows. In the next two sections we describe our conventions and the computational tools we adopted while in the following two we present our numerical results, first for the case of finite width corrections then for interference effects. Finally, we conclude in the last section.

\section{Setup}
\label{sec:setup}

\subsection{Definitions}
To understand the effects of large widths on the signal, we will consider different processes, all leading to the same four-particle final state:

\begin{itemize}

\item\textit{QCD pair production and decay of on-shell VLQs}

This process is usually considered in experimental searches of VLQs. In the NWA it is possible to separate and factorize production and decay of the heavy quarks, thus allowing for a model-independent analysis of the results. The cross-section for this process is given by (hereafter, in our formulae, $Q$ denotes a VLQ):
\begin{equation}
 \sigma_X \equiv \sigma_{2 \to 2}~{\rm BR}(Q)~{\rm BR}(\bar Q)
\end{equation}
where, obviously, $\sigma_{2 \to 2}$ only takes into account pure QCD topologies.

%

\item\textit{Full signal}

In this process all the topologies which contain \textit{at least one} VLQ propagator are taken into account. The only assumption is that the QCD and EW order of the processes are the same as in the processes above, for consistency. The full signal includes the pair production process without the on-shell condition described above. The cross-section of this process will be labelled as $\sigma_S$. Some example topologies for this process which are not included in the previous ones are in Fig.~\ref{fig:fullsignaltopologies}. The full signal contains topologies which are generally subleading in the NWA, but that become more and more relevant as the width of the VLQ increases.

\begin{figure}[H]
\centering\includegraphics[width=.6\textwidth]{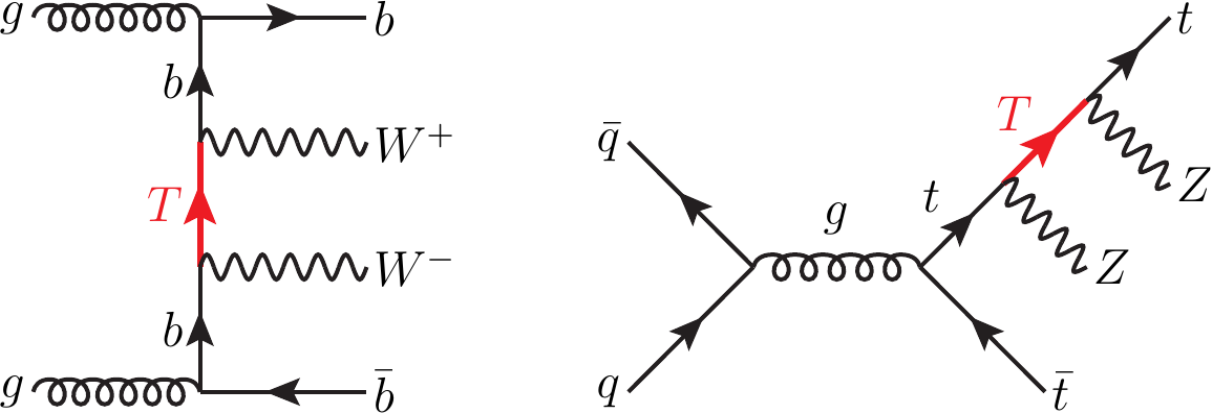}
\caption{Examples of topologies containing only one VLQ propagator for the $P P \to W^+ b W^- \bar b$ and $P P \to Z t Z \bar t$ processes.}
\label{fig:fullsignaltopologies}
\end{figure}

\item\textit{SM irreducible background}

This process trivially corresponds to all the $2 \to 4$ topologies which do not involve any VLQ propagators. The cross-section will be labelled as $\sigma_B$.

\item\textit{Total process}

This process includes the full signal, the SM background and the interference terms. The cross-section will be labelled as $\sigma_T$ and is related to the previous cross-sections by the following relation:

\begin{equation}
 \sigma_T = \sigma_S + \sigma_B + \sigma_\text{interference}
\end{equation}

\end{itemize}

In order to determine the effect of large widths on the cross-section, we will consider a number of variables:

\begin{itemize}
\item $\frac{\sigma_S - \sigma_X}{\sigma_X}$: this ratio takes into account both the off-shell and the subleading contributions given by topologies which contain at least one VLQ propagator. It measures in practice how much the full signal differs from the approximate pair-production-plus-decay signal in the NWA.
\item $\frac{\sigma_T - (\sigma_X + \sigma_B)}{\sigma_X + \sigma_B}$: this ratio measures the correction factor to apply to obtain the full cross-section starting with the pair-production in the NWA and the SM background considered independently.
\item $\frac{\sigma_T - (\sigma_S + \sigma_B)}{\sigma_S + \sigma_B}$: this ratio measures the size of the interference effects between signal and SM background.
\end{itemize}

\subsection{Tools and validation }

Our numerical results at partonic level have been obtained using {\sc MadGraph 5}~\cite{Alwall:2011uj,Alwall:2014hca} with the public VLQ model~\cite{feynrulesVLQ} implemented in {\sc FeynRules}~\cite{Alloul:2013bka}. We have produced events in the five-flavour scheme (5FS), using the {\sc cteq6l1}~\cite{Pumplin:2002vw} PDF set. Hadronisation and parton showering have been obtained through the {\sc Pythia\,8} code~\cite{Sjostrand:2014zea}. To obtain the width-dependent bounds on the VLQ mass we have considered a combination of searches at 8 TeV and an ATLAS search~\cite{TheATLAScollaboration:2016gxs} at 13 TeV. All the searches we considered are present in the database of the code {\sc CheckMATE\,2}~\cite{Dercks:2016npn}, which exploits the {\sc Delphes\,3} framework~\cite{deFavereau:2013fsa}. We stress here that the purpose of our recasting is not to obtain bounds for large width VLQs but to study the performance of sets of cuts currently adopted in searches for pair production of VLQs or optimised for different final states. Determining an optimised set of selection and kinematics cuts to enhance the sensitivity to the kinematics of a $T$ with large width (and therefore determine a reliable bound in the mass-width plane) will be the scope of a future dedicated study.  \\

Furthermore, to fully validate our analysis of the NWA results versus the off-shell ones, we developed a separate code where the
Dirac function is obtained as the appropriate limit of the Breit-Wigner distribution,
we have also prepared a dedicated $2\to 6$ program (hence also including the fermionic decays of the bosons stemming from the two $T$ decays, which are SM-like), wherein we have adopted  a suitable mapping of the integrand function, via the standard change of variable
\begin{equation}
p^2-M^2=M\Gamma\tan\theta,
\end{equation}
where $p^2$ is the (squared) moment flowing through a resonance with mass $M$ and width $\Gamma$. This factorises the Jacobian
\begin{equation}
dp^2=\frac{1}{M\Gamma}[(p^2-M^2)^2+M^2\Gamma^2]d\theta,
\end{equation}
which thus incorporates the resonant behaviour in the sampling of the phase space itself, thereby rendering the multi-dimensional numerical 
integration (done via importance sampling) very efficient. Finally, upon multiplying the integrand function by $\Gamma/\Gamma_{\rm tot}$,
where $\Gamma_{\rm tot}$ is the decaying particle's intrinsic total width,  and taking the limit $\Gamma\to0$, we obtain self-consistently the above transition from the
off-shell to the NWA results. The results obtained this way closely match those obtained through MadGraph 5 for the aforementioned
$2\to2$ (on-shell, times ${\rm BR}$) and   $2\to4$ (off-shell) processes.

\begin{figure}[H]
\centering
\includegraphics[width=0.8\textwidth]{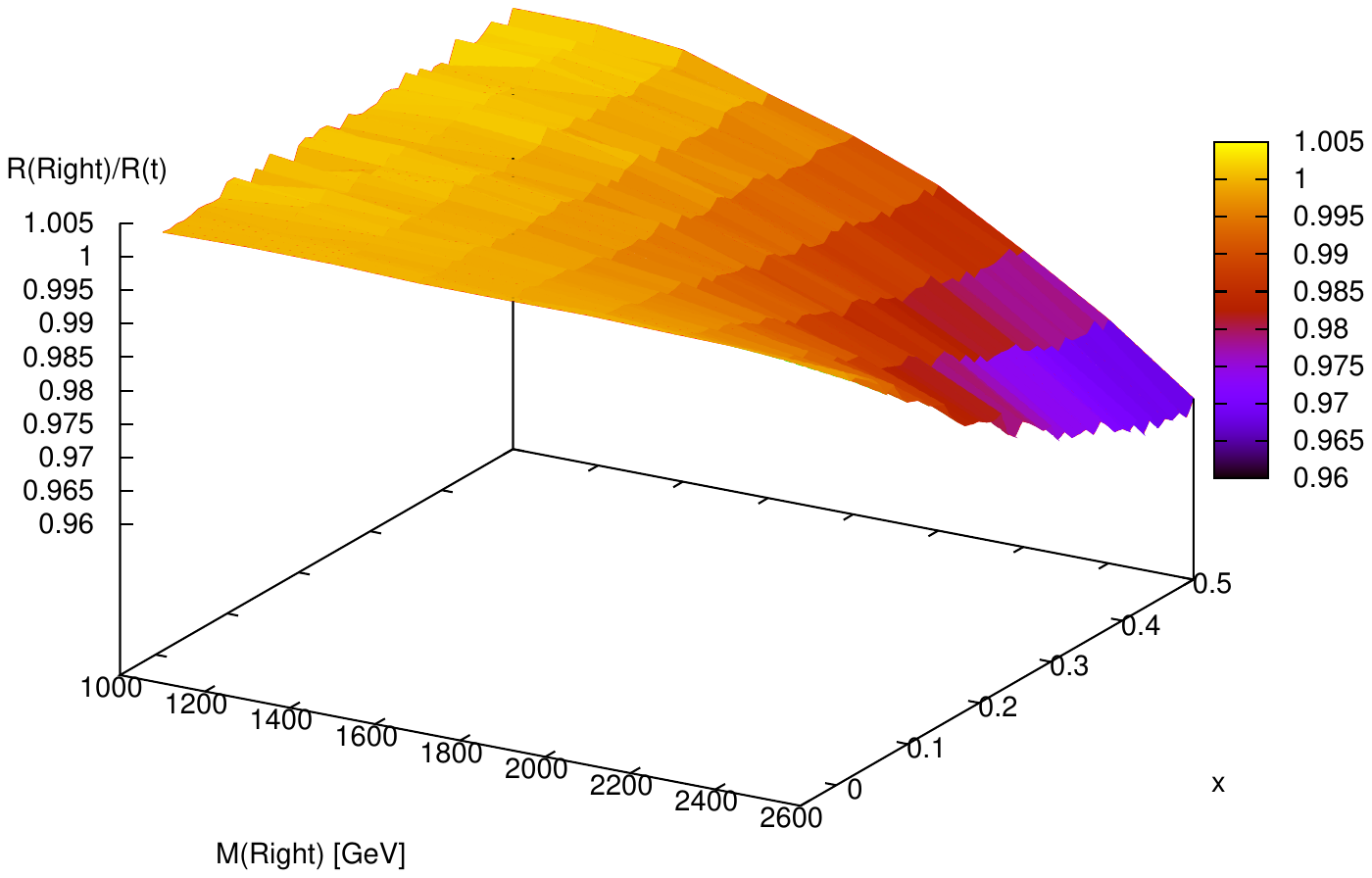}
\vspace{-1cm}
\caption{\label{fig:3D}Ratio of FW corrections with respect to the NWA relative to the $V-A$ case of a 
$V+A$ charged decay current.}
\end{figure}

As the SM top quark, $t$, and the heavy quark with same Electro-Magnetic (EM) charge, $T$, have a common decay channel, i.e., $bW^+$, as a preliminary exercise meant to address the impact of a potentially very different chiral structures in the transitions $t\to bW^+$ and $T\to bW^+$, we have defined the following quantity
\begin{equation}
R(X)=\frac{\sigma(pp\to X\to bW^+\bar bW^-\to 6~{\rm fermions})_{\rm FW}}
                 {\sigma(pp\to X\to bW^+\bar bW^-\to 6~{\rm fermions})_{\rm NWA}},
\end{equation} 
which measures inclusively the effect of a Finite Width (FW) for the cases $X=t$ (a heavy quark with pure $V-A$ couplings,
i.e., top-like) and $X=$~Right (heavy quark with pure $V+A$ couplings). Clearly, these are extreme coupling choices, as 
an interaction eigenstate of a VLQ would have an admixture of $V-A$ and $V+A$ couplings.
However, it should be recalled that VLQ  couplings have always a dominant chirality: this has been demonstrated in  Refs.\cite{delAguila:2000rc,Buchkremer:2013bha}.
In Fig.~\ref{fig:3D} we plot the ratio $R({\rm Right})/R(t)$ mapped as a function of the heavy quark mass $M_{\rm VLQ}$ and relative width $x=\Gamma_{\rm VLQ}/M_{\rm VLQ}$ over the ranges [1000 GeV, 2500 GeV] (i.e., up to the typical mass reach of the LHC for pair production)
and [0, 0.5] (i.e., up to the width limit beyond which the VLQ can no longer be considered a resonance), respectively. One can see that differences are phenomenologically irrelevant.

\vspace{-1cm}

\section{Benchmarks and constraints\label{sec:Benchmarks}} 

In the present analysis we will consider the processes of production of a heavy top-like quark $T$. In principle, from a model-independent point of view, the $T$ quark is allowed to interact with all SM quark generations, but to evaluate the effects of large widths in different scenarios, only specific interactions will be switched on in the different examples we will consider. 

Since the purpose of this analysis is to evaluate the effects of large widths on channels commonly explored by experimental analysis, we will consider only final states allowed by $T$ pair production and decay. The full set of channels in which a pair-produced $T$ quark can decay is given by the following matrix:
\begin{eqnarray}
\small
T\bar T\to\left(
\begin{array}{ccc|ccc|ccc}
Wd W\bar d & Wd Z\bar u & Wd H\bar u & Wd W\bar s & Wd Z\bar c & Wd H\bar c & Wd W\bar b & Wd Z\bar t & Wd H\bar t \\
Zu W\bar d & Zu Z\bar u & Zu H\bar u & Zu W\bar s & Zu Z\bar c & Wd H\bar c & Zu W\bar b & Zu Z\bar t & Zu H\bar t \\ 
Hu W\bar d & Hu Z\bar u & Hu H\bar u & Hu W\bar s & Hu Z\bar c & Wd H\bar c & Hu W\bar b & Hu Z\bar t & Hu H\bar t \\ 
\hline                                                                                                     
Ws W\bar d & Ws Z\bar u & Ws H\bar u & Ws W\bar s & Ws Z\bar c & Wd H\bar c & Ws W\bar b & Ws Z\bar t & Ws H\bar t \\ 
Zc W\bar d & Zc Z\bar u & Zc H\bar u & Zc W\bar s & Zc Z\bar c & Wd H\bar c & Zc W\bar b & Zc Z\bar t & Zc H\bar t \\ 
Hc W\bar d & Hc Z\bar u & Hc H\bar u & Hc W\bar s & Hc Z\bar c & Wd H\bar c & Hc W\bar b & Hc Z\bar t & Hc H\bar t \\ 
\hline                                                                                                     
Wb W\bar d & Wb Z\bar u & Wb H\bar u & Wb W\bar s & Wb Z\bar c & Wd H\bar c & Wb W\bar b & Wb Z\bar t & Wb H\bar t \\ 
Zt W\bar d & Zt Z\bar u & Zt H\bar u & Zt W\bar s & Zt Z\bar c & Wd H\bar c & Zt W\bar b & Zt Z\bar t & Zt H\bar t \\ 
Ht W\bar d & Ht Z\bar u & Ht H\bar u & Ht W\bar s & Ht Z\bar c & Wd H\bar c & Ht W\bar b & Ht Z\bar t & Ht H\bar t 
\end{array}
\right)
\label{eq:finalstates}
\end{eqnarray}
We will focus on two blocks of this matrix, the top-left (corresponding to a $T$ interacting with the first SM generation) and the bottom-right ($T$ interacting with the third SM generation). As we are interested in the width dependence of ratios of cross-sections and of mass bounds, we expect that the scenario of mixing with the second generation will not give sizably different results with respect to the mixing with first generation, so we will not consider it in this analysis. Performing the analysis by selecting specific final states doesn't mean that we are assuming that the $T$ quark only interacts with first or third generation. Effects of large width are different depending on the kinematics of the process and by selecting representative scenarios it is possible to reconstruct intermediate configurations (VLQs interacting partly with heavy and partly with light SM generations).

\begin{figure}[H]
\centering\includegraphics[width=.5\textwidth]{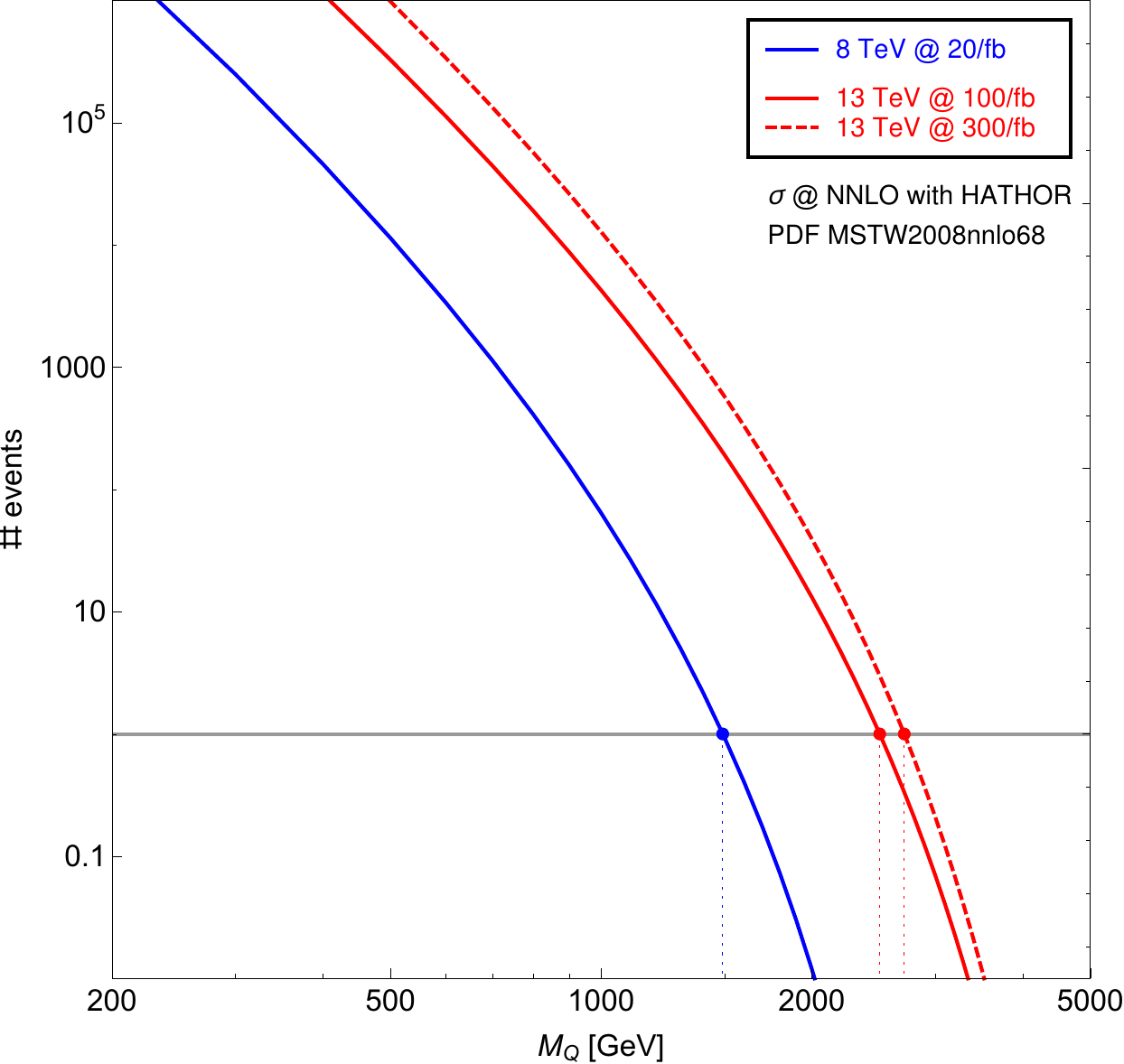} 
\caption{\label{fig:Xsigma}Number of events at partonic level for $Q \bar Q$ pair production and for different LHC energies and luminosities. The corresponding cross-sections have been computed using HATHOR\cite{Aliev:2010zk} with MSTW2008nnlo68 PDFs\cite{Martin:2009bu}.}
\end{figure}

This analysis is of phenomenological interest only for mass values for which the number of final events is (ideally) larger than 1. In Fig.~\ref{fig:Xsigma} we show the number of events for different LHC luminosities for the $X$ channel, which is common to all scenarios. The number of events in Fig.~\ref{fig:Xsigma} has been computed considering a NNLO cross-section, however the results in the next sections will correspond to LO cross-sections, as we are assuming that for processes of pair production the kinematics won't change appreciably and all the differences can be factorised through a K-factor. From Fig.~\ref{fig:Xsigma} it is possible to see that the ideal practical validity of our results is limited to mass values of around 1500 GeV for LHC@8TeV, 2500 GeV (2700 GeV) for LHC@13TeV with 100/fb (300/fb) integrated luminosity. Of course we are not considering here effects due to experimental acceptances and efficiencies.

\subsection{How large can the width be?}

\begin{figure}[H]
\centering
\includegraphics[width=.32\textwidth]{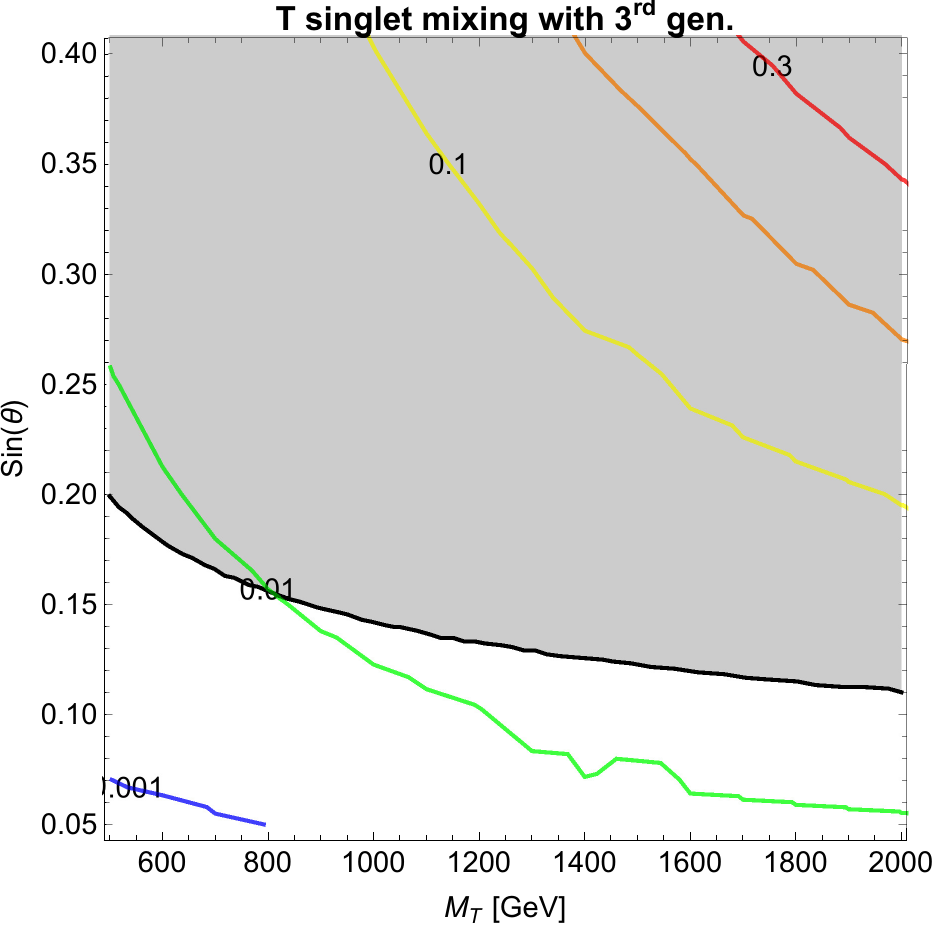}\hfill
\includegraphics[width=.32\textwidth]{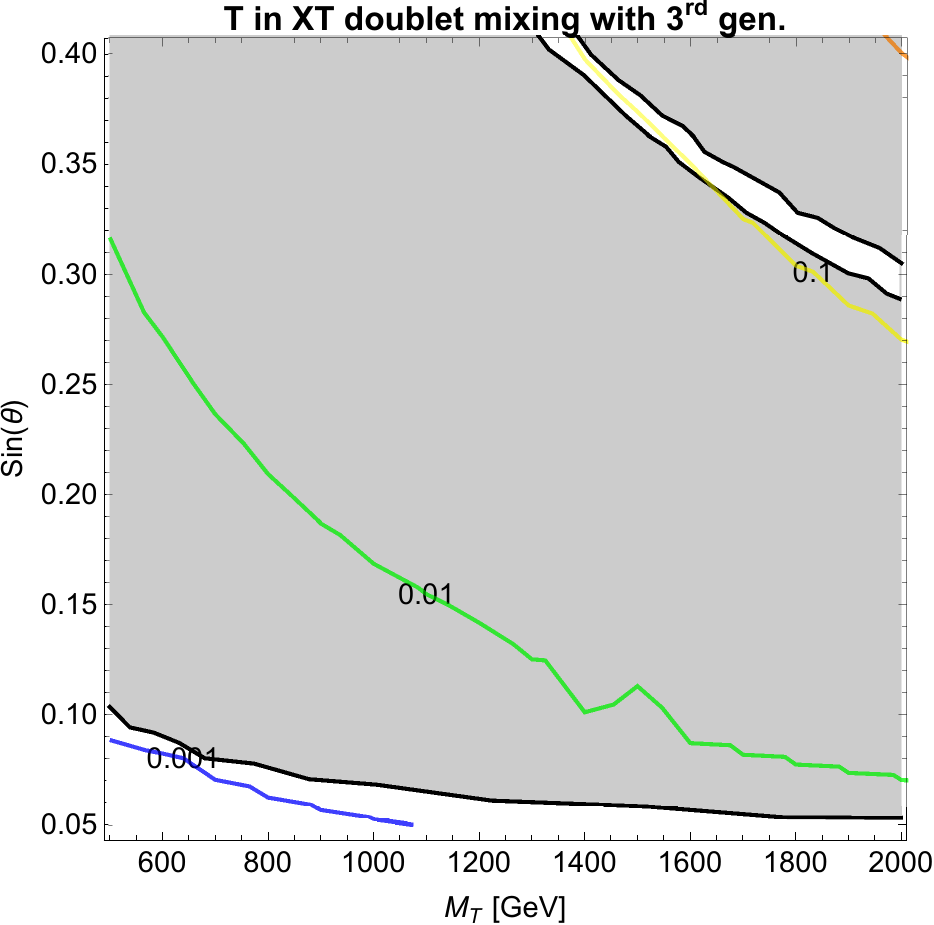}\hfill
\includegraphics[width=.32\textwidth]{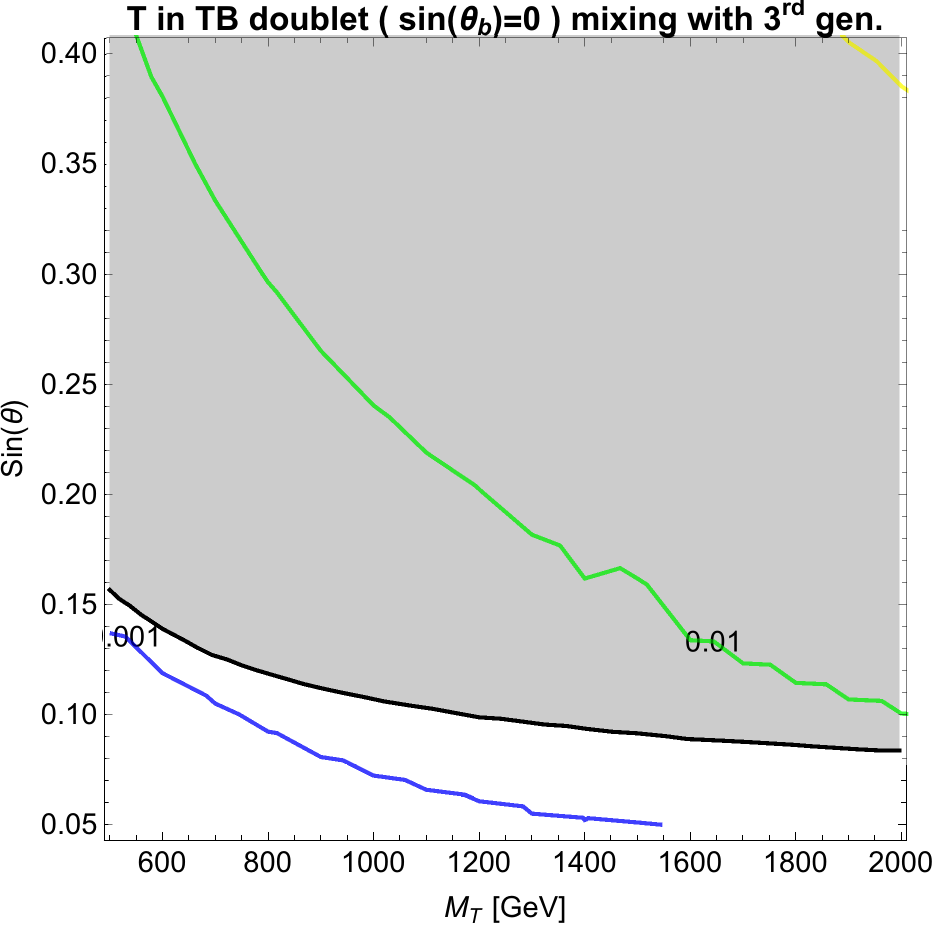}
\caption{\label{fig:Tsimplifiedbounds}Contours with constant $\Gamma/M$ ratio as function of $T$ mass and mixing angle for $T$ belonging to different representations and with different mixing hypotheses. The excluded (shaded) regions from~\cite{Chen:2017hak} have been superimposed.}
\end{figure}

In a simplified model where the SM is only augmented by the presence of a VLQ representation containing a $T$ quark the couplings of the VLQ are constrained by different observables \cite{Okada:2012gy}. In contrast, a $T$ VLQ with a large width in such a scenario can only be obtained if its couplings are large. It is therefore important to determine how large the width can be in simplified scenarios if constraints on the $T$ couplings are saturated to the current bounds. Such bounds depend on the specific representation the $T$ state belongs to.
We will consider here as representative scenarios a $T$ singlet and a $T$ as part of a doublet (both $(X,~T)$ and $(T,~B)$). In both cases the Branching Ratios (BRs) depend on both mass and width, but for the singlet the couplings are dominantly left-handed, while for the doublet the couplings are dominantly right-handed.
In Fig. \ref{fig:Tsimplifiedbounds} we show the contours with constant $\Gamma/M$ ratio for different values of the $T$ mass and mixing angle with the SM top quark, to which we have superimposed the excluded regions from EWPTs and $Zbb$ constraints, borrowed from Ref.\cite{Chen:2017hak}. 

Clearly, simplified models where the SM is extended with one VLQ representation containing a $T$ with large mixing are strongly constrained, and therefore the $T$ width cannot become larger than few \% of the mass (at best). The scenarios are even more constrained for $T$ quarks mixing with light generations, for which the bounds are tighter~\cite{Cacciapaglia:2011fx,Okada:2012gy}.
Therefore, to keep a model-independent perspective we must assume that the width of the $T$ can become large because of the presence of further (yet undiscovered) new states lighter than the $T$ VLQ, which results in a larger number of decay channels into further BSM particles, and/or because of mixing with other VLQs, which may relax constraints from flavour or precision observables because of cancellations of effects~\cite{Cacciapaglia:2015ixa}. Hence, for the purposes of this analysis, the {\it total width} of the $T$ will be considered as a free parameter, limited to be less than the extreme value of 50\% of the mass of the VLQ. In practice, we will consider values up to 40\% of the $T$ mass for our numerical evaluations.

\section{Extra $T$ quark mixing with third generation SM quarks}

\subsection{Large width effects on the signal at parton level}

\begin{figure}[H]
\centering
\includegraphics[width=.3\textwidth]{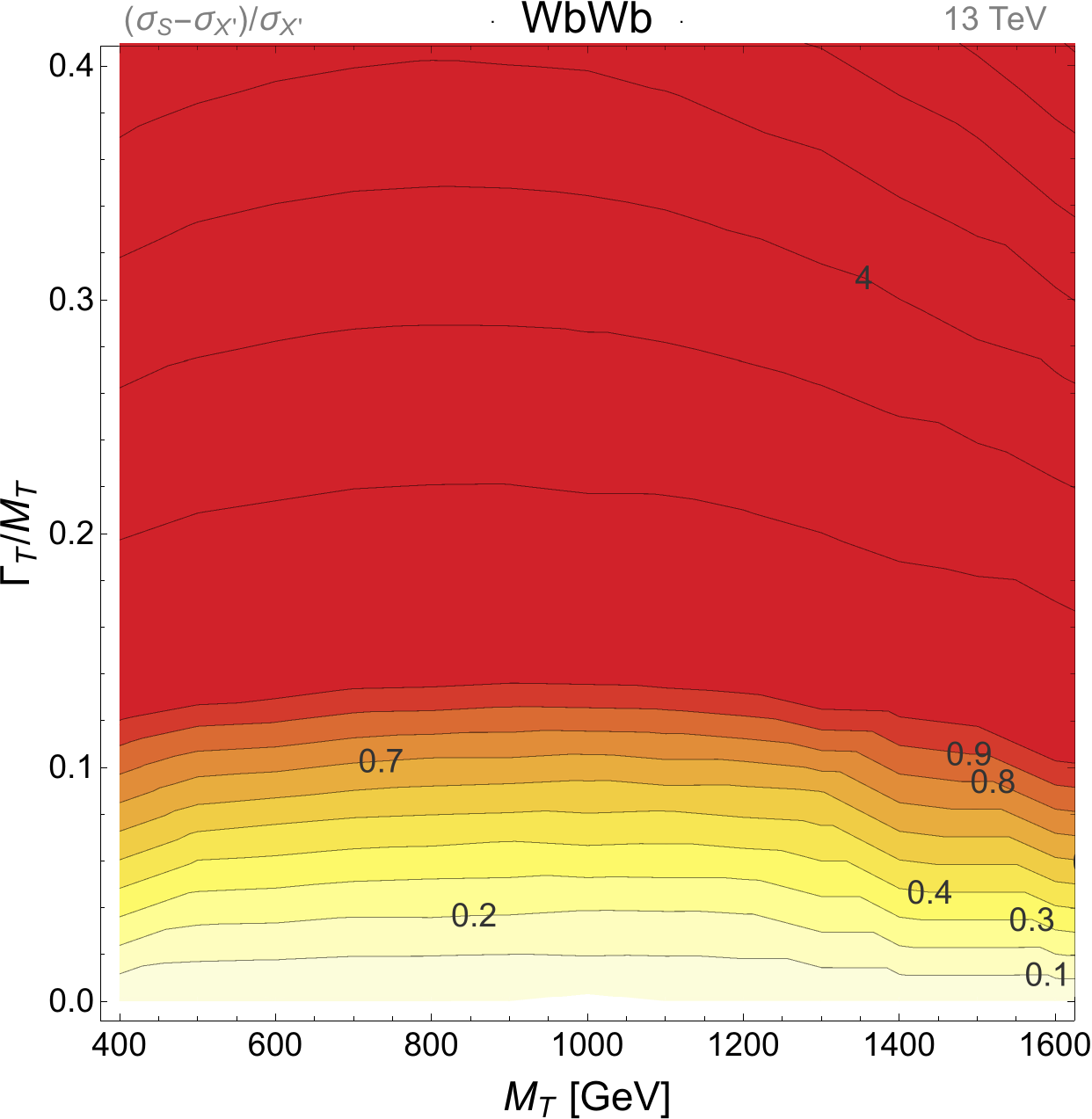}
\includegraphics[width=.3\textwidth]{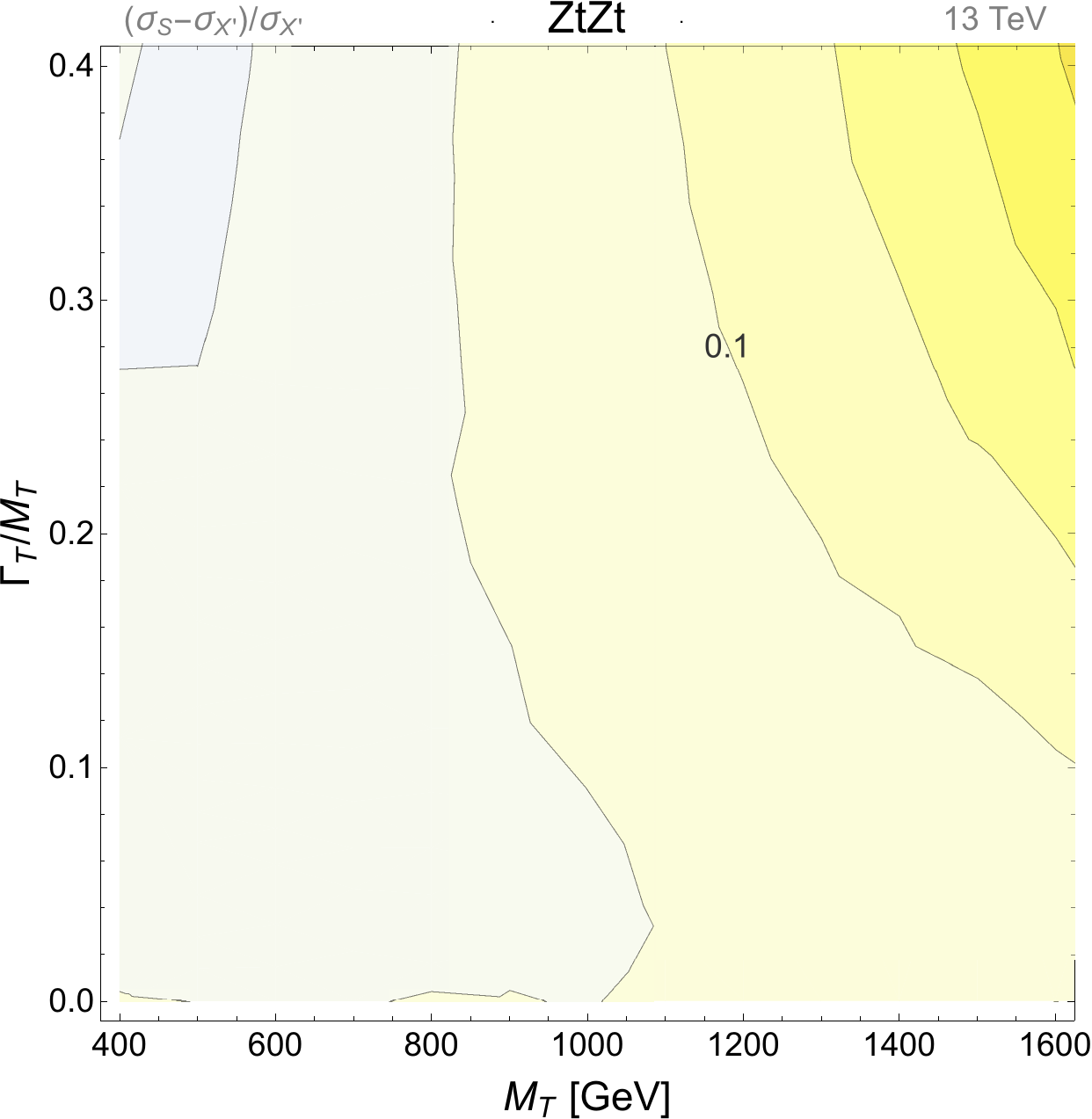} 
\includegraphics[width=.3\textwidth]{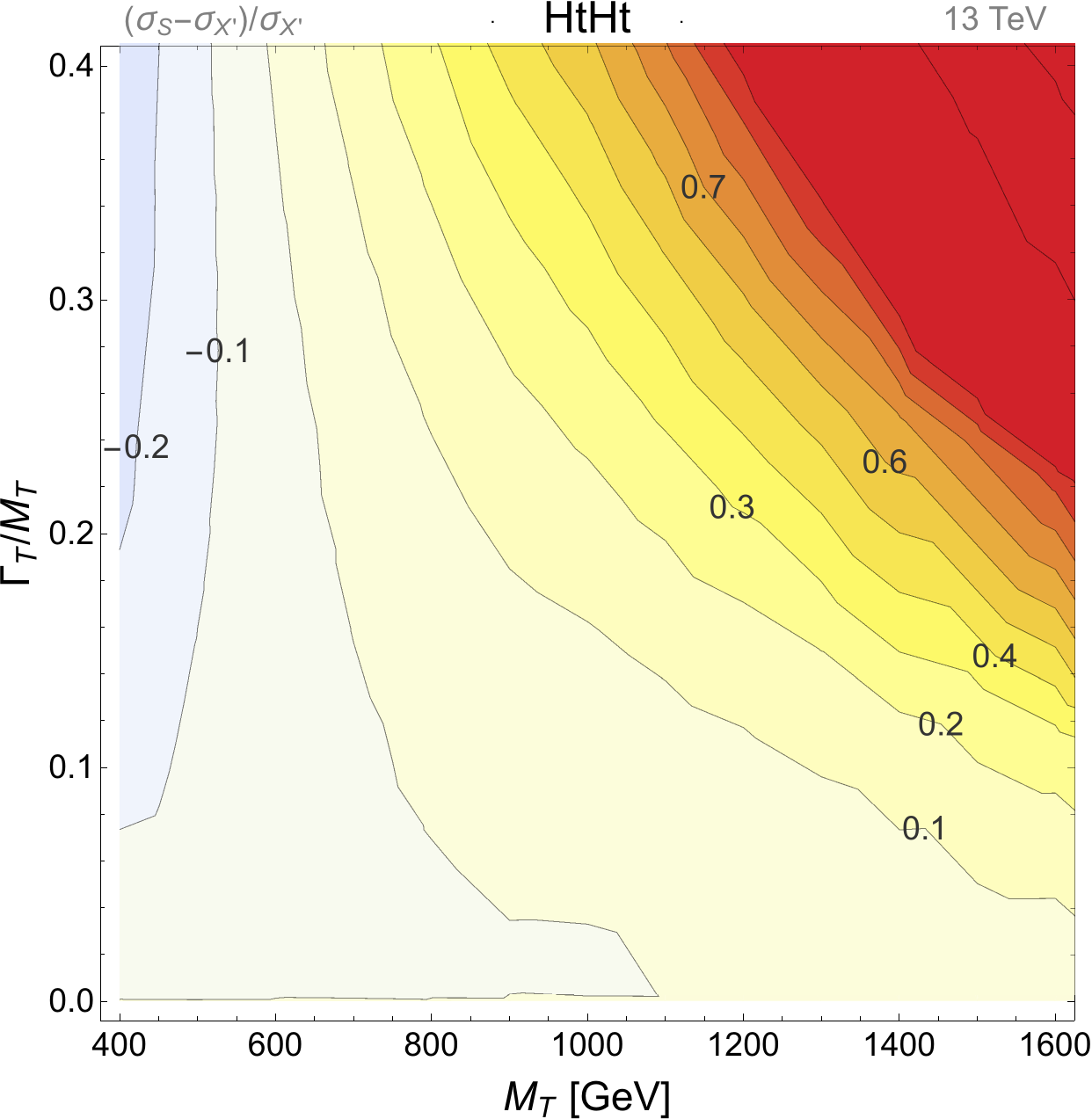}\\

\includegraphics[width=.3\textwidth]{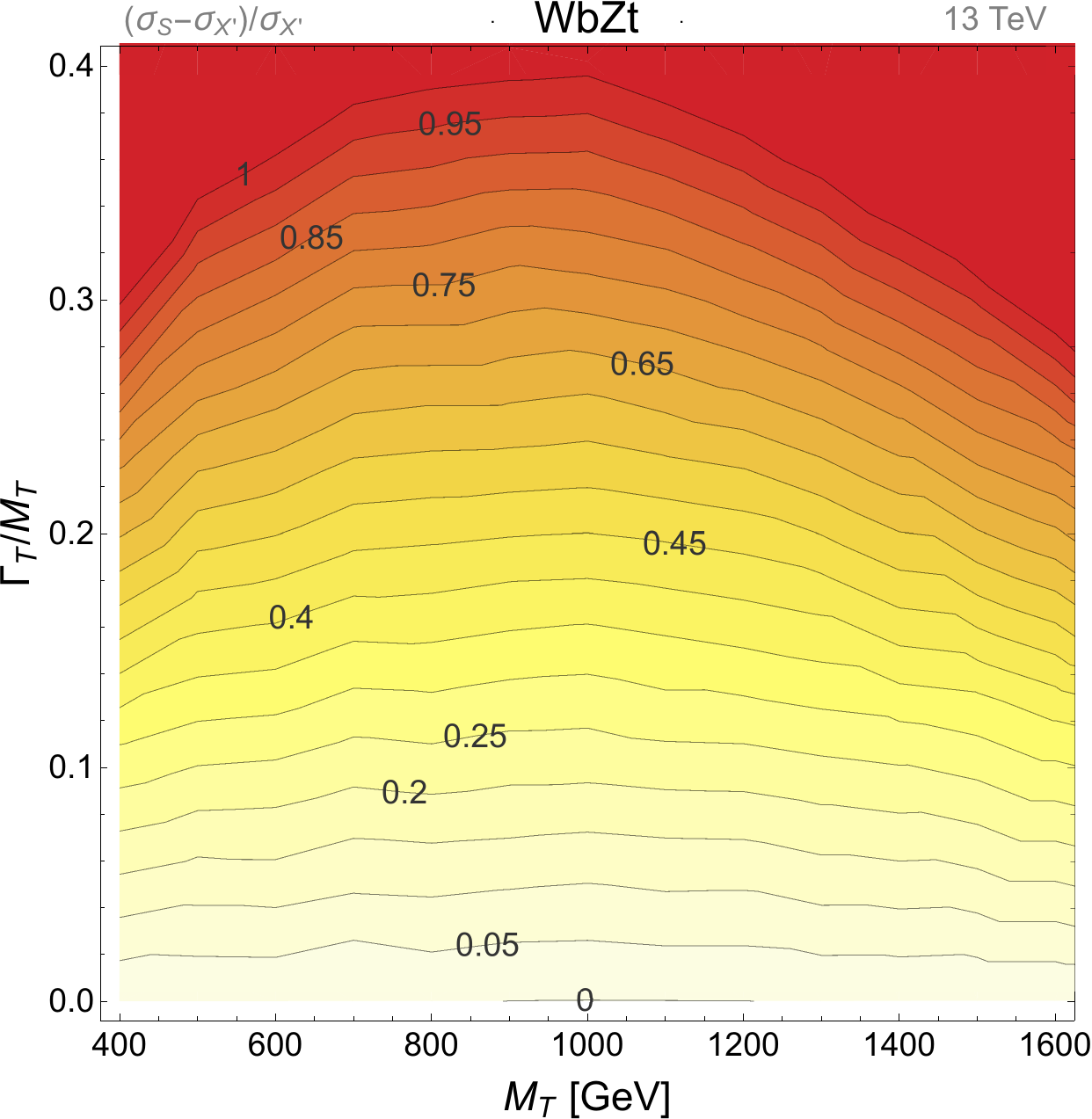} 
\includegraphics[width=.3\textwidth]{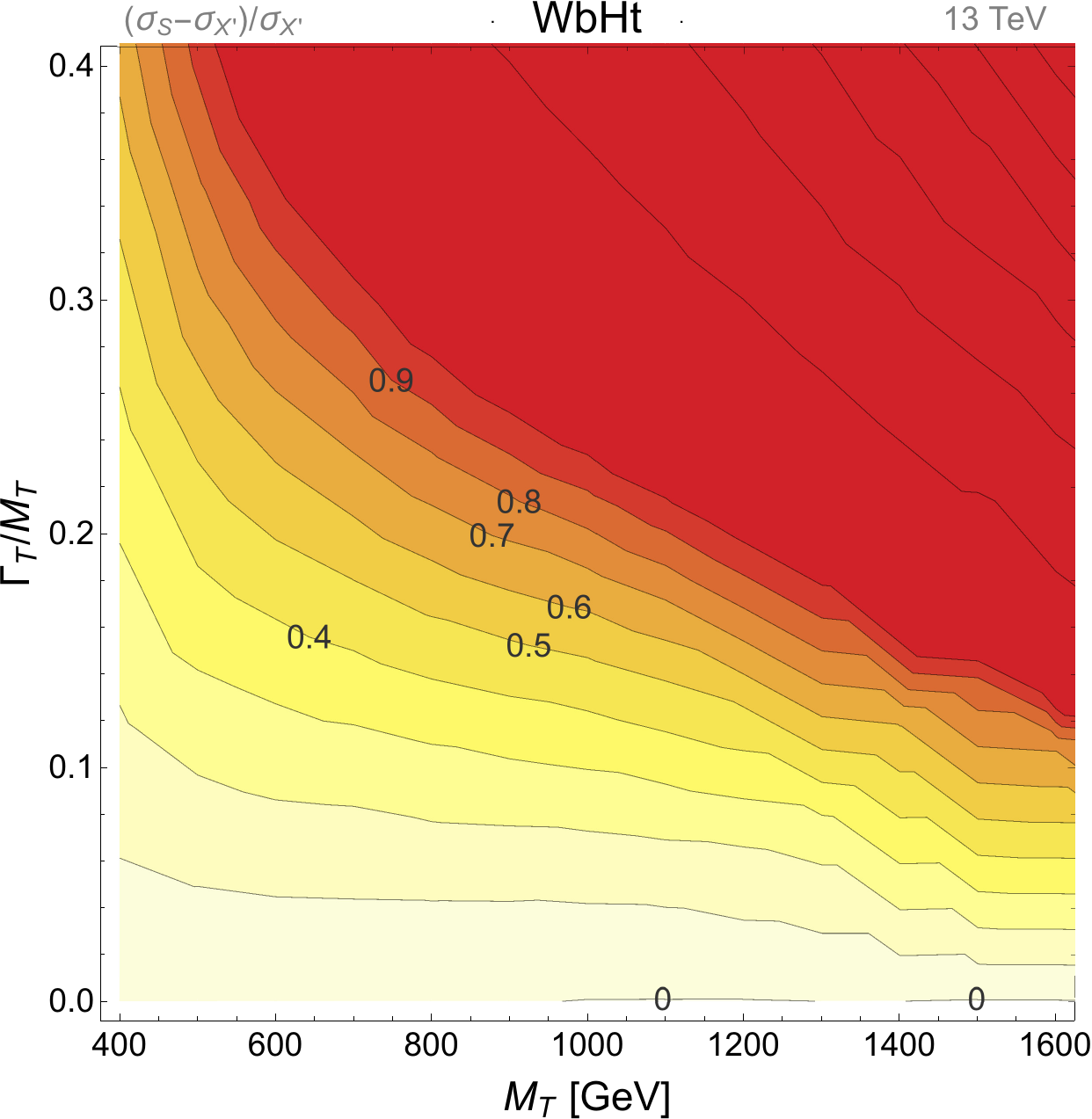} 
\includegraphics[width=.3\textwidth]{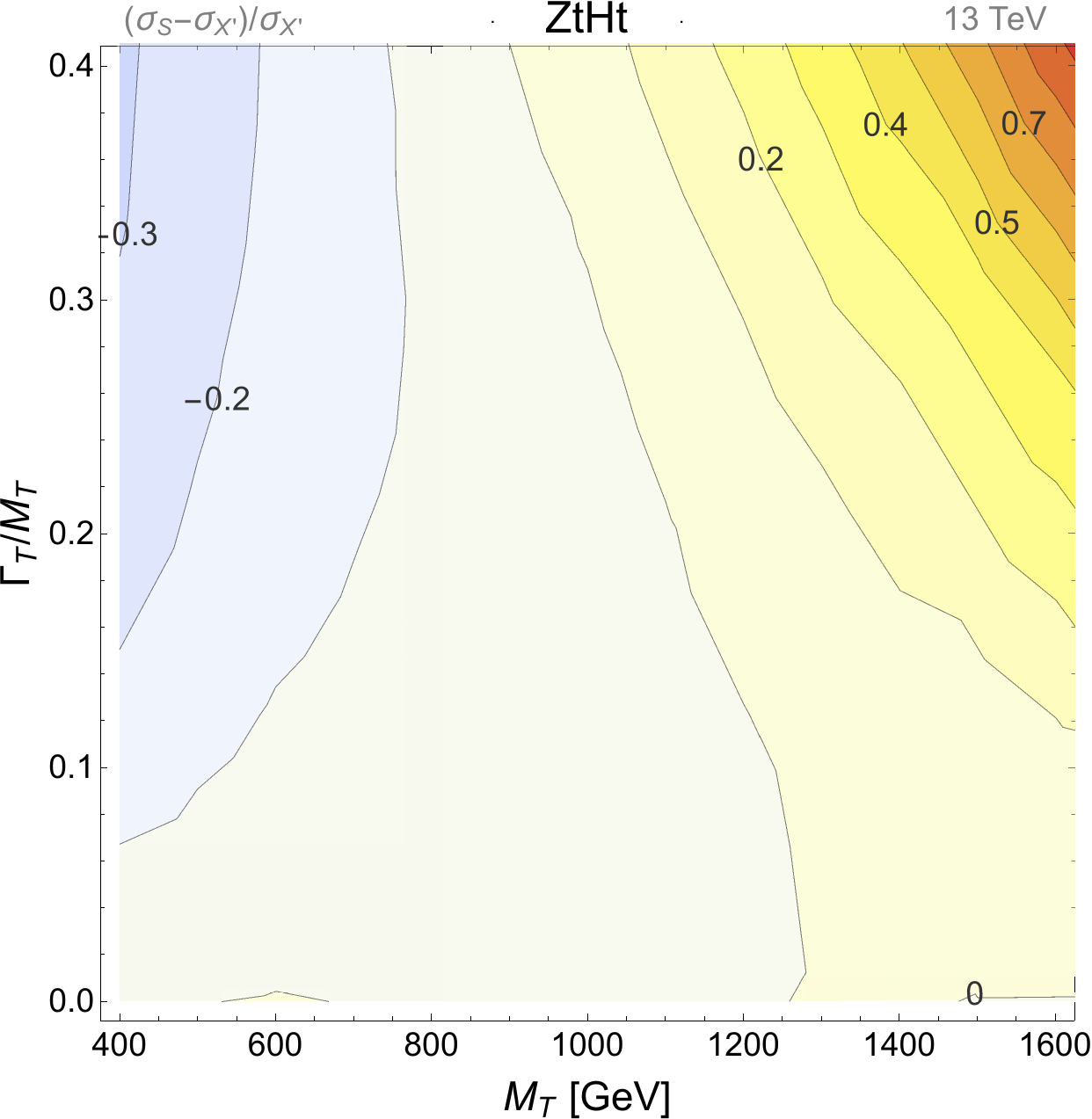} 

\caption{\label{fig:SXthird} Relative difference between the full signal cross-section and the cross-section of QCD pair production for $T$ mixing with the SM top quark.}
\end{figure}

The effect of a large width in the cross-section due to off-shell contributions and to topologies which are absent in the NWA limit is shown in Fig.~\ref{fig:SXthird}. At parton level we will only show results at 13 TeV. We verified that the results at 8 TeV are qualitatively similar.

As a first sanity check of our calculations we observe that, as expected, in the NWA limit the off-shell contributions are negligible. The contributions of off-shellness and new topologies become more and more relevant as the width of the $T$ increases and the cross-section may eventually become several factors larger than in the NWA for some final states. The large increase of the cross-section even for small $T$ masses for channels with the bottom in the final state is explained by the presence of diagrams where the $b$-jets are radiated directly from the initial state or generated by gluon splittings:  such topologies are enhanced by collinear divergences. We will not explore this aspect further, as the isolation and kinematics cuts applied at analysis level usually remove such enhanced contributions, independently of the $T$ mass and width as we will show in Sec. \ref{sec:detector3}.

\begin{figure}[H]
\centering
\includegraphics[width=.45\textwidth]{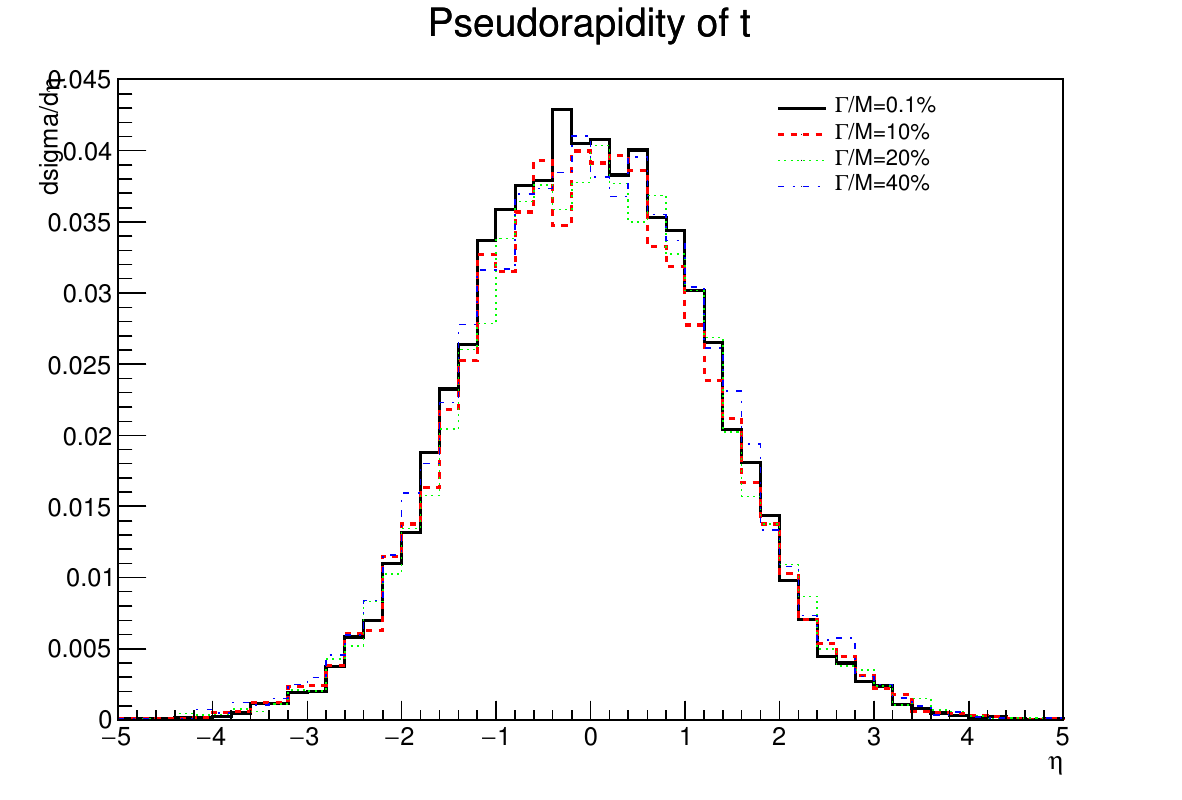}
\includegraphics[width=.45\textwidth]{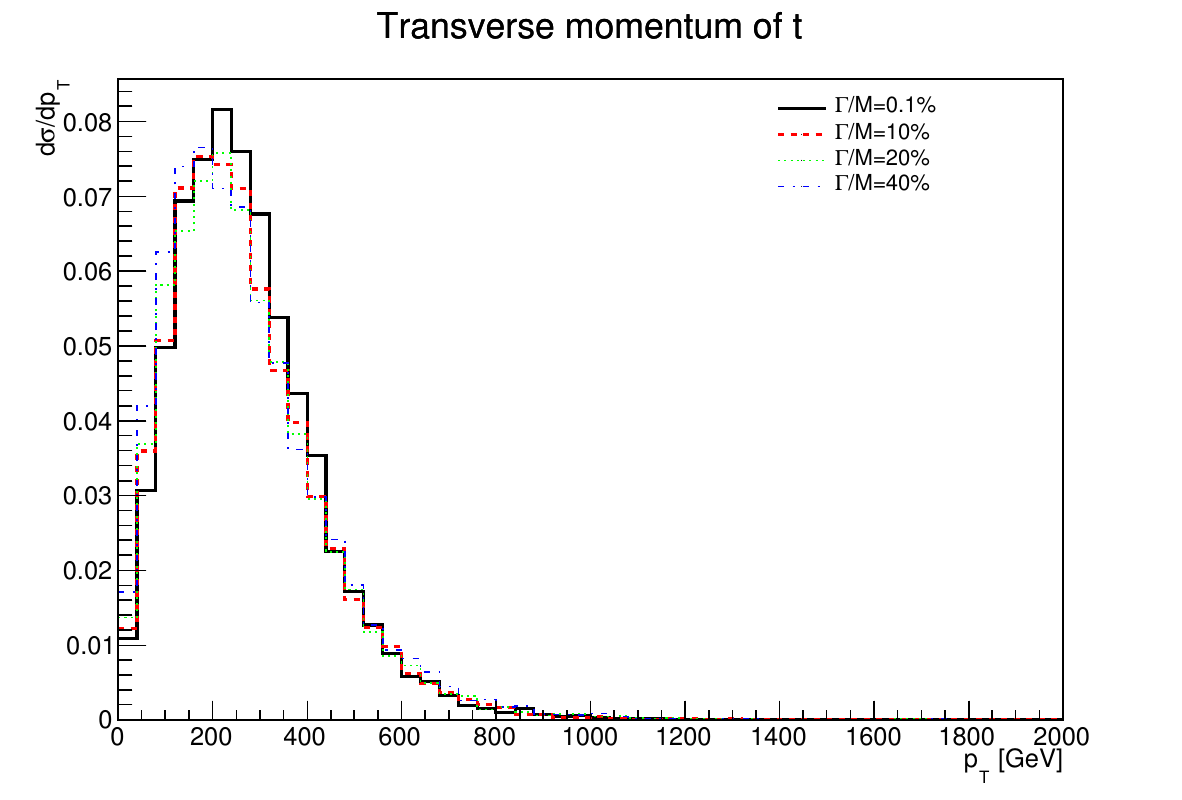}\\
\includegraphics[width=.45\textwidth]{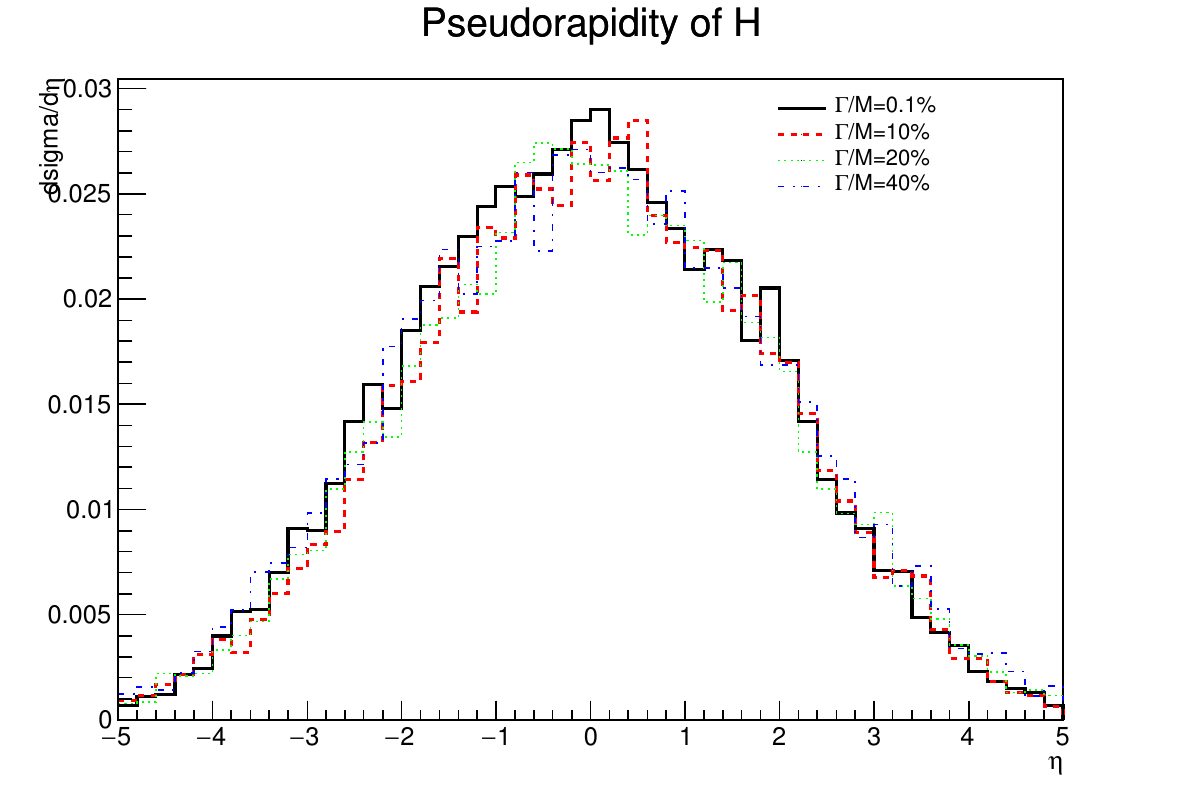}
\includegraphics[width=.45\textwidth]{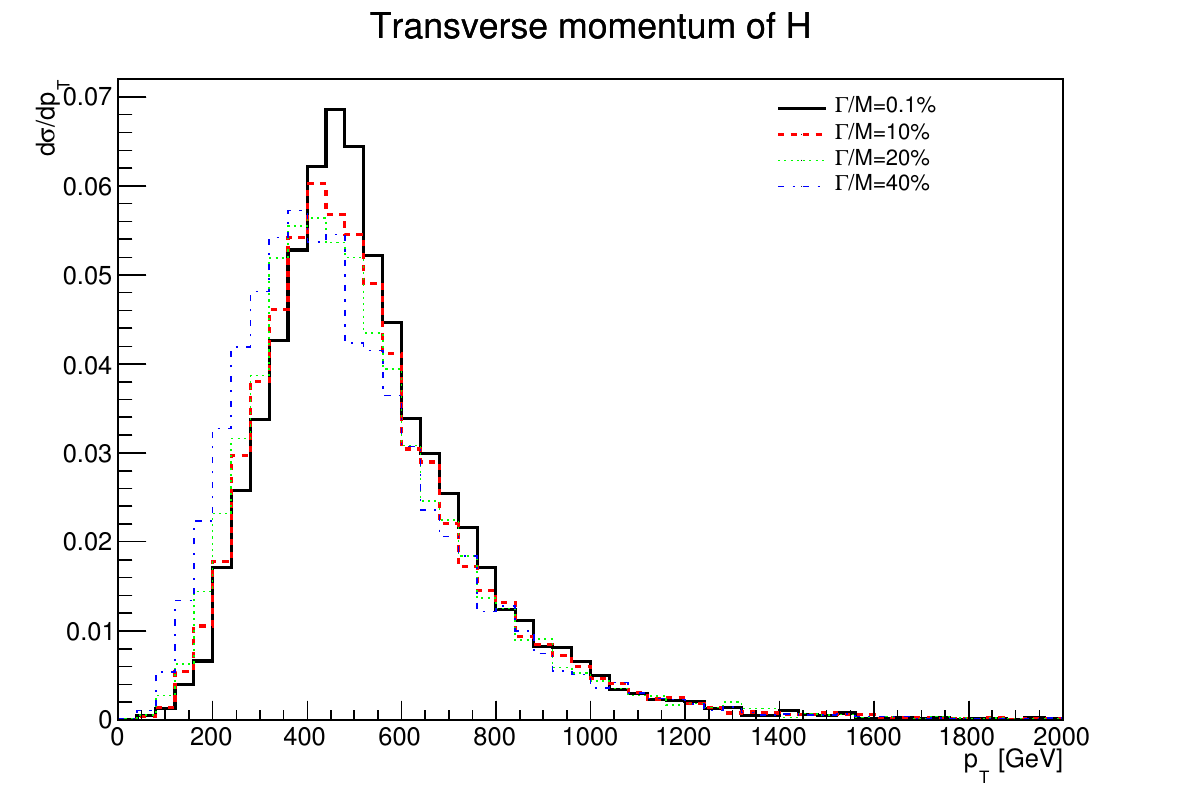}
\caption{\label{fig:disththt}Partonic level differential cross-sections for the $HtHt$ channel. From left to right and top to bottom: $\eta_t$, $p_{Tt}$, $\eta_H$ and $p_{TH}$. All distributions correspond to a $T$ mass of 600 GeV, for which $\sigma_S\sim\sigma_X$ almost independently of the $T$ width.}
\end{figure}

\begin{figure}[H]
\centering
\includegraphics[width=.45\textwidth]{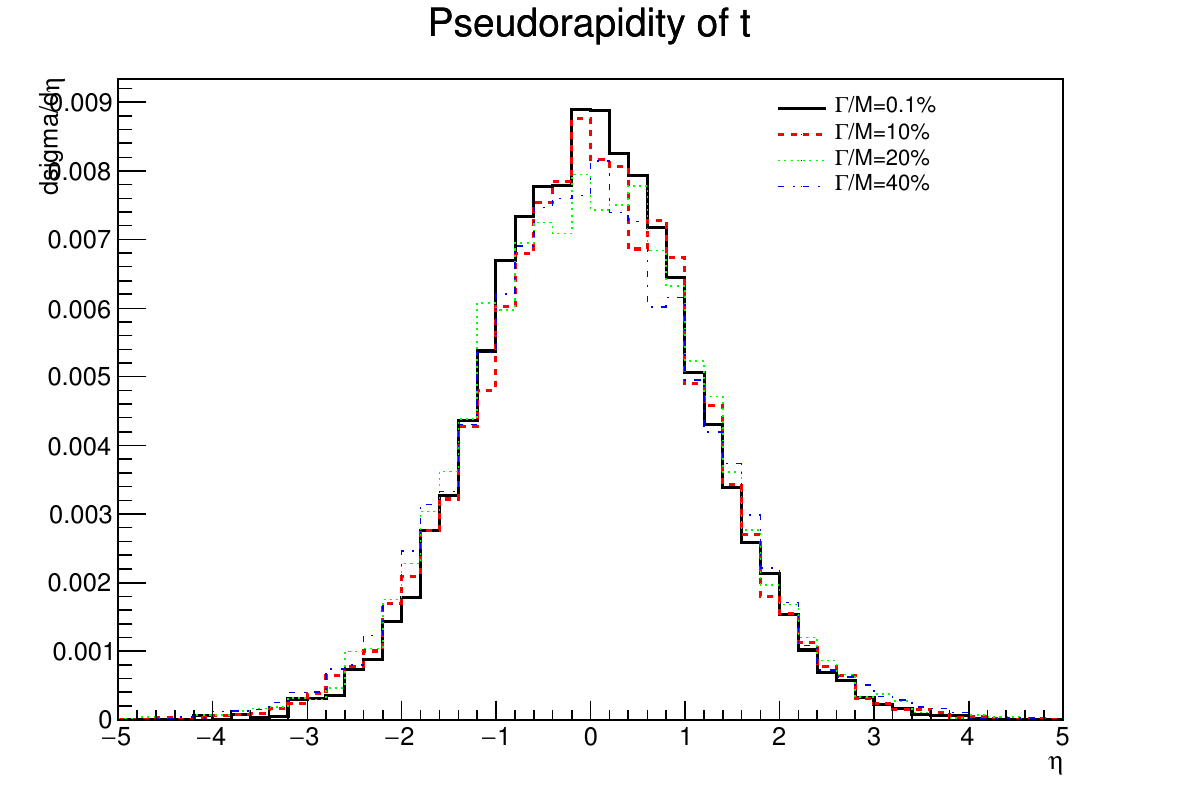}
\includegraphics[width=.45\textwidth]{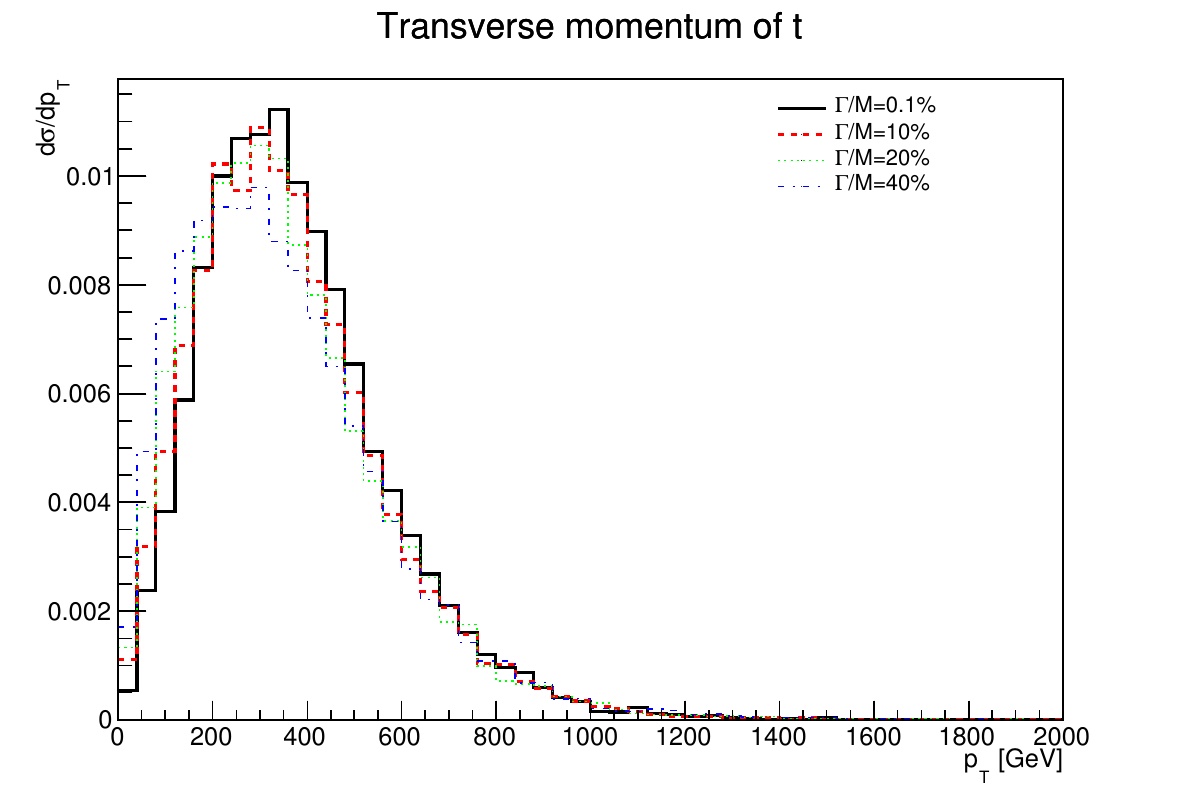}\\
\includegraphics[width=.45\textwidth]{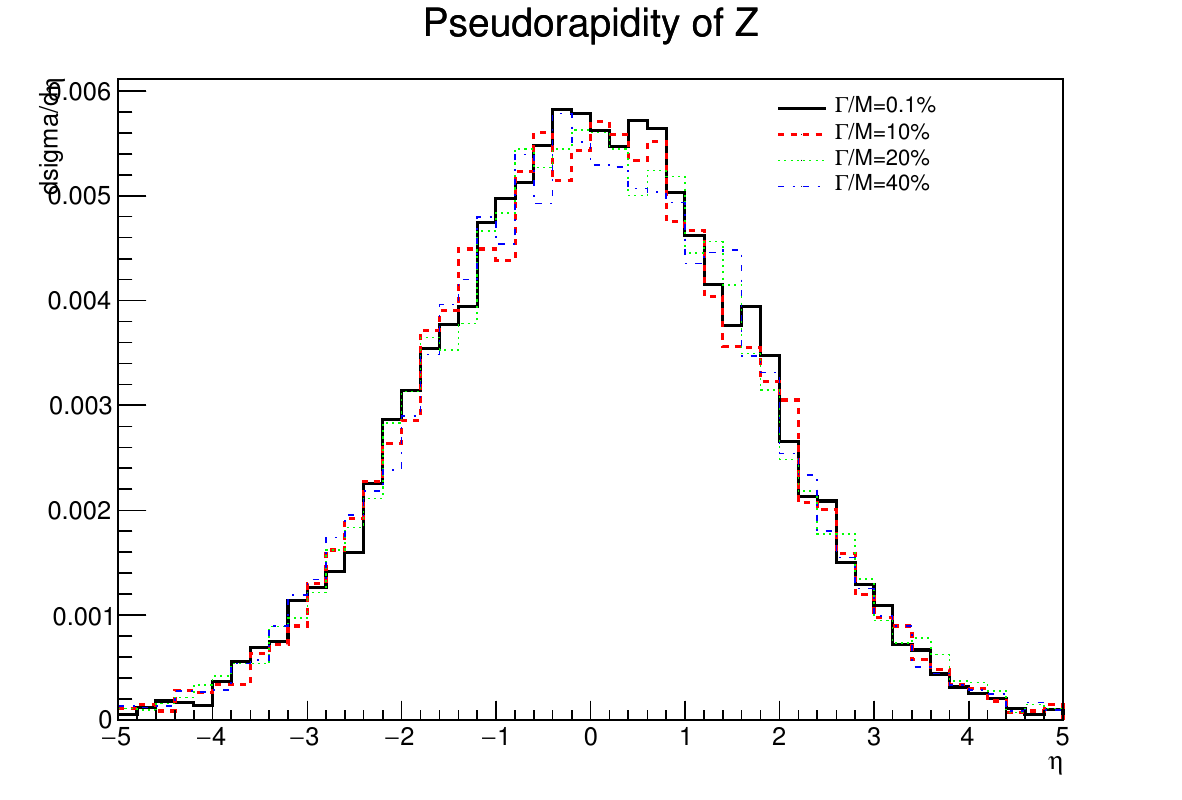}
\includegraphics[width=.45\textwidth]{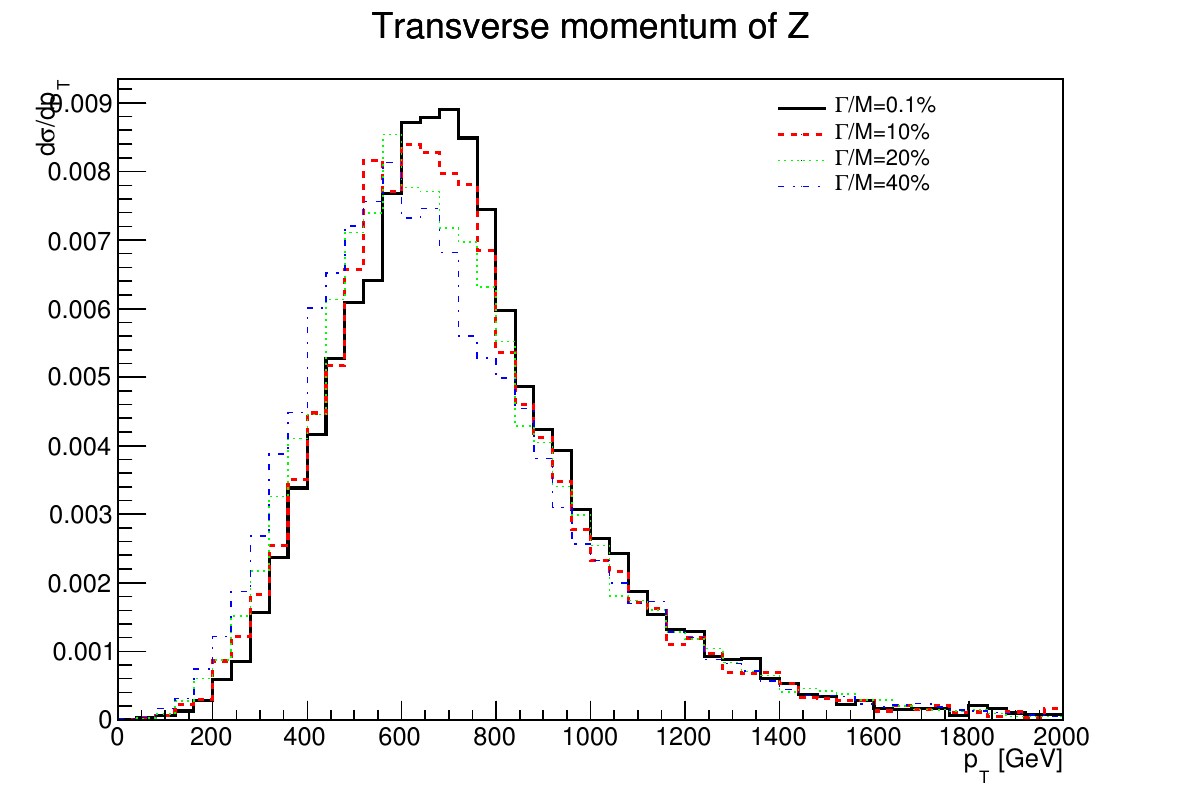}
\caption{\label{fig:distztzt}Partonic level differential cross-sections for the $ZtZt$ channel. From left to right and top to bottom: $\eta_t$, $p_{Tt}$, $\eta_Z$ and $p_{TZ}$. All distributions correspond to a $T$ mass of 800 GeV, for which $\sigma_S\sim\sigma_X$ almost independently of the $T$ width.}
\end{figure}

For some channels it is possible to notice a cancellation of effects which makes the QCD pair production cross-section similar to the cross-section including off-shell contributions even for large values of the width. The cancellations appear at different values of the $T$ mass, depending on the channel and for processes involving the bottom quark in the final state they are partially masked by the large increase of the cross-section due to the collinear divergences caused by topologies where the bottom quarks arise from gluon splitting, as the one shown in Fig.~\ref{fig:fullsignaltopologies}. Such cancellations are due to the different scaling of phase space between the large and narrow width regimes. Indeed, if the $T$ VLQ has a large width, the transferred momentum of the process can have values in a larger range than in the NWA case, where it is constrained by the resonant production of the $T$ pair: this means in turn that the PDFs are sampled at different scales and therefore the cross-section receives a non-trivial mass- and width-dependent contribution which results in the observed behaviour. Of course, this does not necessarily mean that the NWA approximation can be used along the cancellation regions. Sample kinematical distributions of the decay products of the $T$ in different width regimes are shown in Fig.~\ref{fig:disththt} for the $HtHt$ channel and $M_T=600$ GeV and in Fig.~\ref{fig:distztzt} for the $ZtZt$ channel and $M_T=800$ GeV. In both cases, while the $\eta$ distribution does not change significantly as the width increases, the $p_T$ distributions exhibits a visible shift towards the softer region. 

\subsection{Interference with SM background}

When considering processes of pair production of heavy quarks in the NWA, interferences with the SM background are zero by construction, but if the width of the heavy quark is large, it is crucial to explore the relevance of interference terms in the determination of the total number of events. Moreover, understanding this contribution for regions which are not usually explored in experimental analyses may be useful in the determination of sets of kinematical cuts for the optimisation of future searches, if any hint of a VLQ with large width  appears in the data.

\begin{figure}[H]
\centering
\includegraphics[width=.3\textwidth]{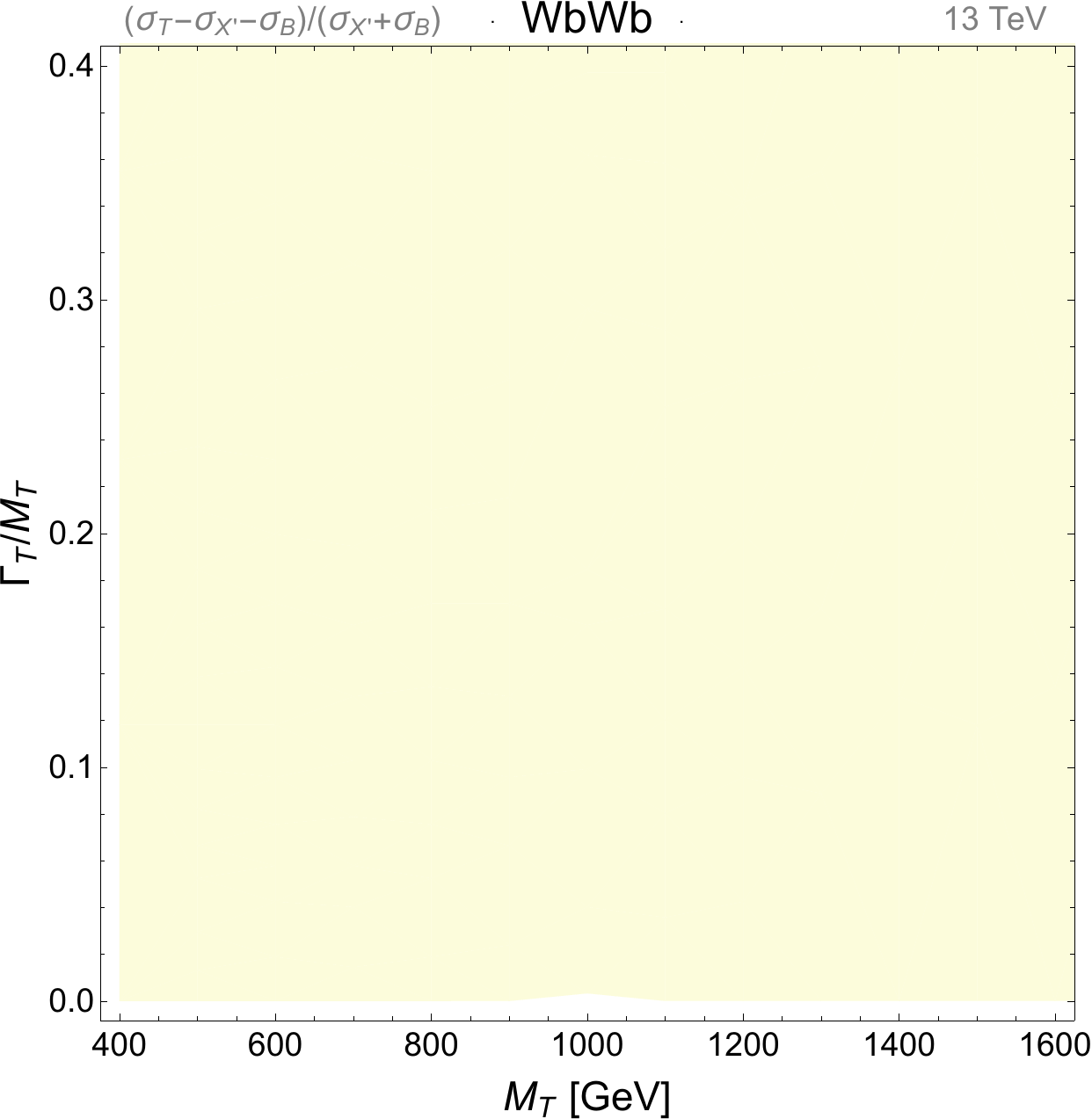}
\includegraphics[width=.3\textwidth]{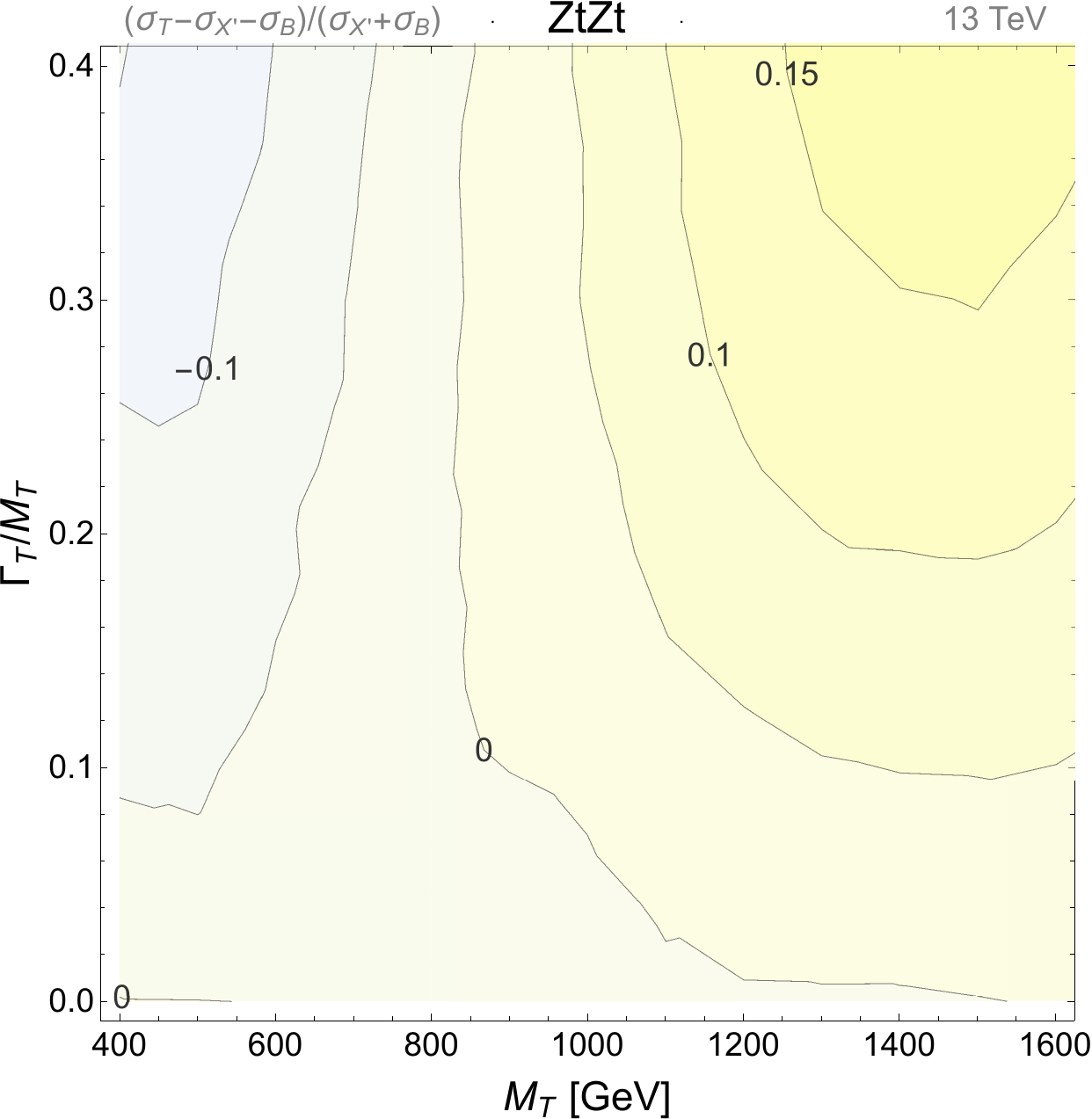}
\includegraphics[width=.3\textwidth]{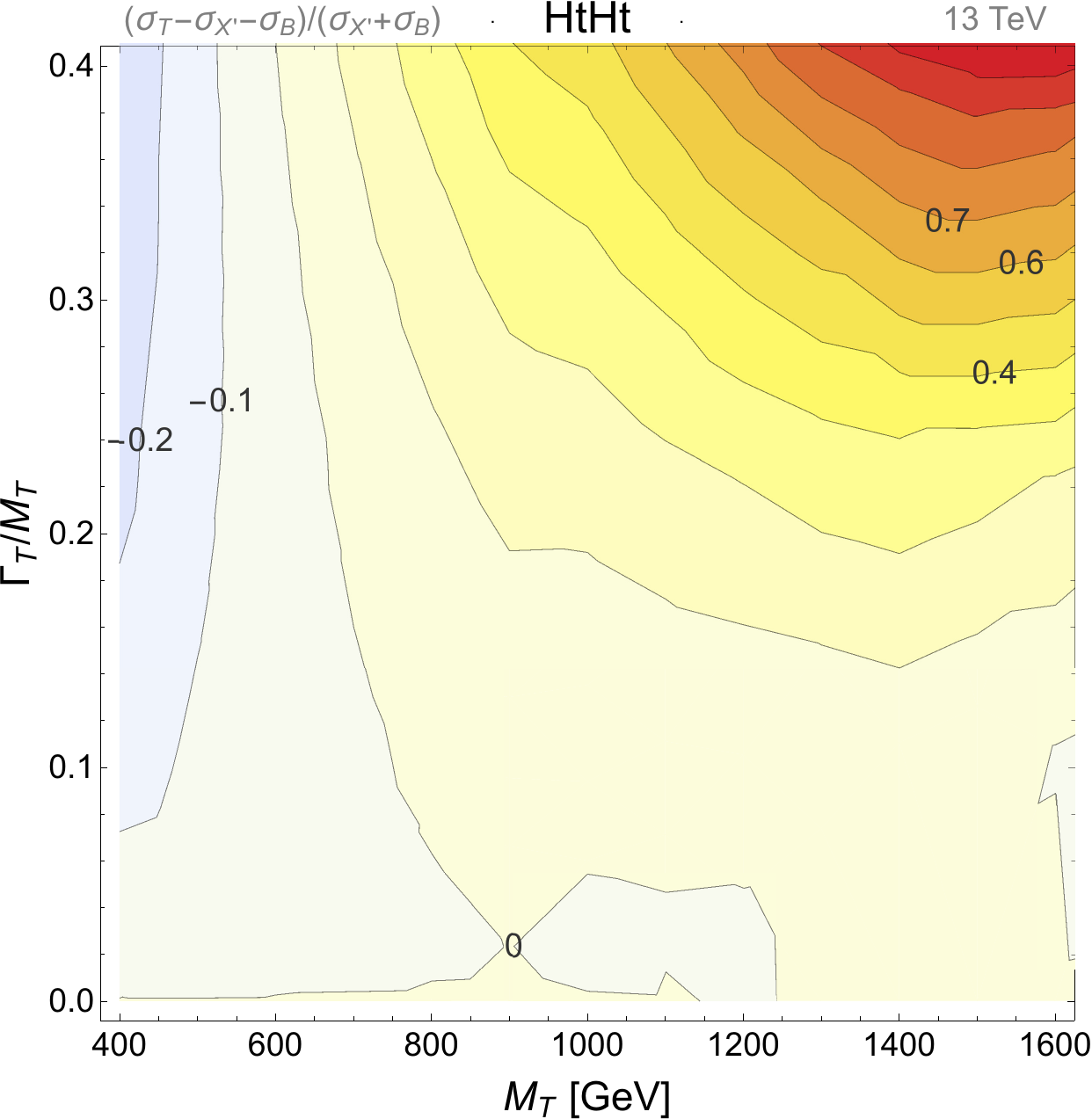}\\
\includegraphics[width=.3\textwidth]{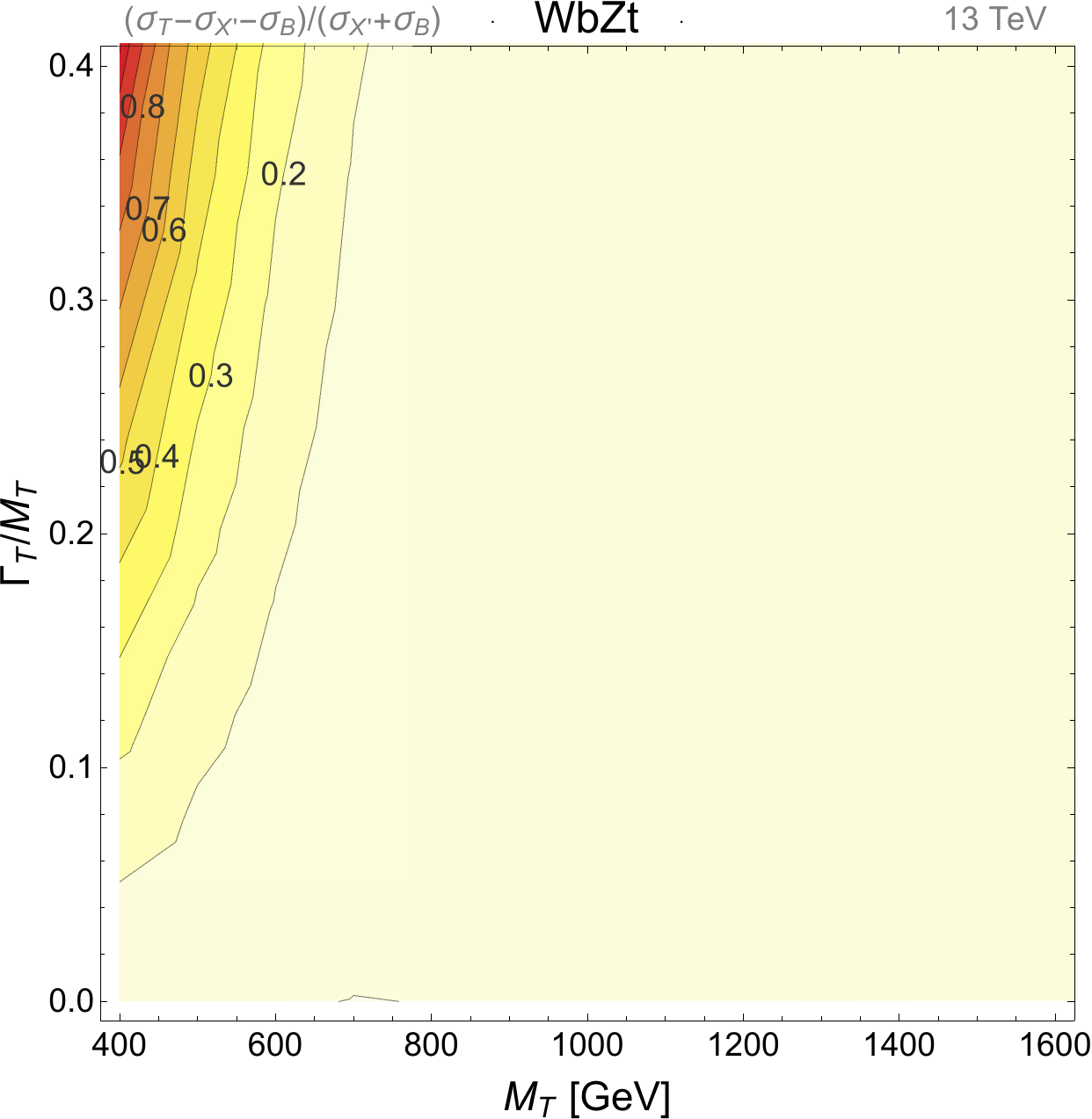}
\includegraphics[width=.3\textwidth]{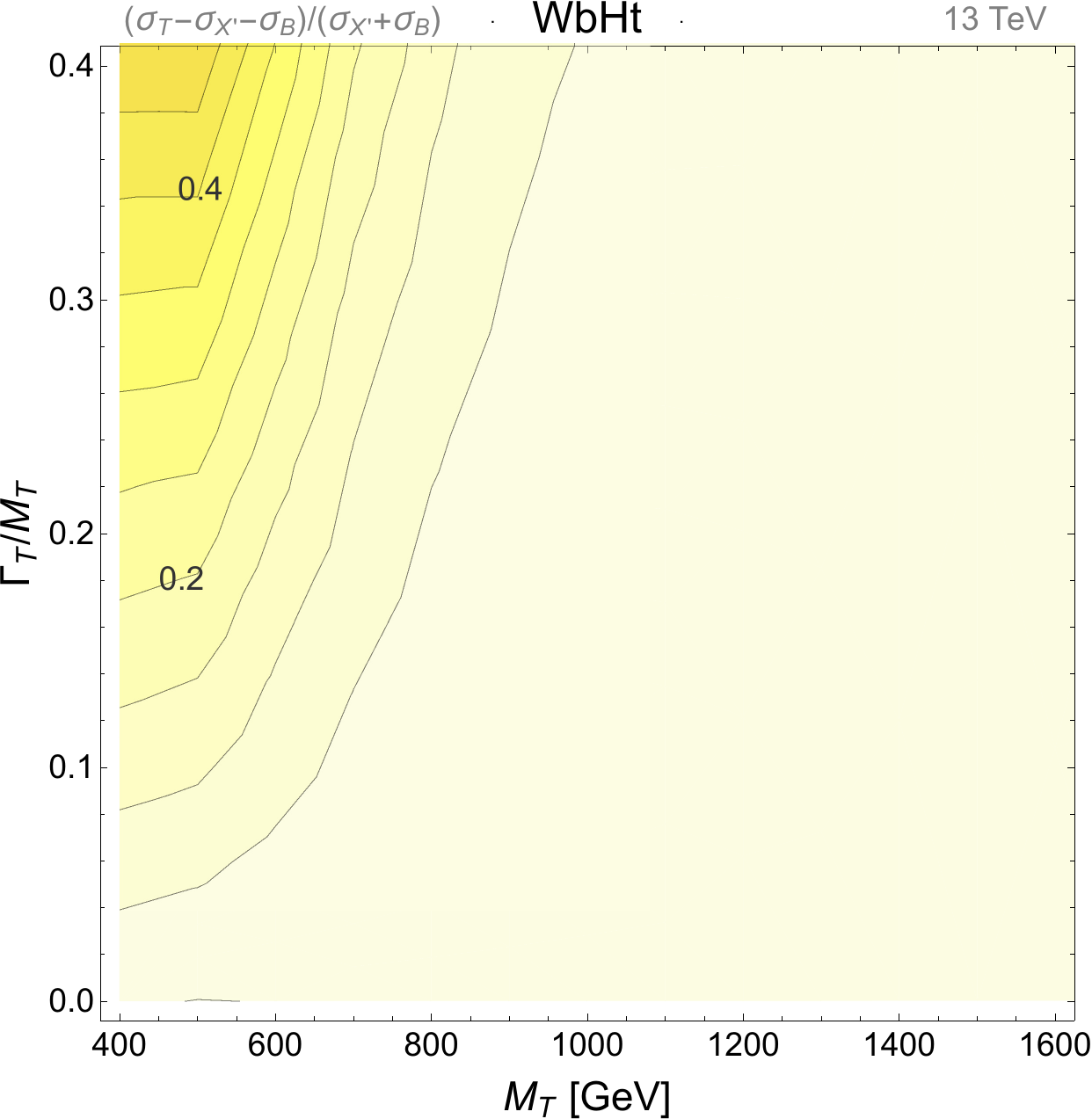}
\includegraphics[width=.3\textwidth]{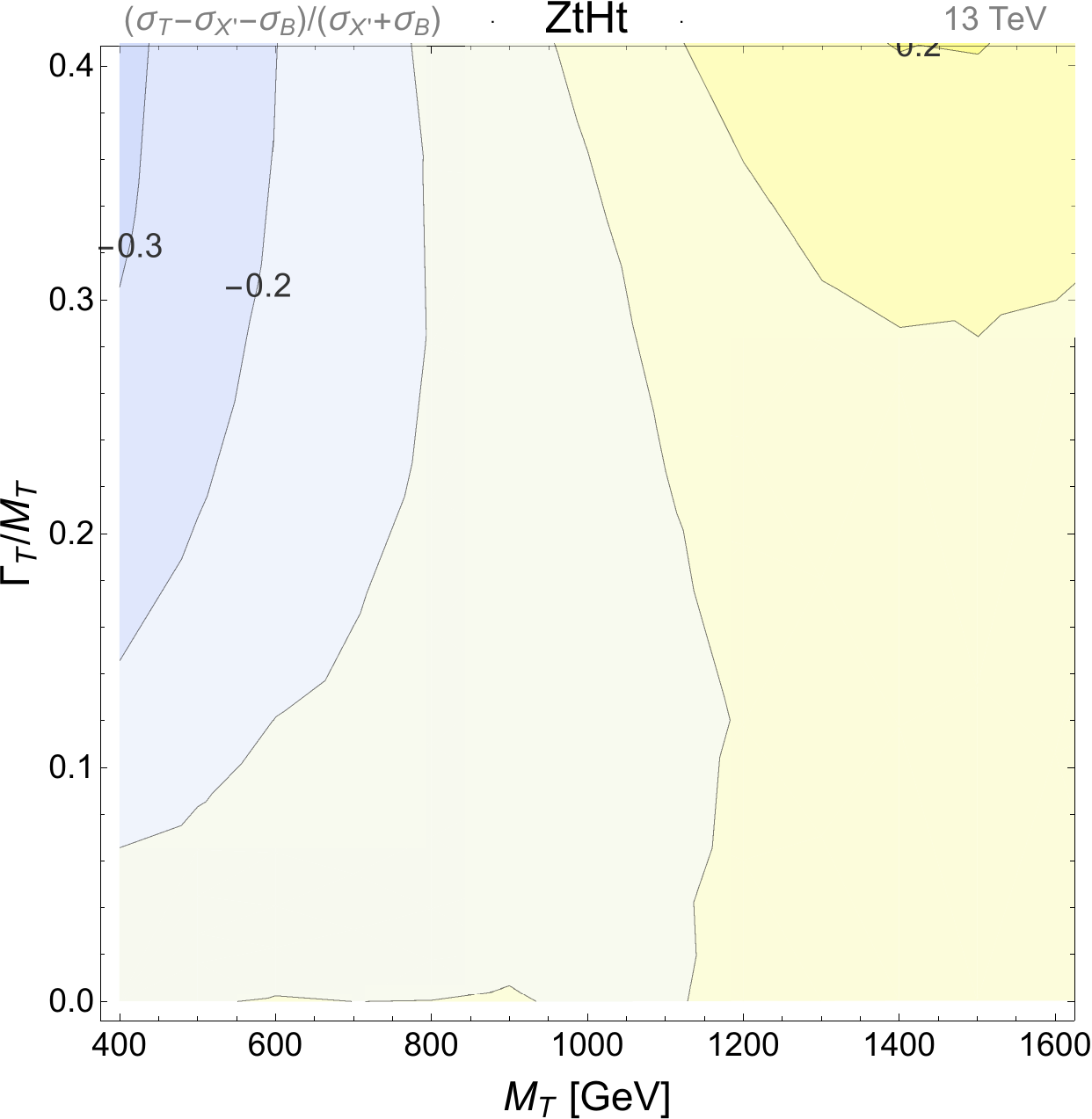}
\caption{\label{fig:TXBthird}Relative difference in cross-section between the total $2\to4$ process, including the SM background and the sum of QCD pair production and SM backgrounds. Top row: final states on the diagonal of the matrix in Eq. \ref{eq:finalstates} (third generation mixing); bottom row: off-diagonal final states (third generation mixing).}
\end{figure}

The correction factor between the total cross-section and the sum of NWA pair production and SM backround cross-section is plotted in Fig.~\ref{fig:TXBthird}. Such correction factors depend on the relative weight of the SM background contribution in the determination of the total cross-section: they are almost negligible in the whole parameter space where the background is the dominant contribution to the total signal, while they are become larger where the new physics signal has a more relevant role. This can easily be understood by considering what affects the various terms of the ratio. Herein,  $\sigma_B$ is a constant term (for fixed final state), $\sigma_X$ only depends on the $T$ mass and $\sigma_T$ is the only term which depends on the both the $T$ mass and width. For the $WbWb$ case, however, $\sigma_T$ is almost entirely dominated by the SM background contribution (mostly by the top pair production process) and therefore the contribution of the $T$ is just a small correction, which does not produce relevant effects in the whole range of masses and widths we have explored. For the $ZtZt$ and $HtHt$ scenarios, on the contrary, the SM background is comparable or negligible with respect to the signal contribution, and therefore the dependence on the $T$ mass and width is much more evident.

\begin{figure}[H]
\centering
\includegraphics[width=.3\textwidth]{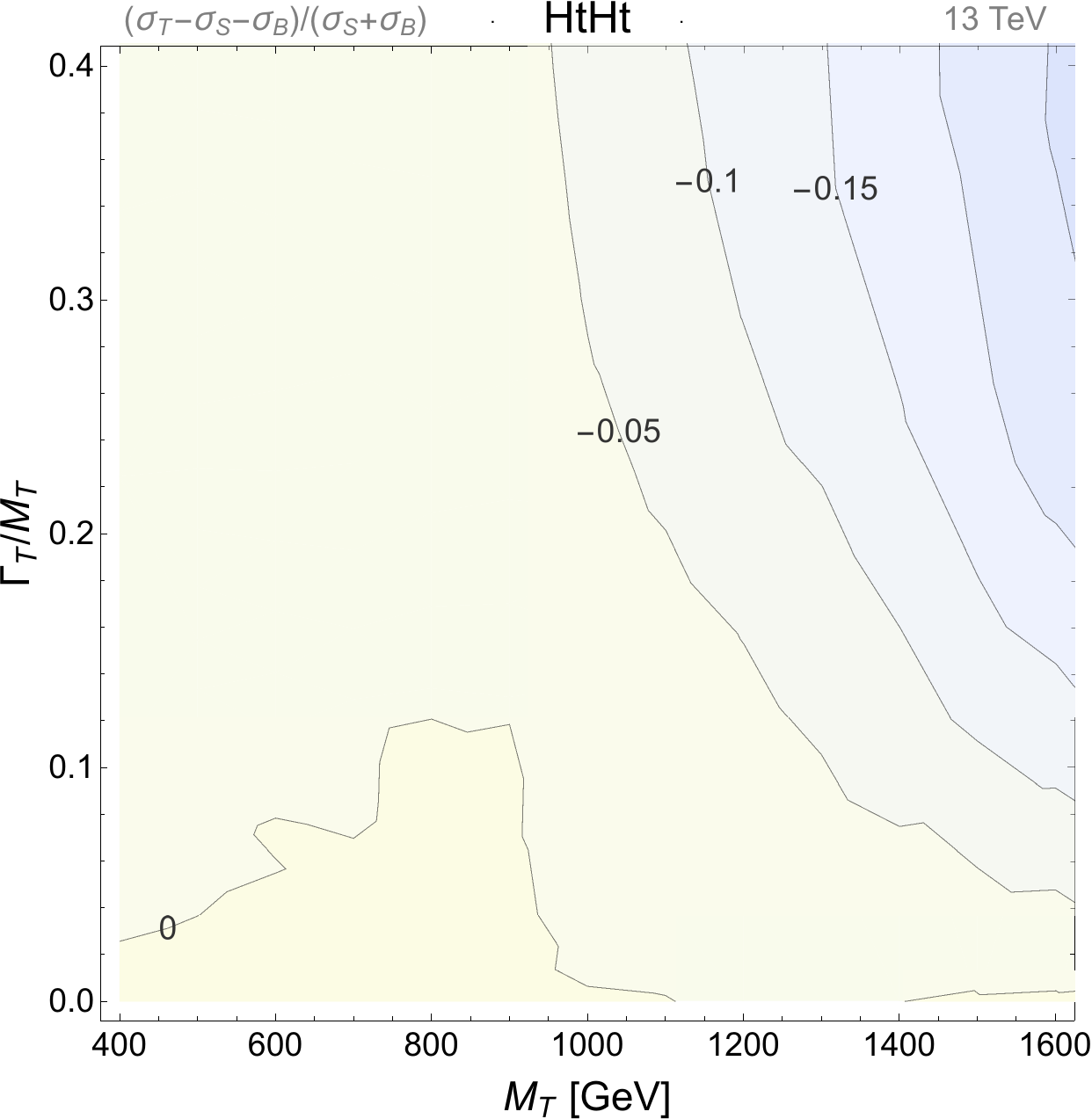}
\caption{\label{fig:TSBthird}Relative contribution of the interference between the full signal and the SM background. $HtHt$ is the only channel for which this contribution can reach values above 10\% in size.}
\end{figure}

The full contribution of interference terms, considering the full signal instead of the signal in the NWA, is always numerically negligible. In Fig.~\ref{fig:TSBthird} we have shown the only channel for which the contribution can become larger than 10\% in absolute value. The inclusion of single-resonance effects, therefore, changes the picture in a substantial way, showing that interference effects between the full signal and the SM background are always negligible, except for the $HtHt$ channel in the large width and large $M_T$ region. This has to be expected because the kinematical properties of signal and background are usually different. However, this can only be seen by taking into account the full signal contribution. This means that, if searches for VLQs with large width are designed, considering the full signal instead of rescaling the NWA results would almost in any case automatically kill any contribution from interference with the SM background, especially for scenarios where the SM background is large. 

\subsection{Results at detector level} \label{sec:detector3}

In this section we will study the performance of 8 TeV and 13 TeV searches from both ATLAS and CMS in determining the excluded region in the $\{M_T,\Gamma_T/M_T\}$ plane. We will consider only final states in the diagonal of the matrix of Eq.~\ref{eq:finalstates} because non-diagonal final states would not represent, by themselves, physically valid scenarios. Such final states arise only if the VLQ has non-zero {\rm BR}s in different channels, and a consistent treatment would require the combination of diagonal and off-diagonal final states together. As stated above, the purpose of this study is not to set limits, but to study the performance of experimental searches in regions yet unexplored for these scenarios. Indeed, the set of searches we consider are not necessarily optimised for the discovery of VLQs at the LHC, therefore our recast bounds are not likely to be competitive with current bounds for pair production of VLQs in the NWA, and, in this respect, we will not compare our results with other bounds from direct searches for pair production of VLQs.\\

\begin{figure}[H]
\centering
\includegraphics[width=.3\textwidth]{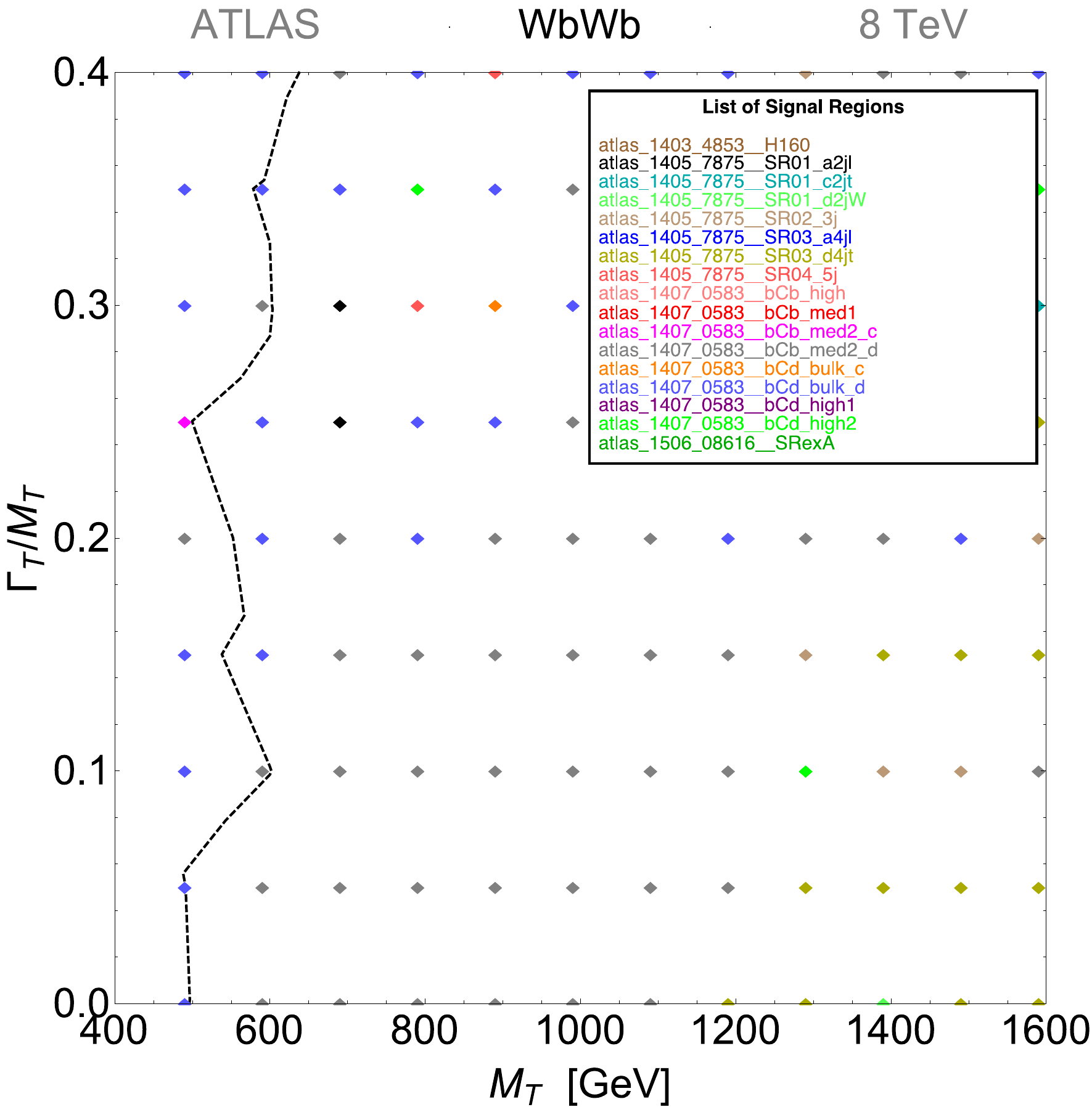}
\includegraphics[width=.3\textwidth]{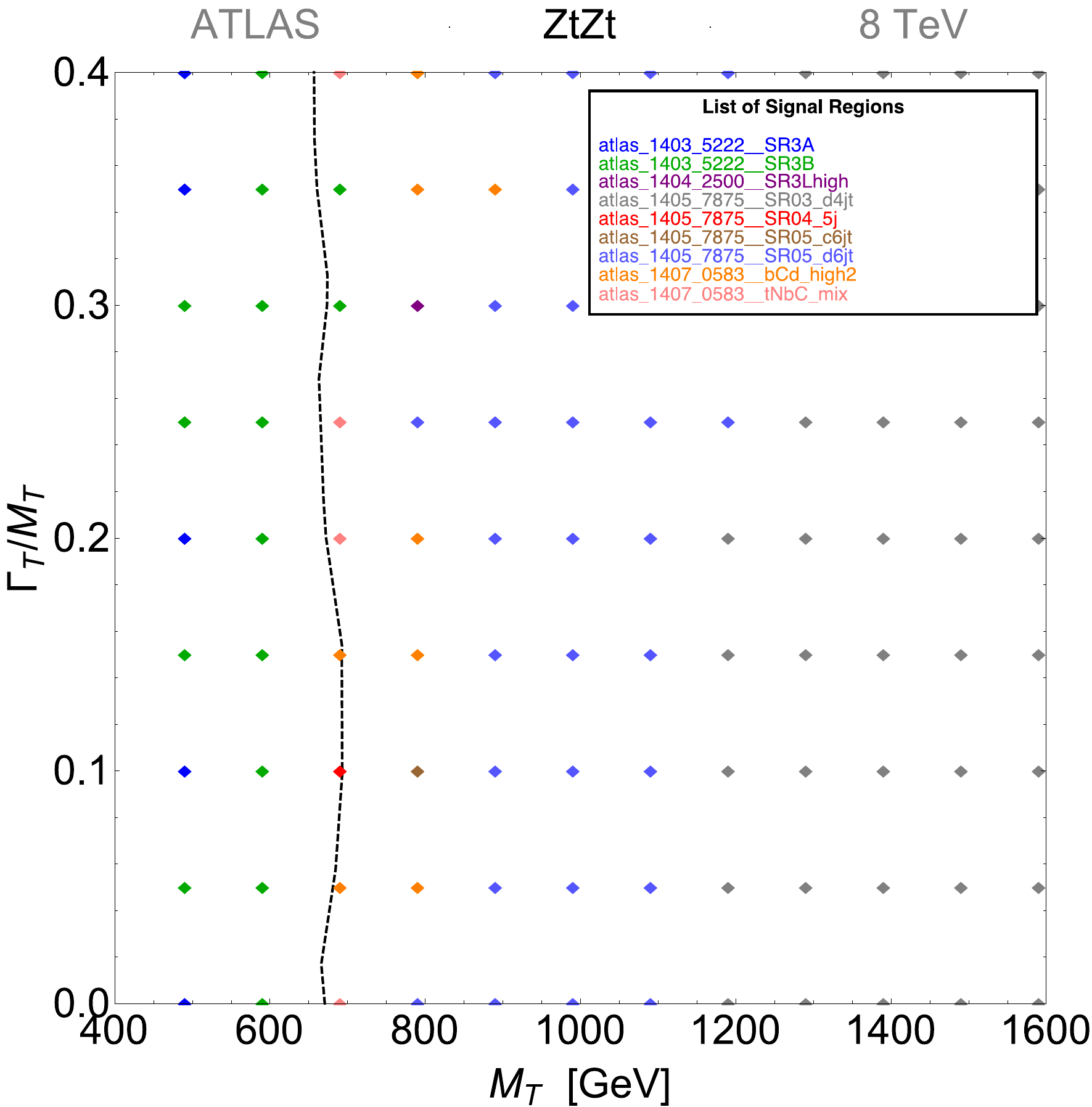}
\includegraphics[width=.3\textwidth]{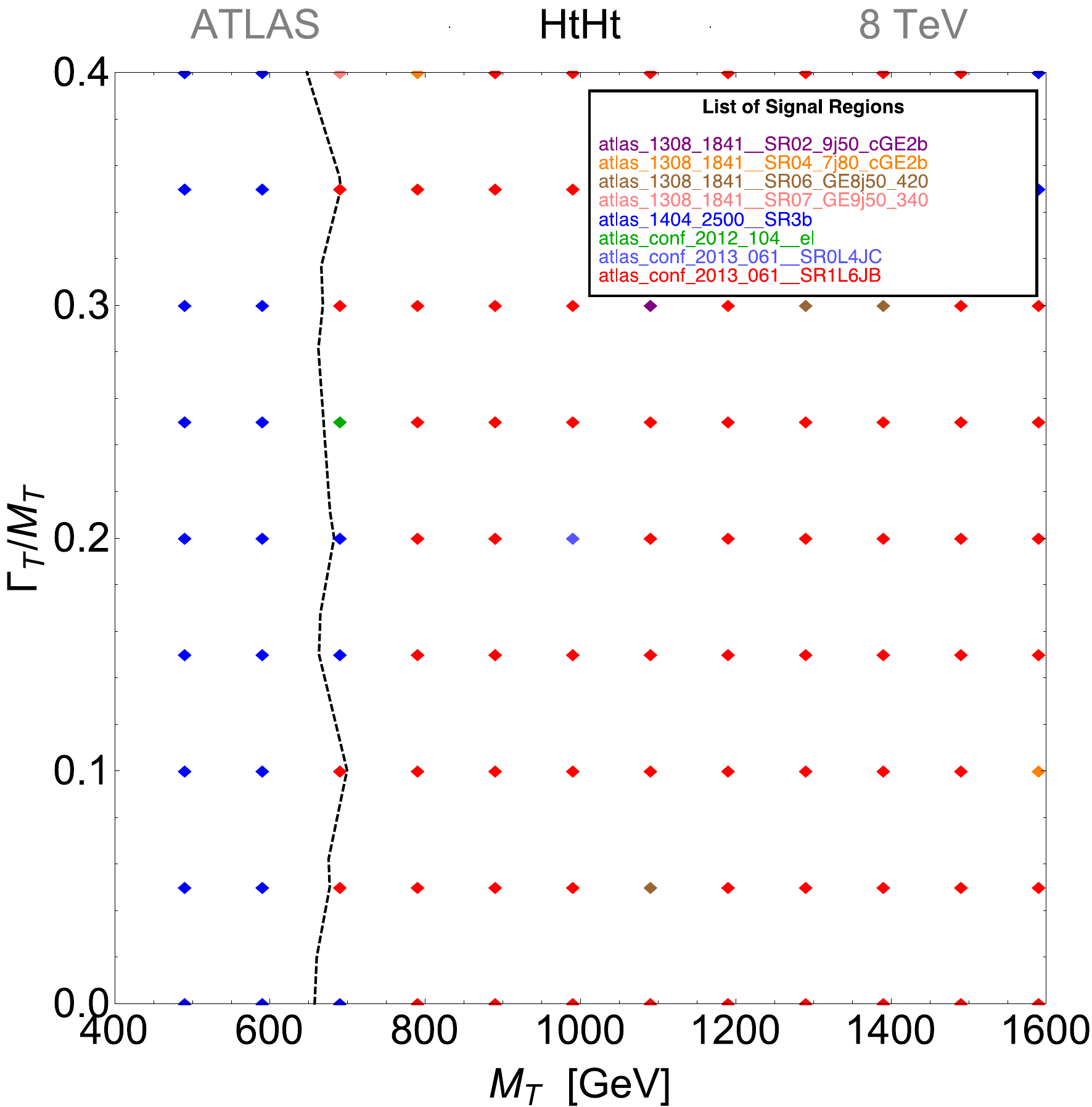}\\
\includegraphics[width=.3\textwidth]{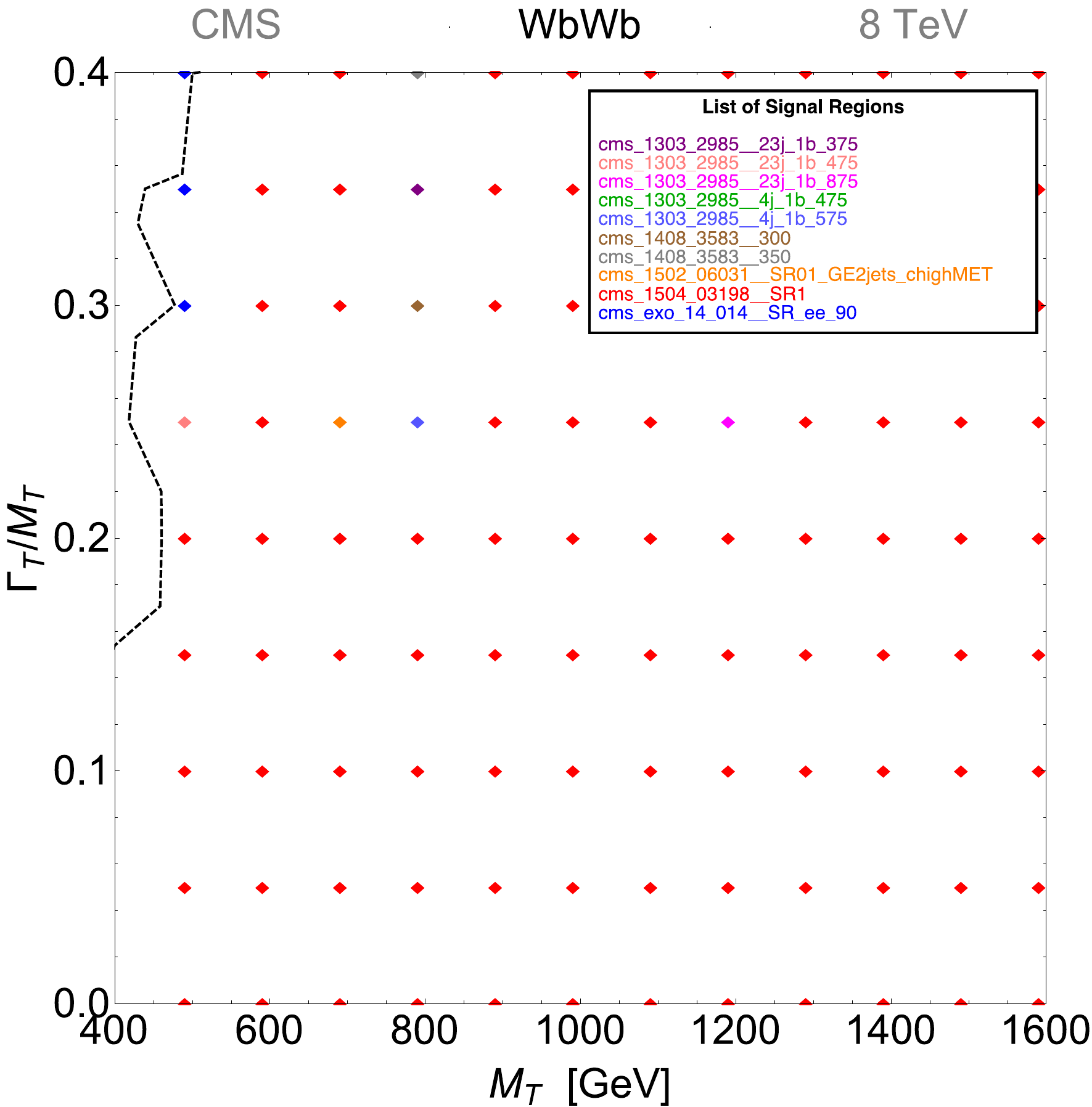}
\includegraphics[width=.3\textwidth]{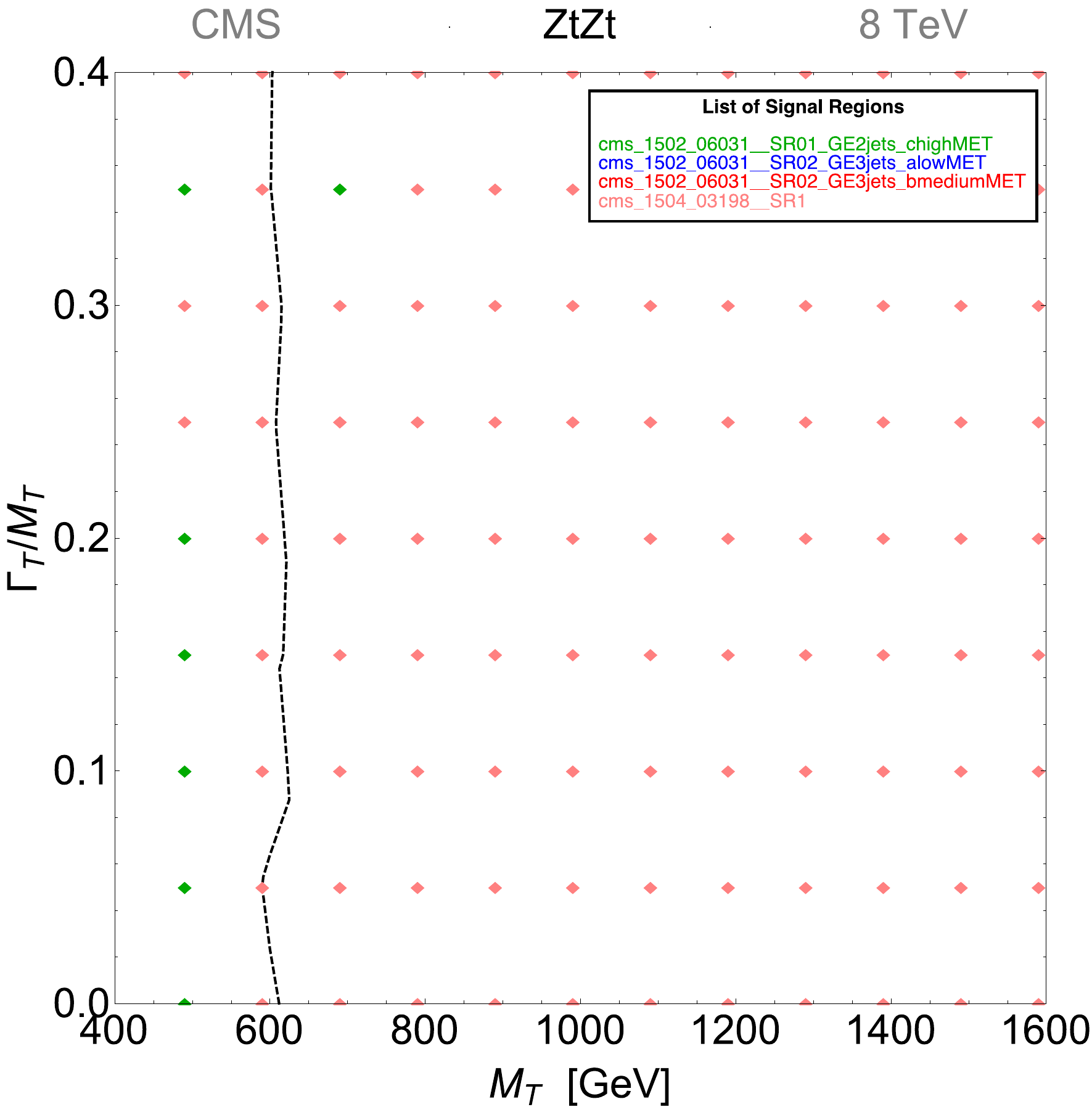}
\includegraphics[width=.3\textwidth]{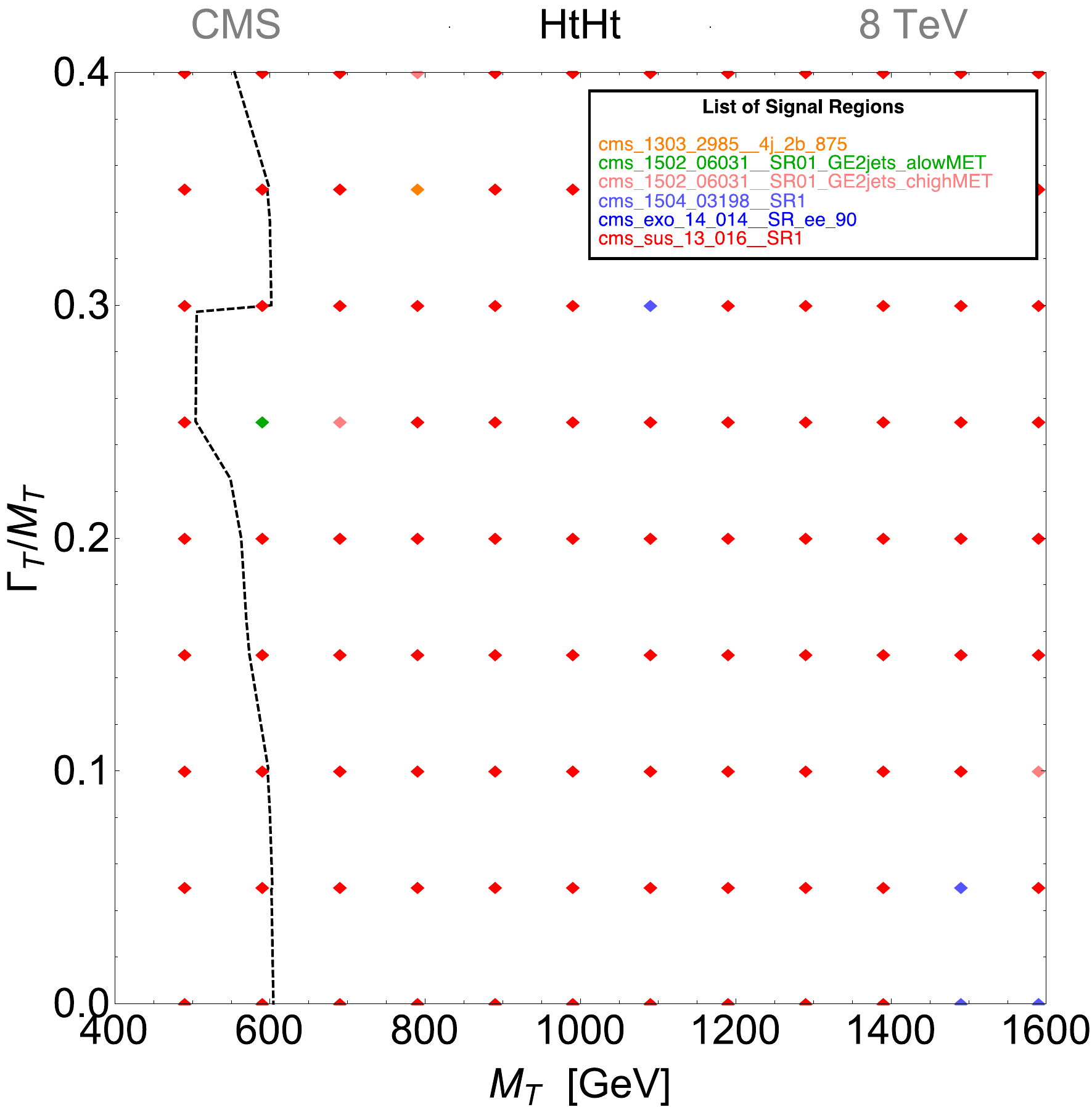}
\caption{\label{fig:Detector8TeV3rdgen} Recast bounds in the ($M_T, \Gamma_T/M_T$) plane with a set of ATLAS (top row) and CMS (bottom row) searches at 8 TeV for diagonal final states.}
\end{figure}

We show in Fig.~\ref{fig:Detector8TeV3rdgen} the exclusion lines for combinations of 8 TeV searches from both ATLAS and CMS for the three diagonal final states compatible with pair production and decay of $T$ VLQs. Our results show that none of the signal regions in the considered searches is sensitive to the large width scenarios: the exclusion bound are, for all final states, analogous to the NWA limit.

\begin{figure}[H]
\centering
\includegraphics[width=.3\textwidth]{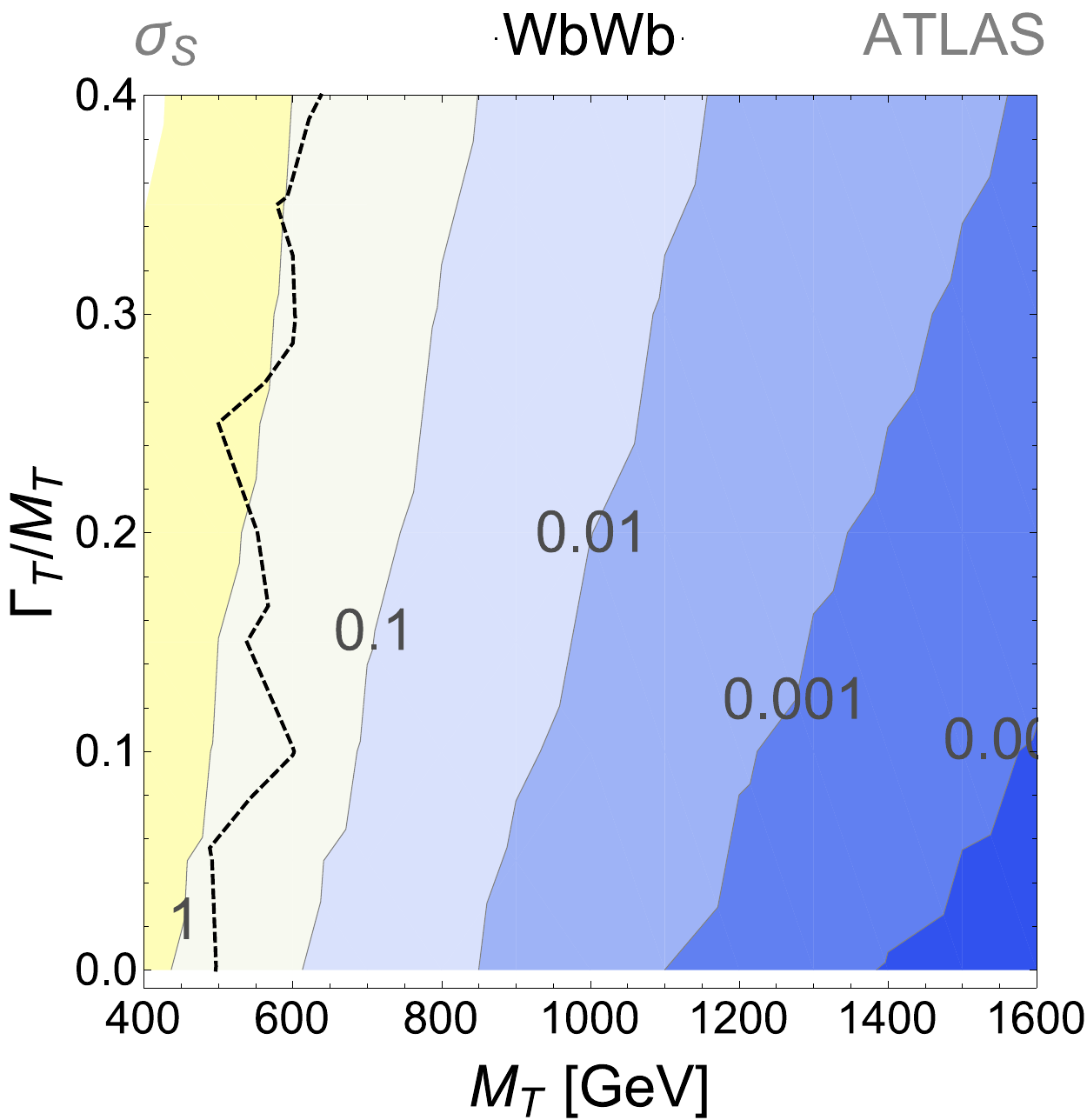}
\includegraphics[width=.3\textwidth]{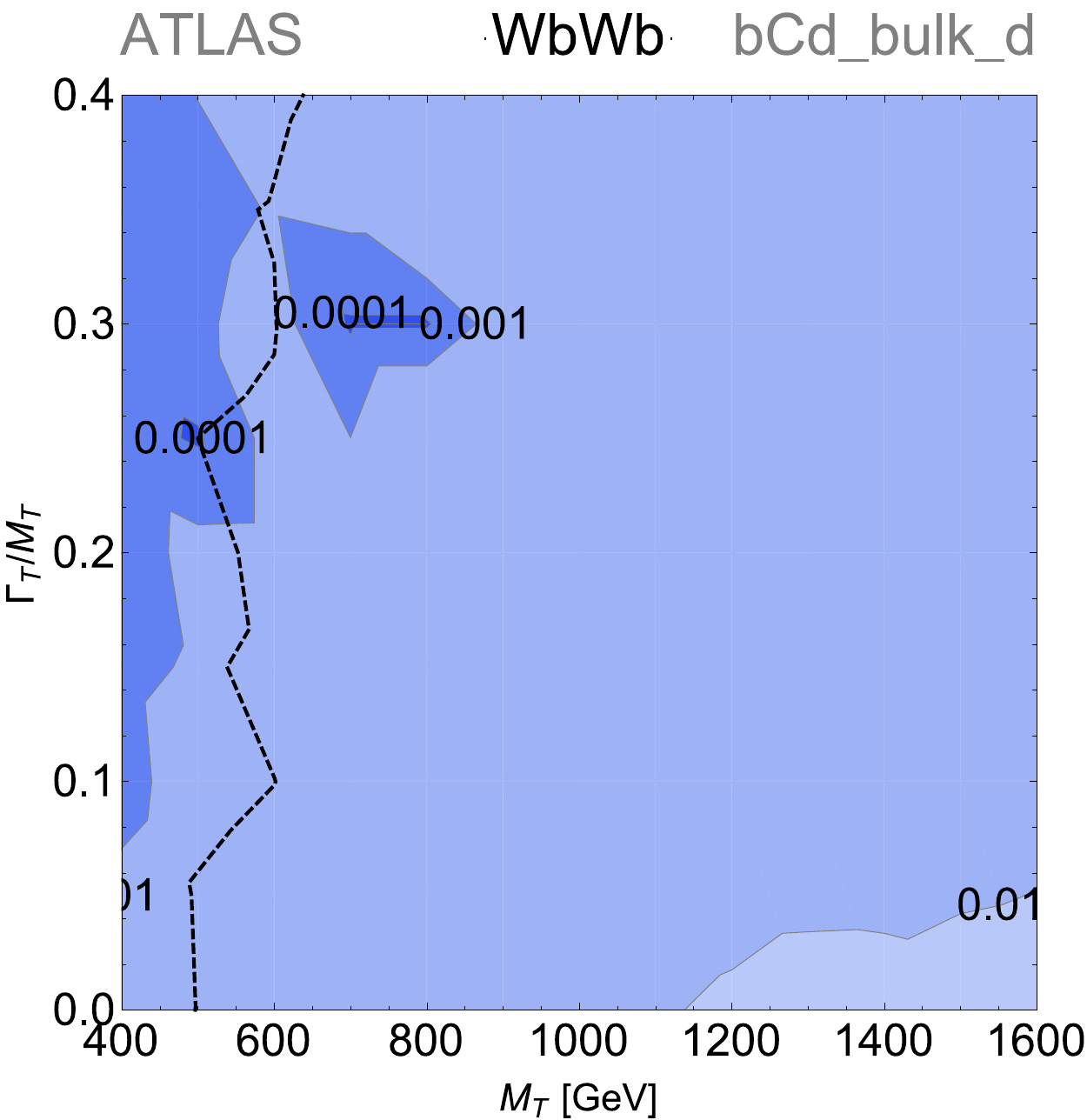}
\caption{\label{fig:Combined8TeV3genWbWb} Cross-section and efficiency of the best ATLAS SR (bCd\_bulk\_d of \cite{Aad:2014kra}) for the $WbWb$ channel, compared with the bound.}
\end{figure}

This can be understood by considering the cross-section of the full signal, $\sigma_S$, and the dependence on the $T$ width of the efficiencies of the Signal Regions (SRs) which is most marked near the bounds. In Fig.~\ref{fig:Combined8TeV3genWbWb} we superimpose the bound from the combination of ATLAS searches at 8 TeV with the cross-section of the full signal for the $WbWb$ channel (the others are qualitatively similar): the dependence on the width of the cross-section is weak in the region where the searches fix the exclusion limit, and becomes slightly stronger for higher (allowed) masses. Moreover, the variation of the kinematics of the final states is not large enough to increase the sensitivity of the search cuts, as can be seen by looking at the efficiency of the the SR  bCd\_bulk\_d of the ATLAS search~\cite{Aad:2014kra}, which depends rather weakly on the width of the $T$. 
                                                                                                 
\begin{figure}[H]
\centering
\includegraphics[width=.3\textwidth]{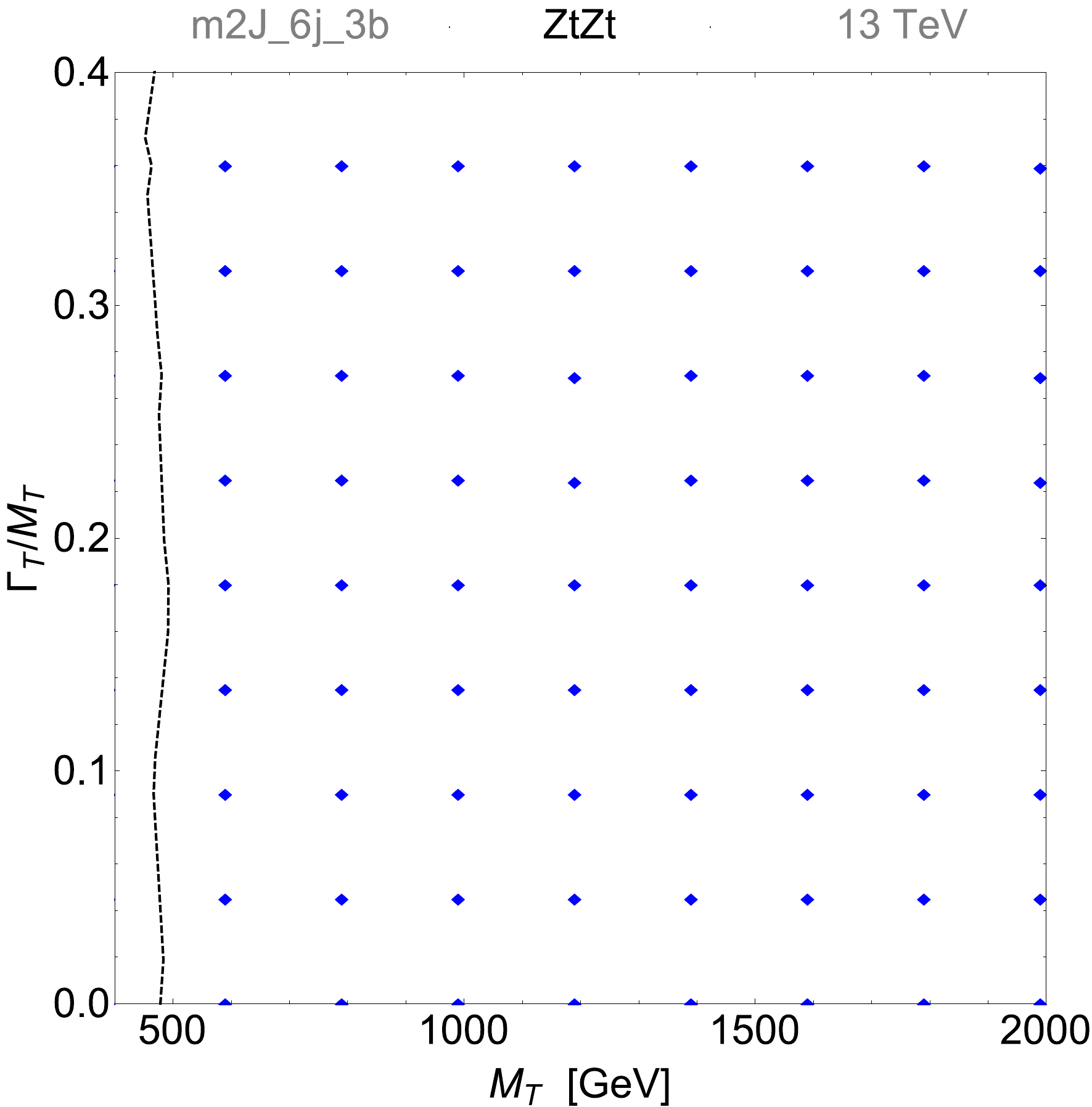}
\includegraphics[width=.3\textwidth]{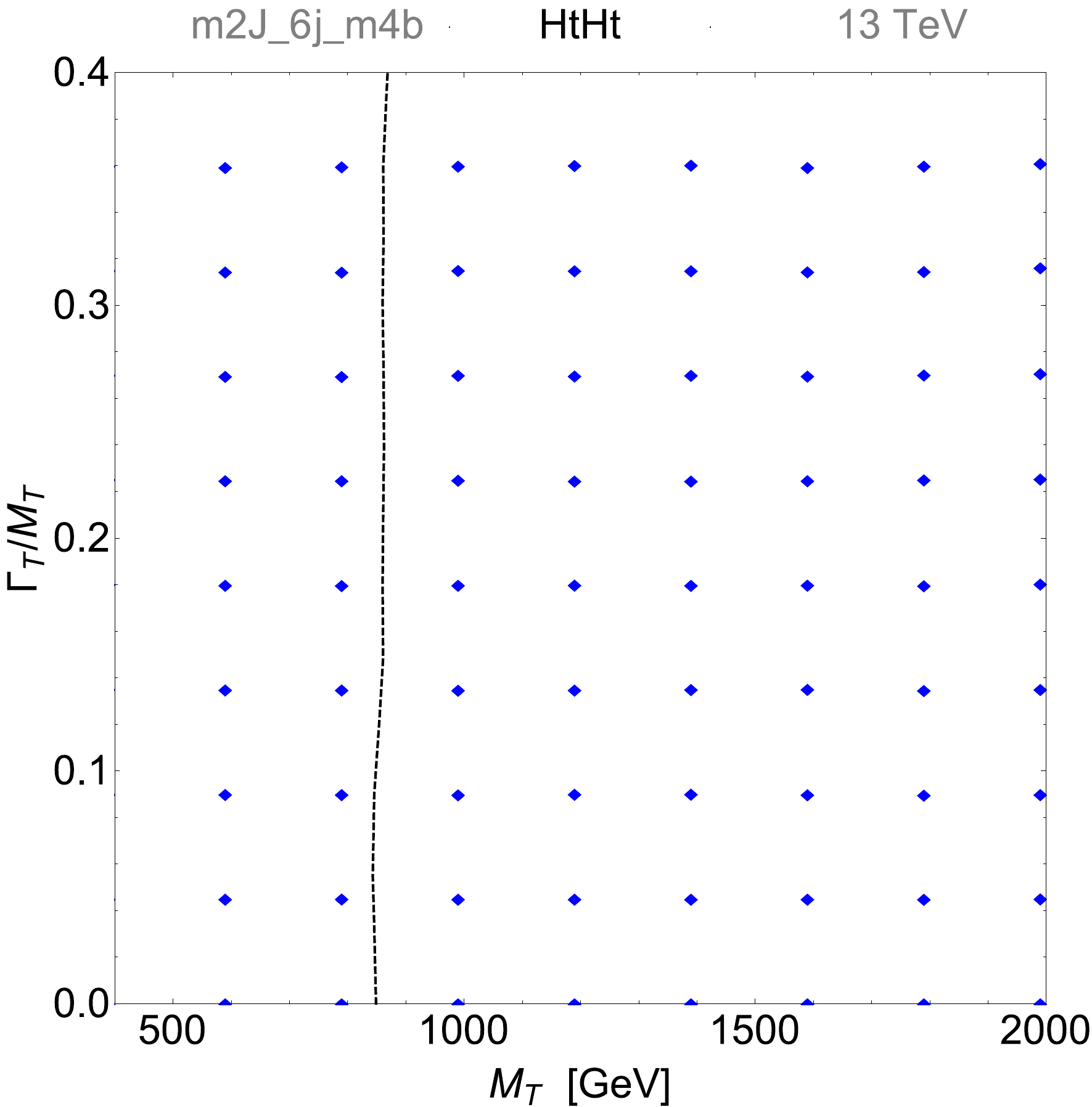}
 \caption{\label{fig:Detector13TeV3rdgen} Same as Fig.~\ref{fig:Detector8TeV3rdgen} for the ATLAS search at 13 TeV~\cite{TheATLAScollaboration:2016gxs} implemented in CheckMATE. The plot for the $WbWb$ channel is not shown because within the explored range the recasting does not set any limit.}
\end{figure}

Our results at 13 TeV have been obtained considering a dedicated search for pair production of a $T$ VLQ~\cite{TheATLAScollaboration:2016gxs} implemented in CheckMATE. The results exhibit a similar behaviour as the set of 8 TeV ones. Our bounds are rather different from those reported in Ref.~\cite{TheATLAScollaboration:2016gxs}. However, we did not rescale the bounds considering different BRs, as we have not factorised the production from decay, and we are mostly interested in the dependence on the width of such bounds. In this respect, the bounds weakly depend on the $T$ width, as can be seen in Fig.~\ref{fig:Detector13TeV3rdgen}. As for the 8 TeV case, the slight increase in cross-section, and relative deformation of kinematics distribution of the final state objects is compensated by an increase of the efficiencies of the SRs cuts. This information can be exploited for the design of future dedicated searches if the discovery of VLQs with large width are among the goals of the studies.

\section{Extra $T$ quark mixing with first generation SM quarks}

\subsection{Large width effects on the signal at parton level}

\begin{figure}[H]
\centering\includegraphics[width=.6\textwidth]{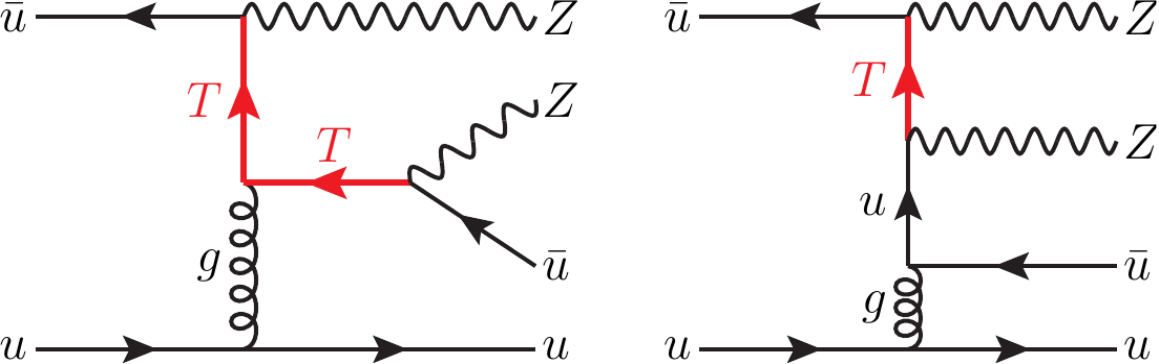}
\caption{\label{fig:firstgentopologies}Examples of neutral-current topologies for heavy quarks with large width mixing with first generation.}
\end{figure}

If the $T$ interacts with first generation SM quarks, topologies where gluons splitting into light quarks increase the cross-section due to collinear enhancements are present also for neutral currents, as shown in Fig.~\ref{fig:firstgentopologies}. In the case of mixing with third generation, such topologies were not present for neutral currents due to the large top mass.

\begin{figure}[H]
\centering
\includegraphics[width=.3\textwidth]{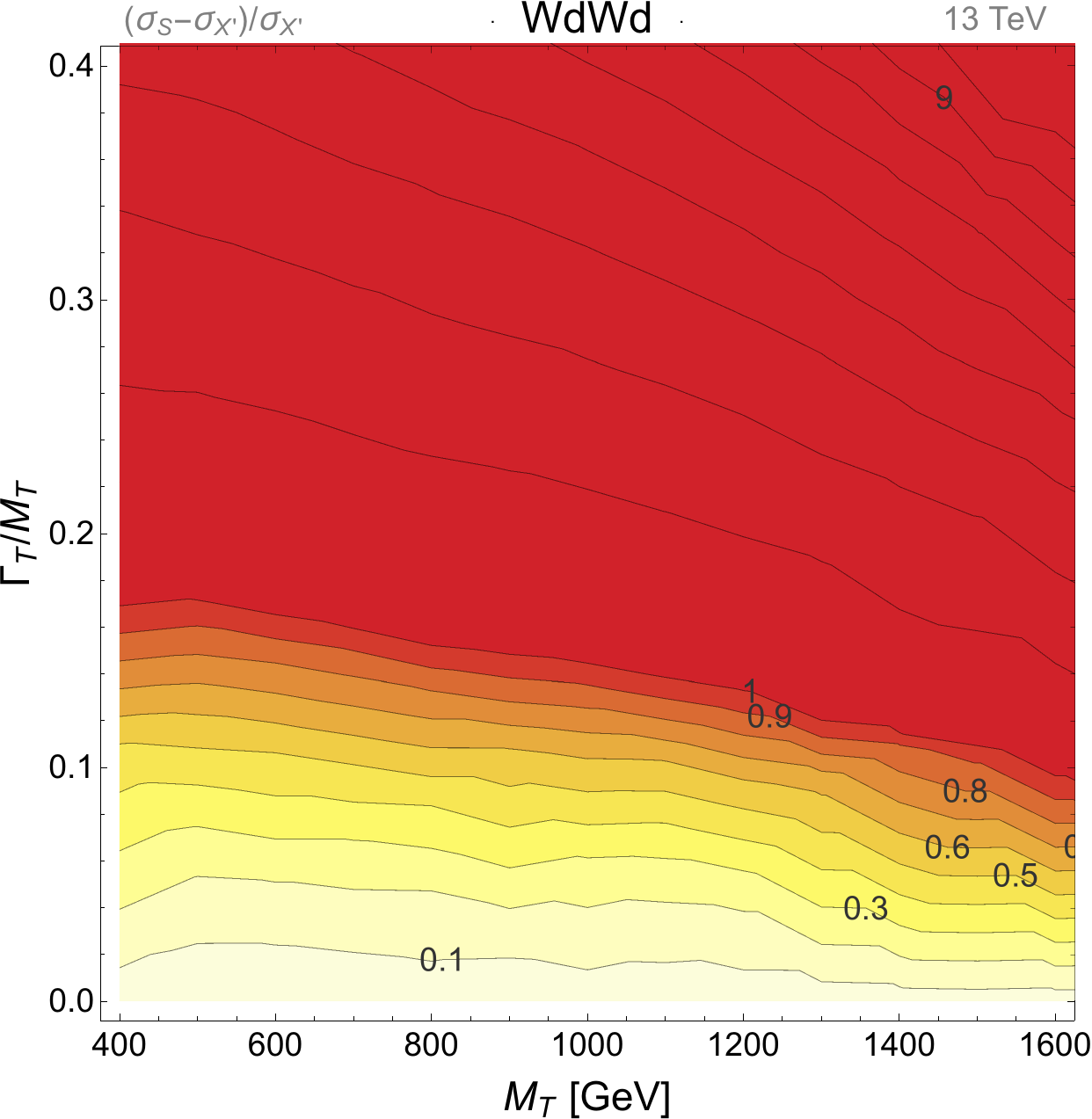}
\includegraphics[width=.3\textwidth]{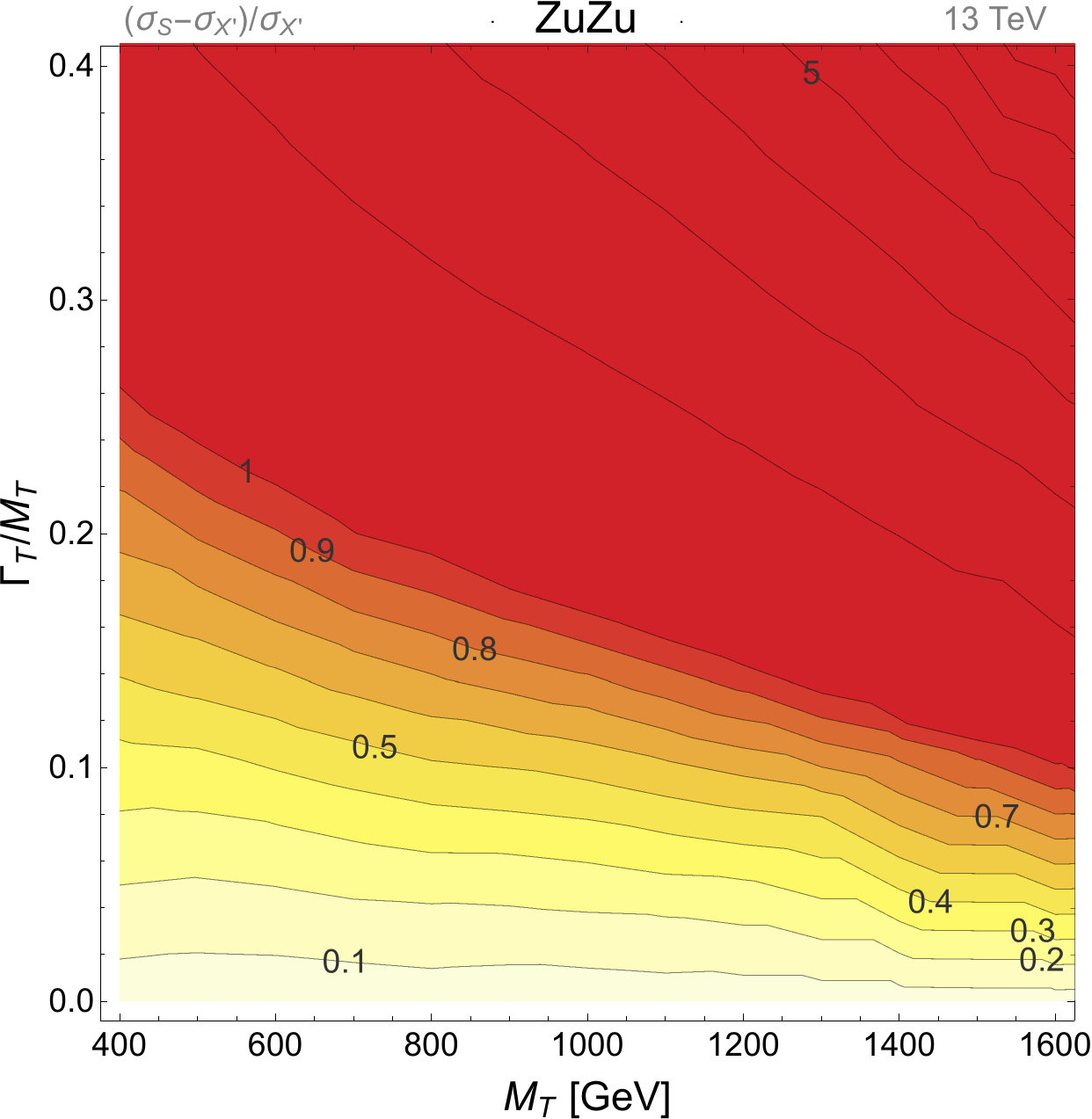} 
\includegraphics[width=.3\textwidth]{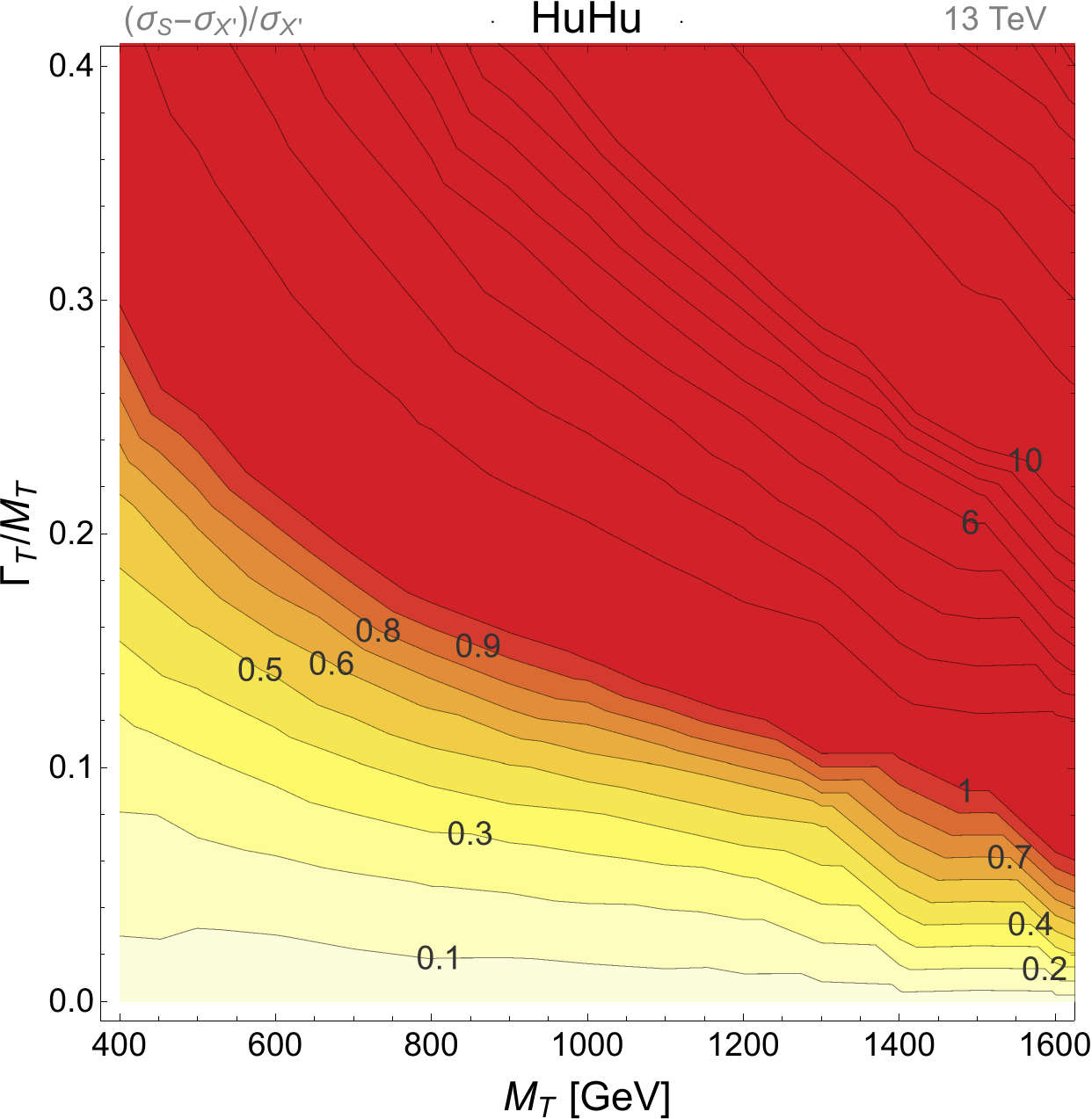}\\
\includegraphics[width=.3\textwidth]{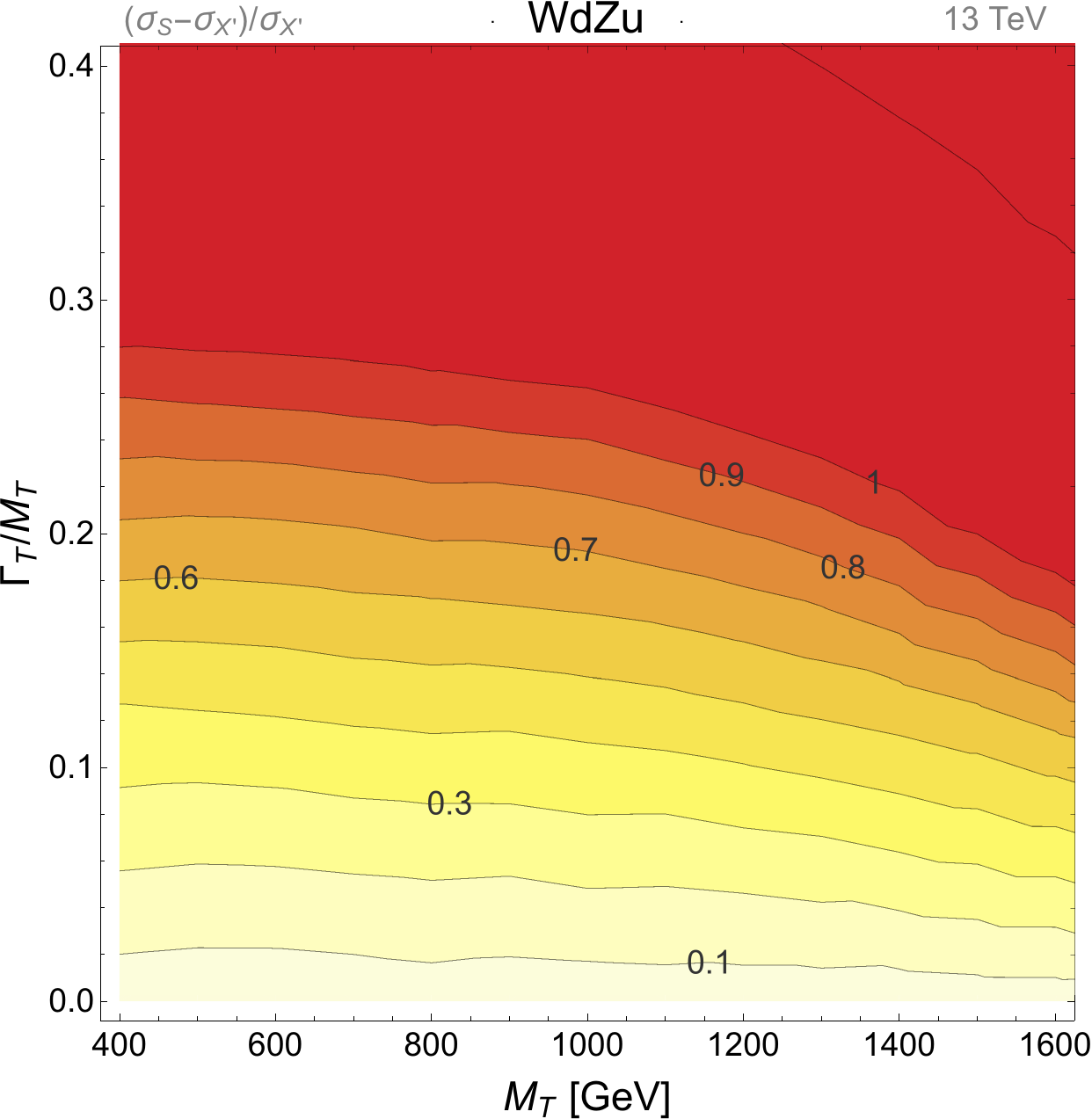} 
\includegraphics[width=.3\textwidth]{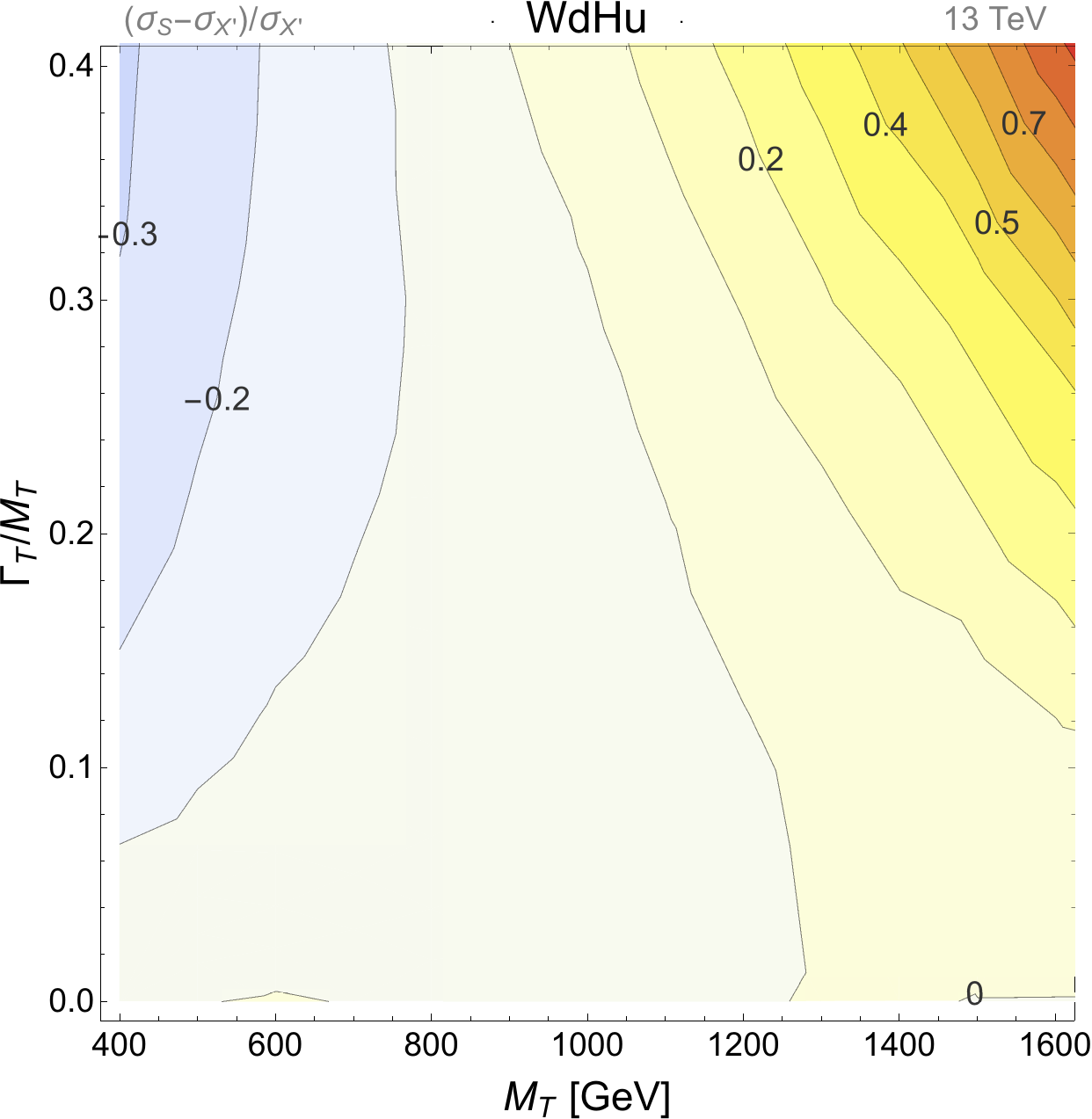} 
\includegraphics[width=.3\textwidth]{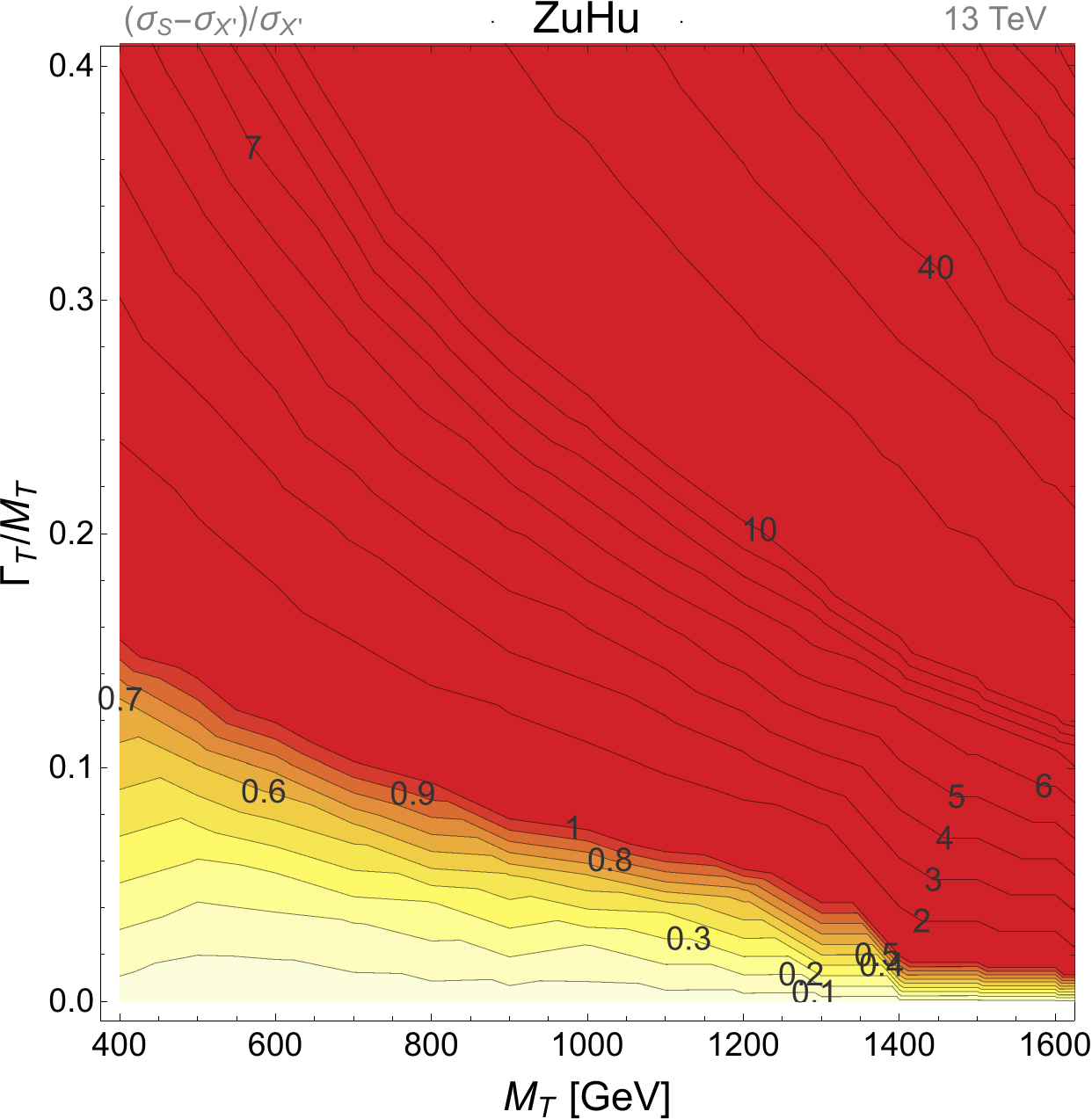} 
\caption{\label{fig:SXfirst}Same as Fig.~\ref{fig:SXthird} for $T$ mixing with first generation.}
\end{figure}

The relative increase of the cross-section with respect to the NWA regime is shown in Fig.~\ref{fig:SXfirst} for an energy of 13 TeV (we have checked that the results at 8 TeV are analogous), where it is possible to notice the large enhancement due to topologies with collinear divergences for all final states.

\subsection{Interference with SM background}

\begin{figure}[H]
\centering
\includegraphics[width=.3\textwidth]{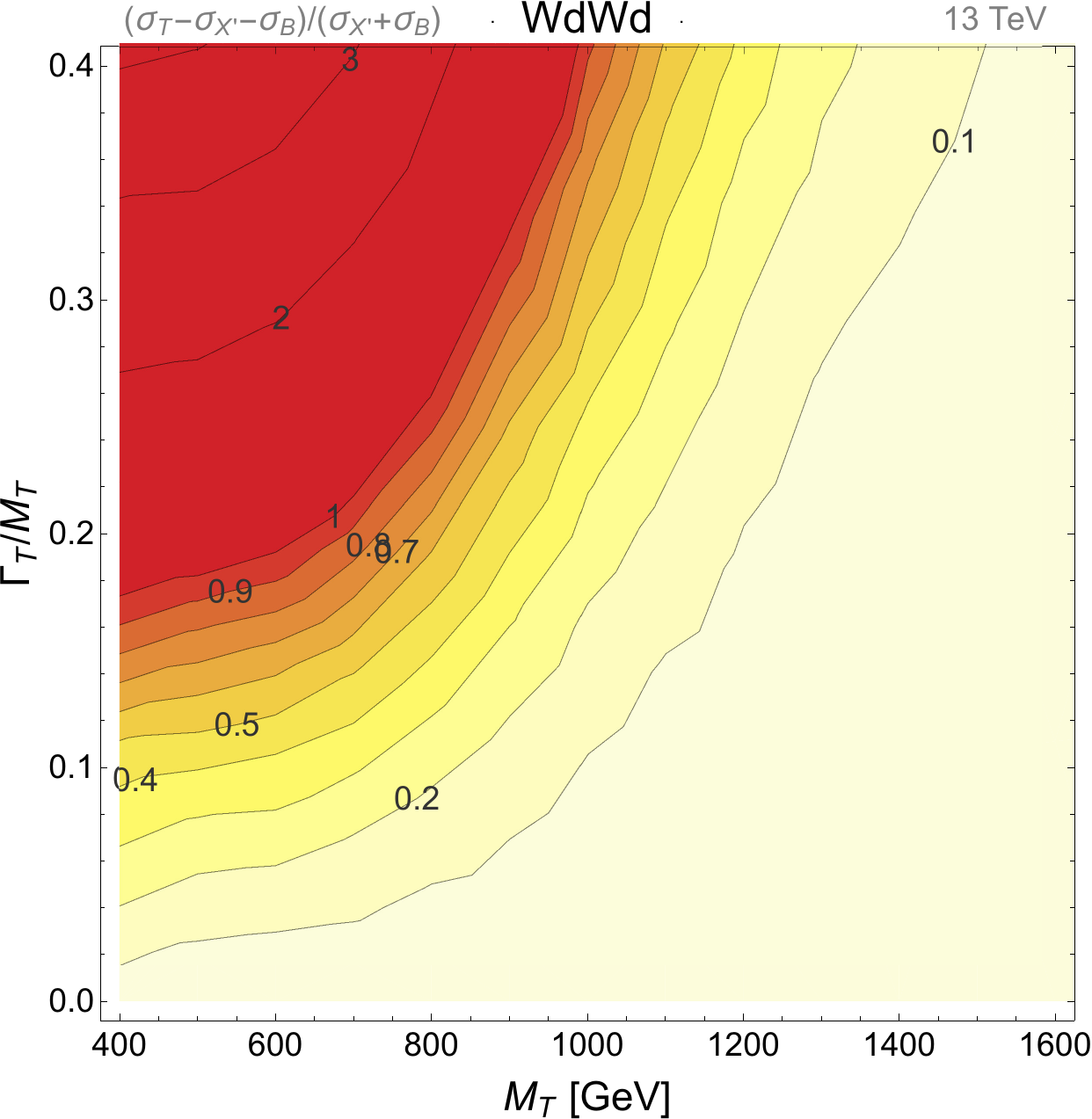}
\includegraphics[width=.3\textwidth]{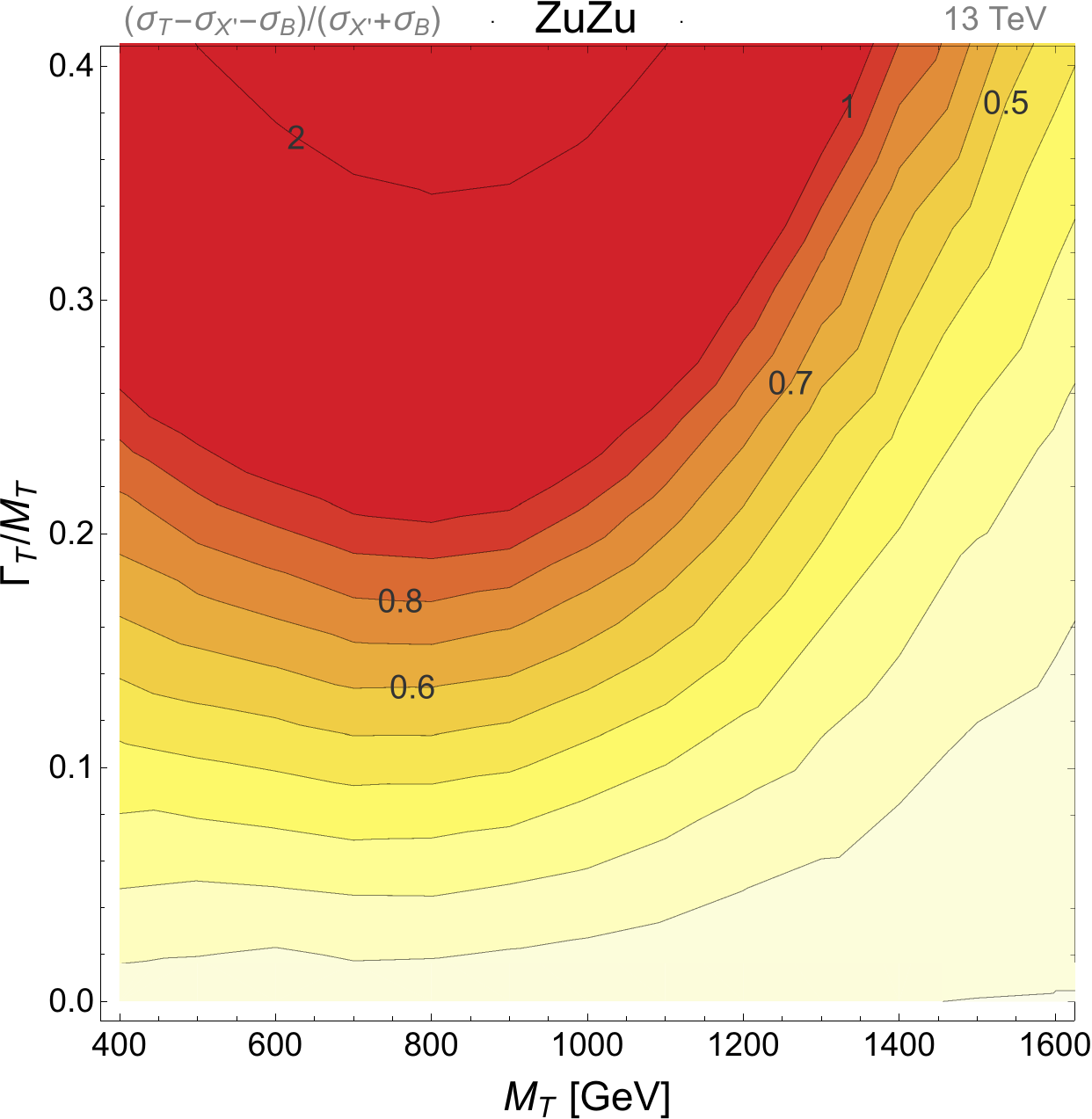}
\includegraphics[width=.3\textwidth]{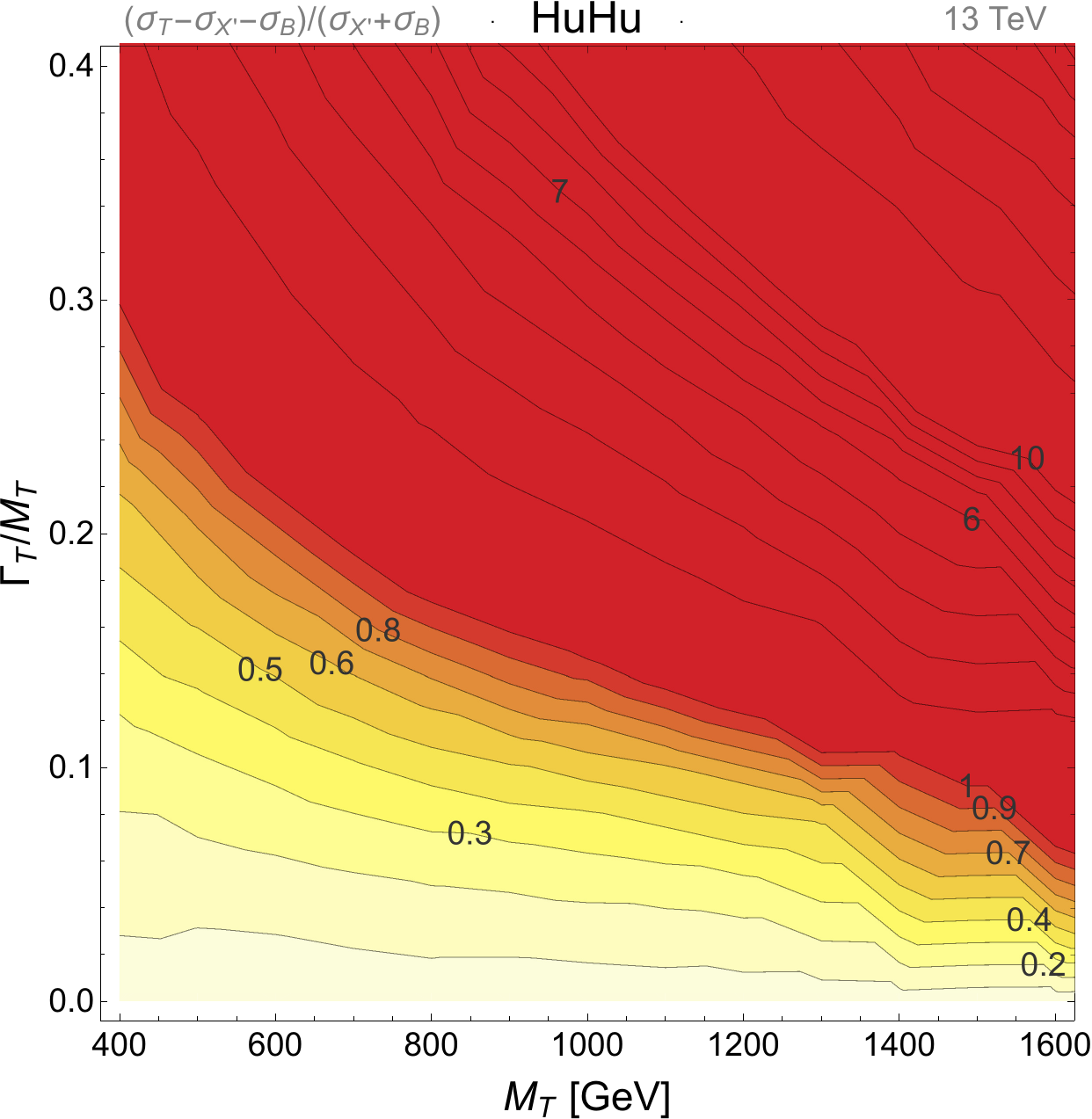}\\
\includegraphics[width=.3\textwidth]{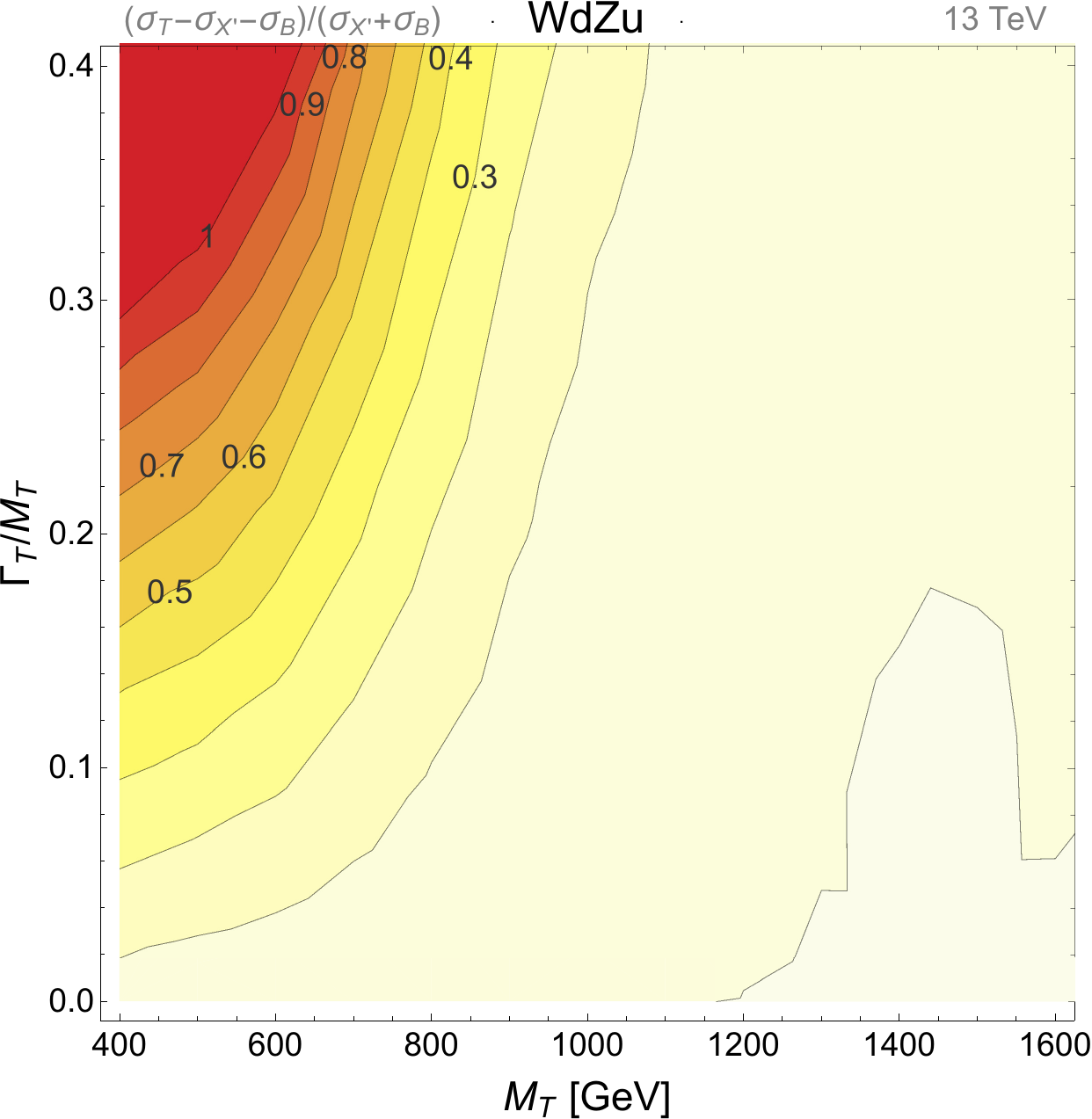}
\includegraphics[width=.3\textwidth]{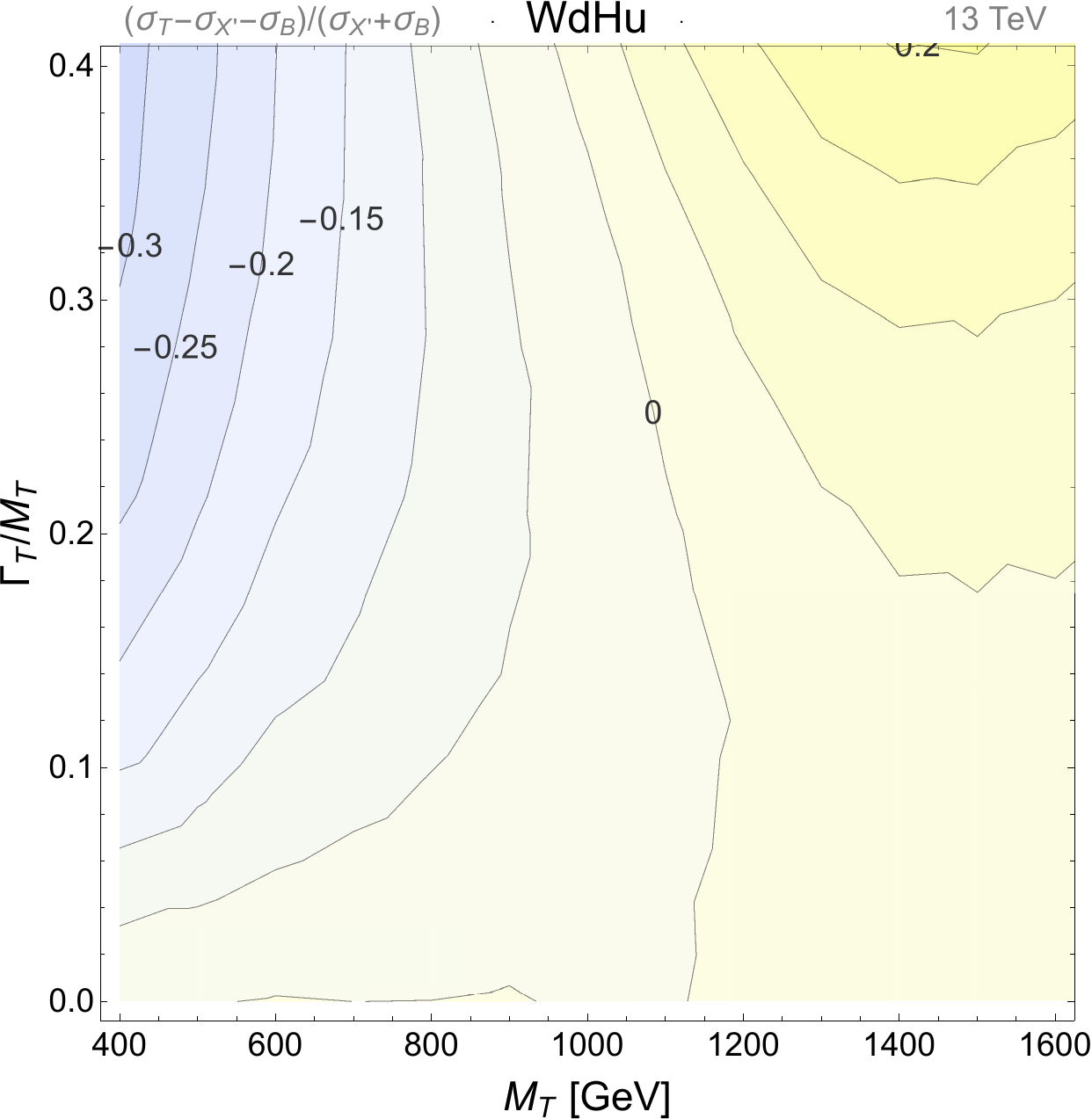}
\includegraphics[width=.3\textwidth]{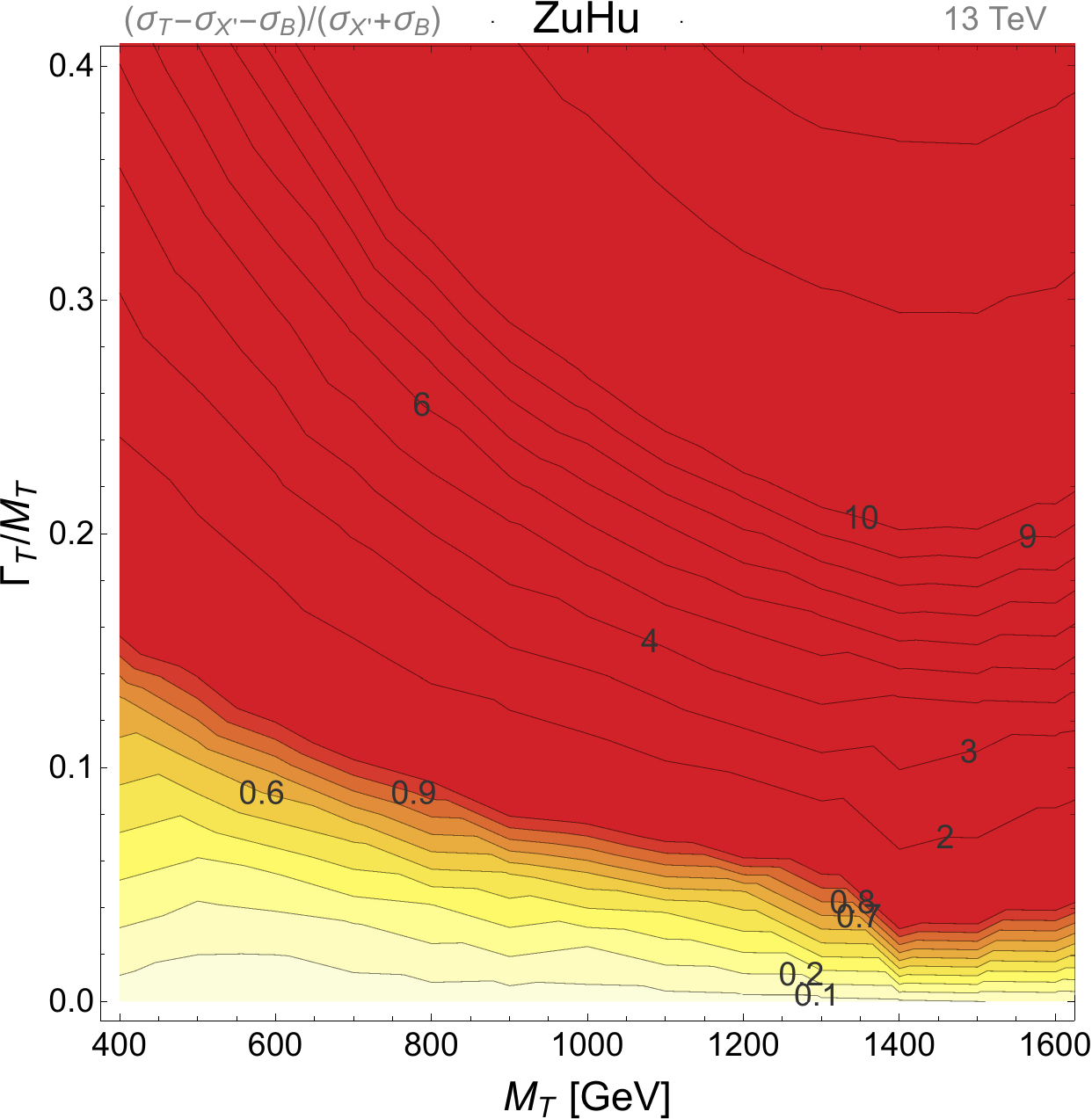}
\caption{\label{fig:TXBfirst}Same as Fig.~\ref{fig:TXBthird} for $T$ mixing with first generation.}
\end{figure}

The correction factors to multiply to the sum of NWA cross-section and SM background to obtain the interference term are plotted in Fig.~\ref{fig:TXBfirst}. For all channels the correction factor becomes quickly large as the $T$ width increases, even if in different fashions depending on the channel. The relative differences between signal and background are small in this case, such that $\sigma_T$ receives a large contribution from the signal. However, when taking into account the full signal, including the large width effects, the interference effects with the SM background become small or negligible in the whole parameter space. As in the case of mixing with third generation, these results show that searches for the exploration of scenarios where the VLQs mix with light generations and have a large width would be significantly more accurate by considering the full signal instead than reinterpreting the NWA results.

\begin{figure}[H]
\centering
\includegraphics[width=.3\textwidth]{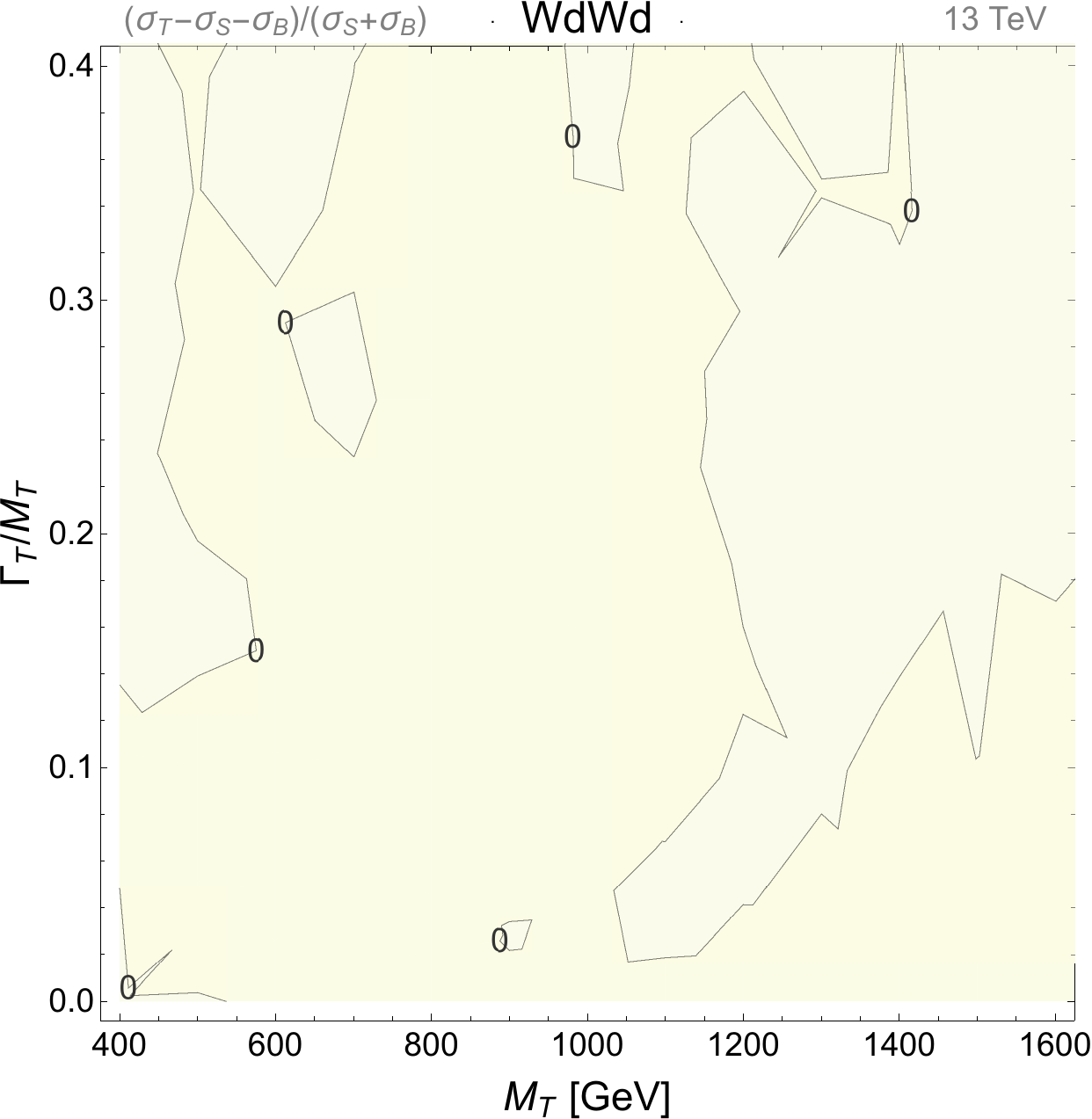}
\includegraphics[width=.3\textwidth]{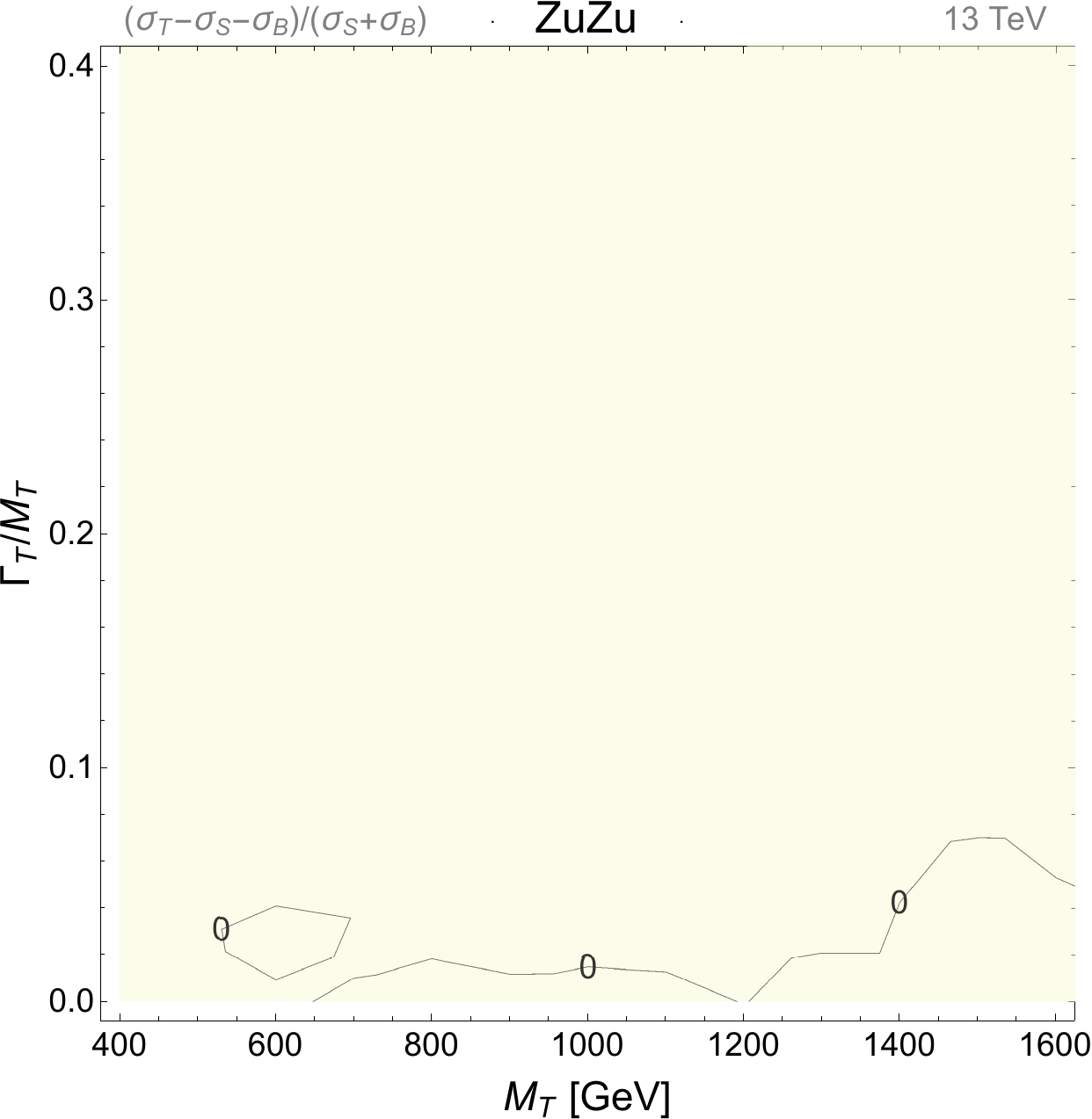}
\includegraphics[width=.3\textwidth]{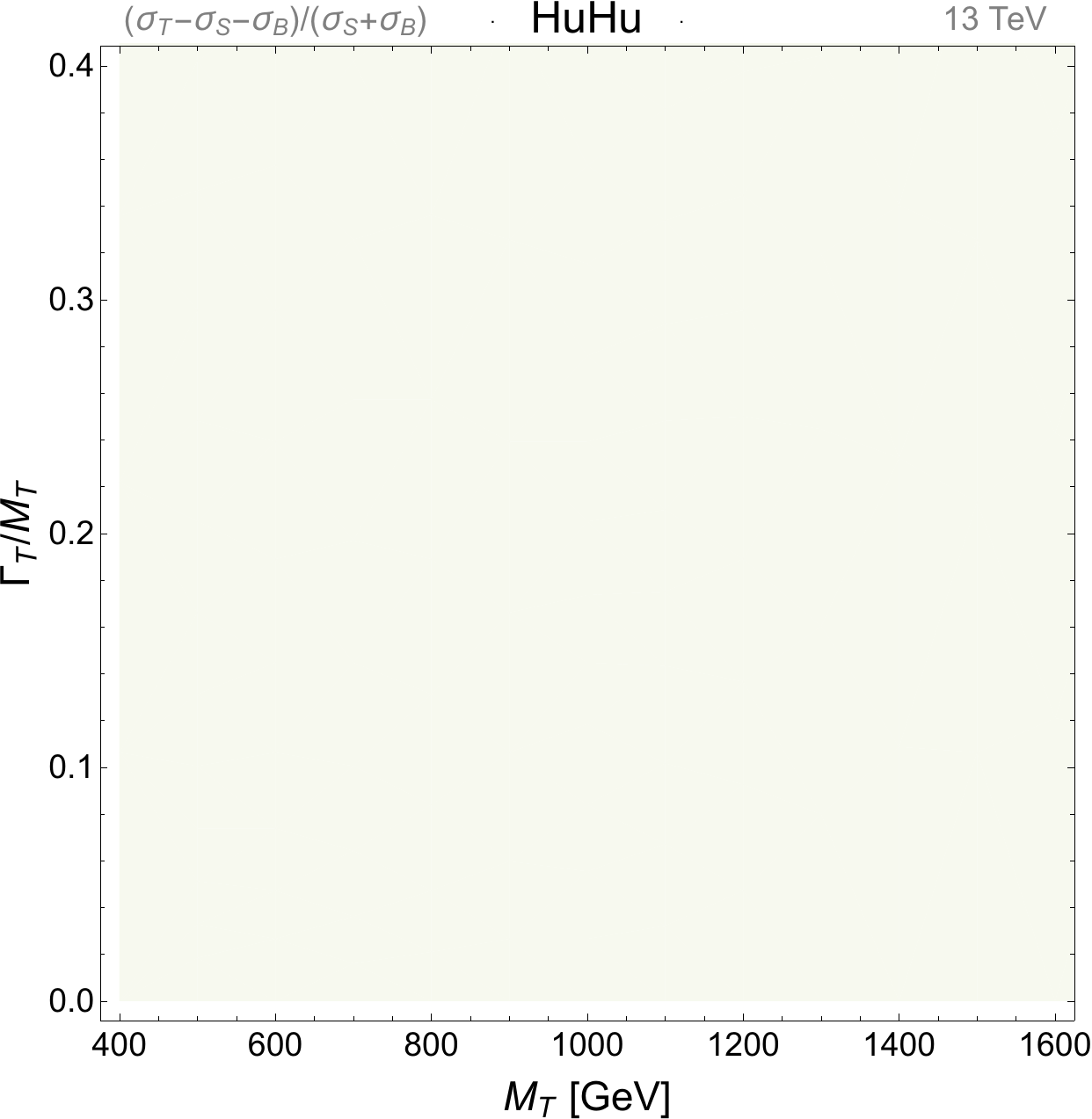}\\
\includegraphics[width=.3\textwidth]{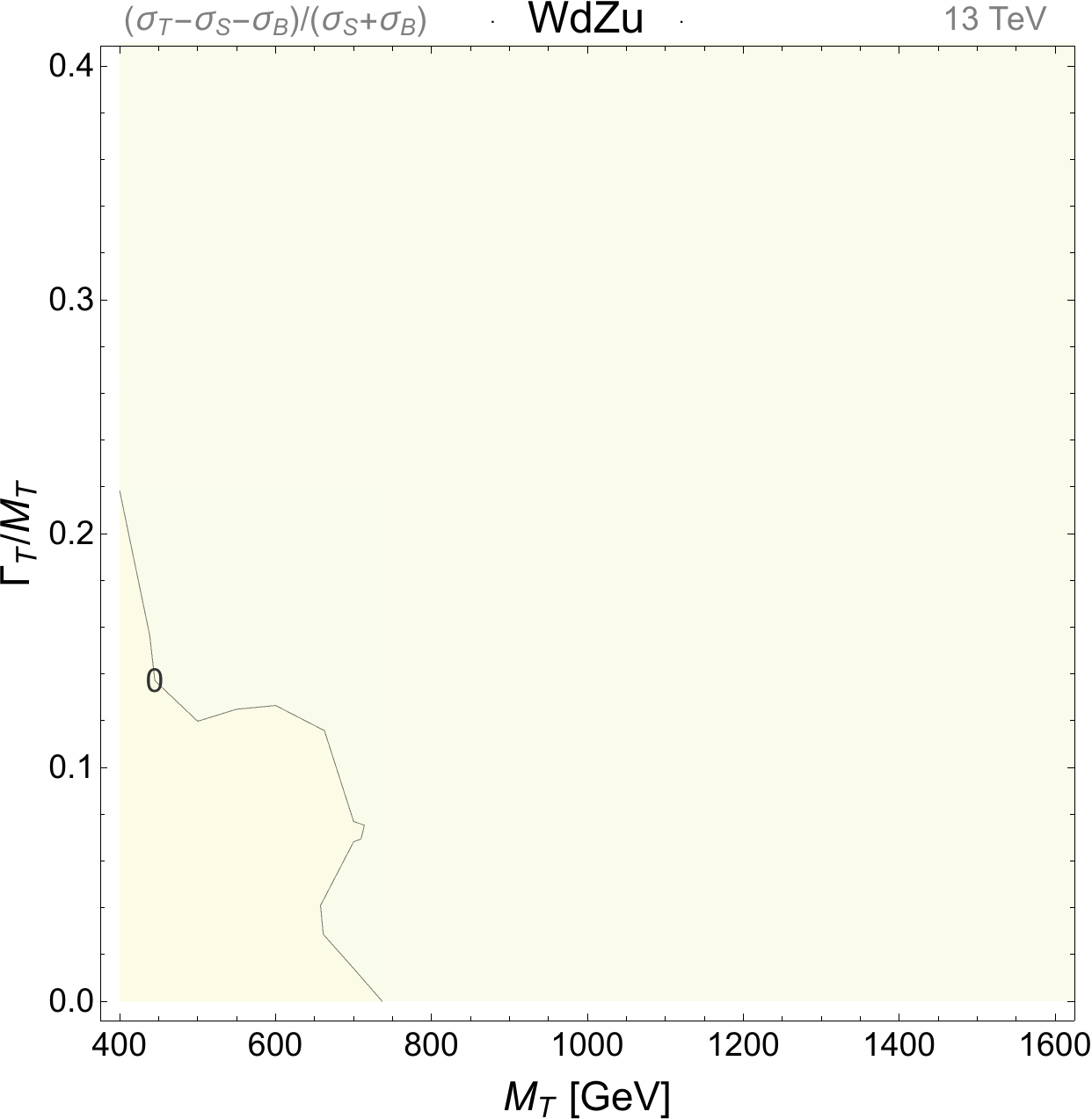}
\includegraphics[width=.3\textwidth]{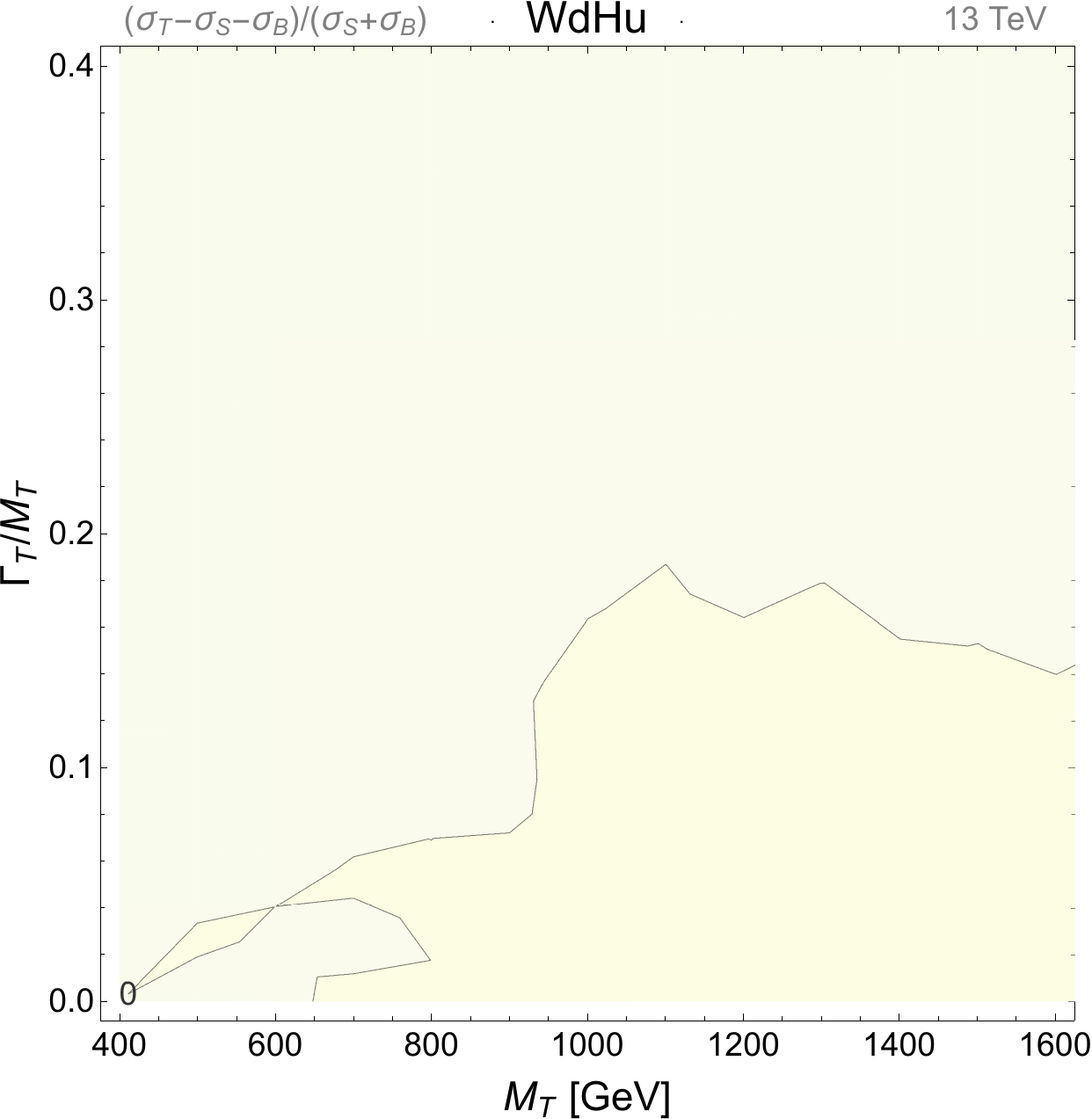}
\includegraphics[width=.3\textwidth]{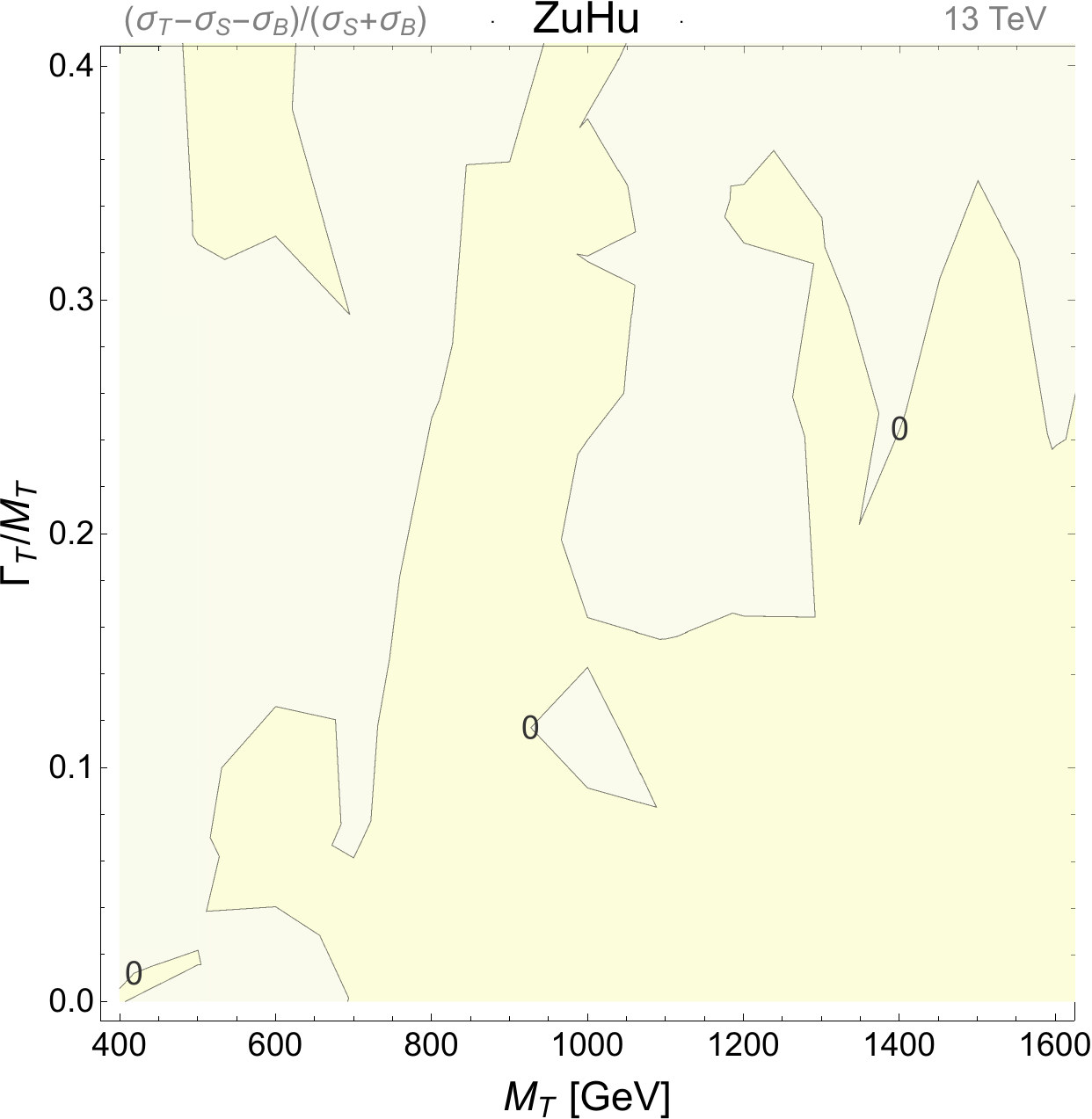}
\caption{\label{fig:TSBfirst}Same as Fig.~\ref{fig:TSBthird} for $T$ mixing with first generation.}
\end{figure}

\subsection{Results at detector level}

\begin{figure}[H]
\centering
\includegraphics[width=.3\textwidth]{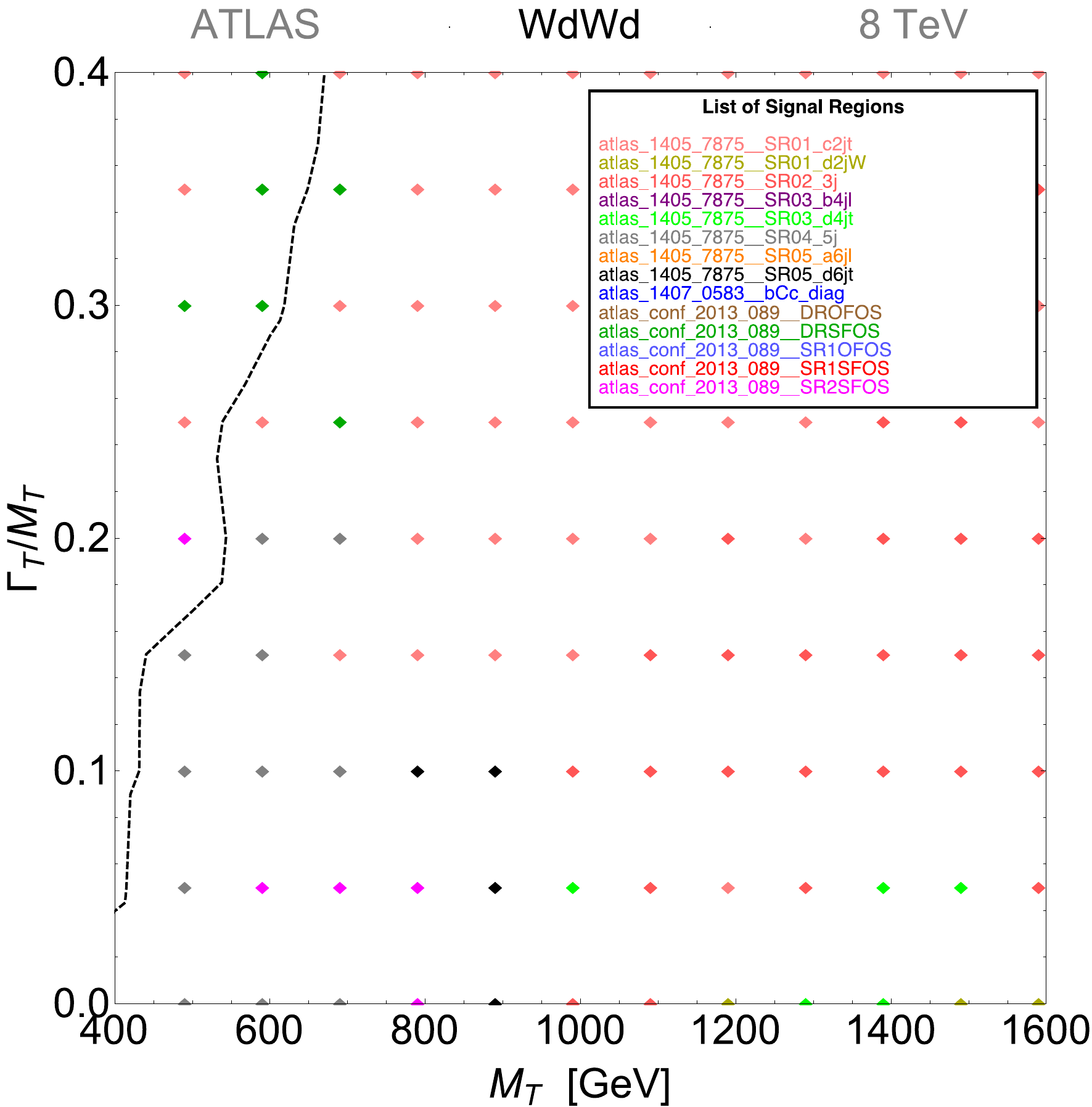}
\includegraphics[width=.3\textwidth]{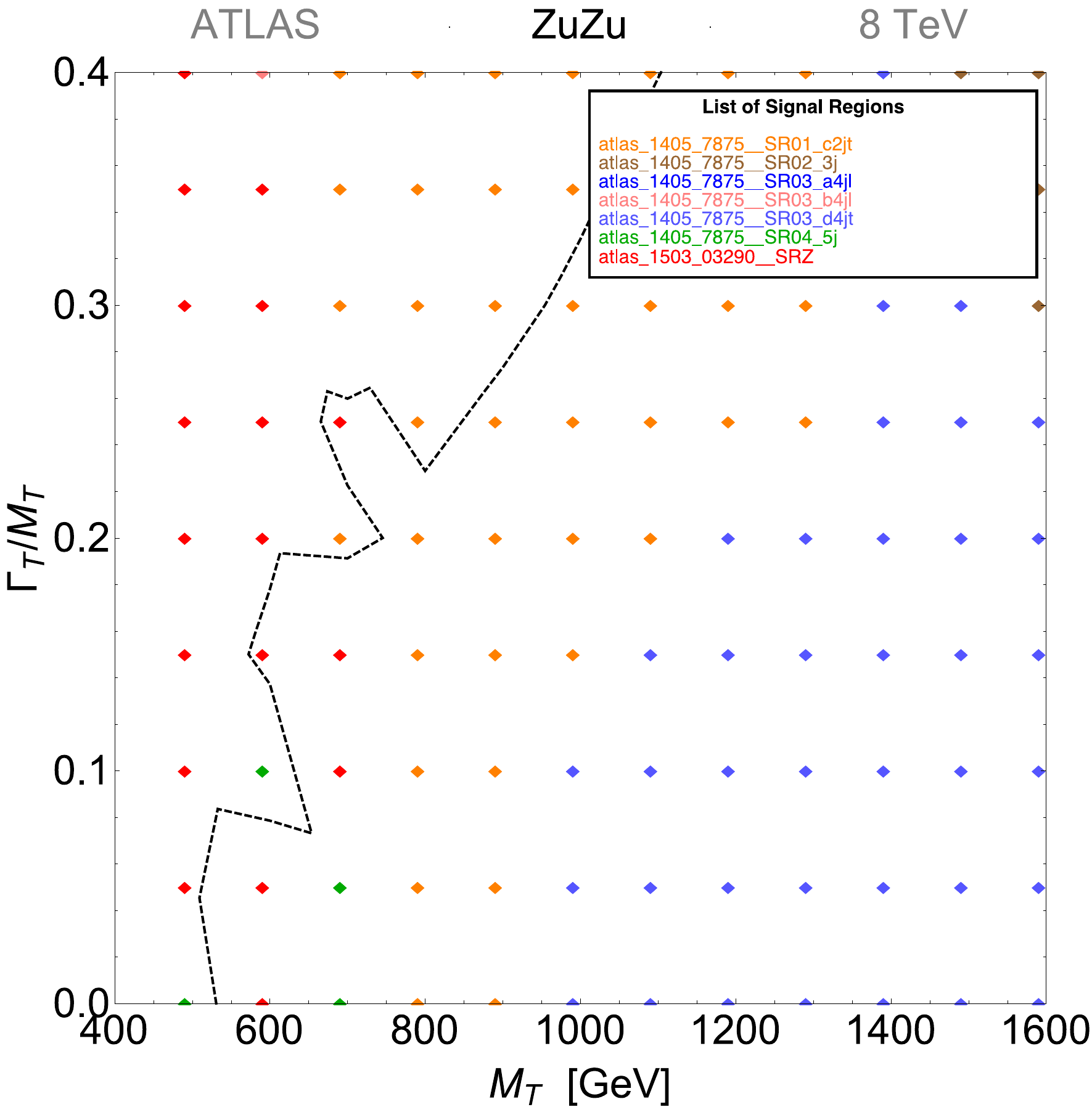}
\includegraphics[width=.3\textwidth]{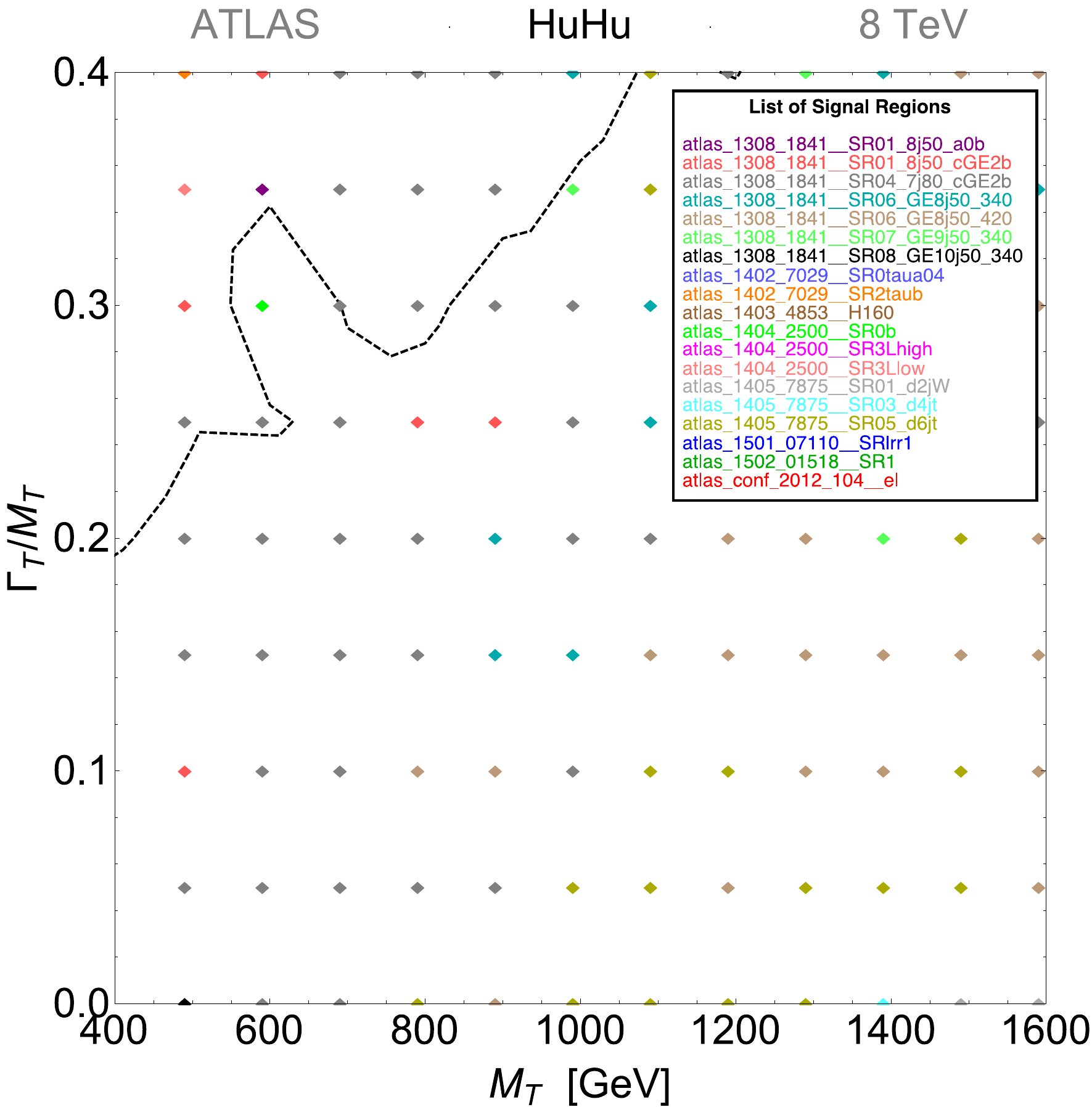}\\
\includegraphics[width=.3\textwidth]{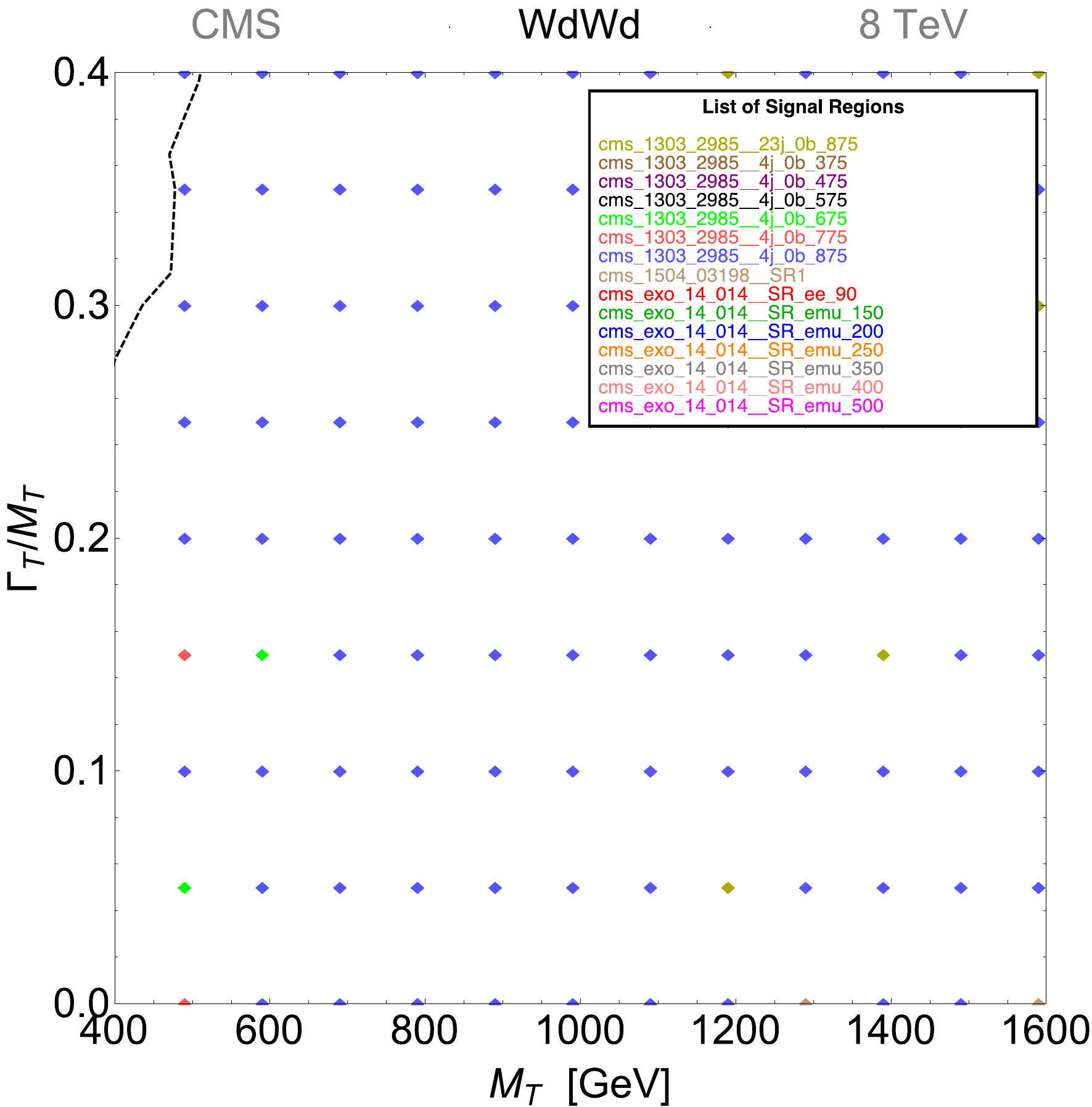}
\includegraphics[width=.3\textwidth]{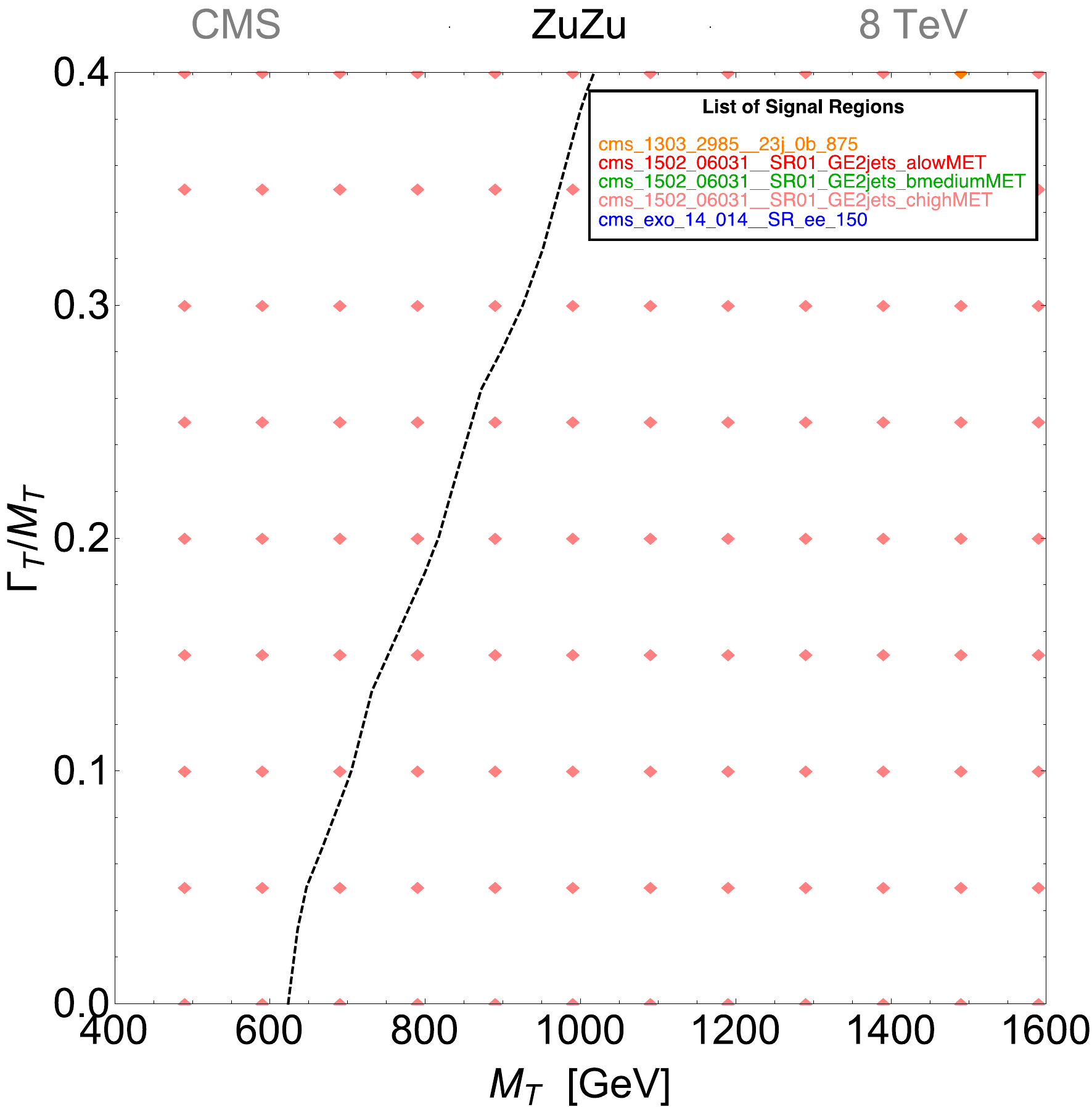}
\includegraphics[width=.3\textwidth]{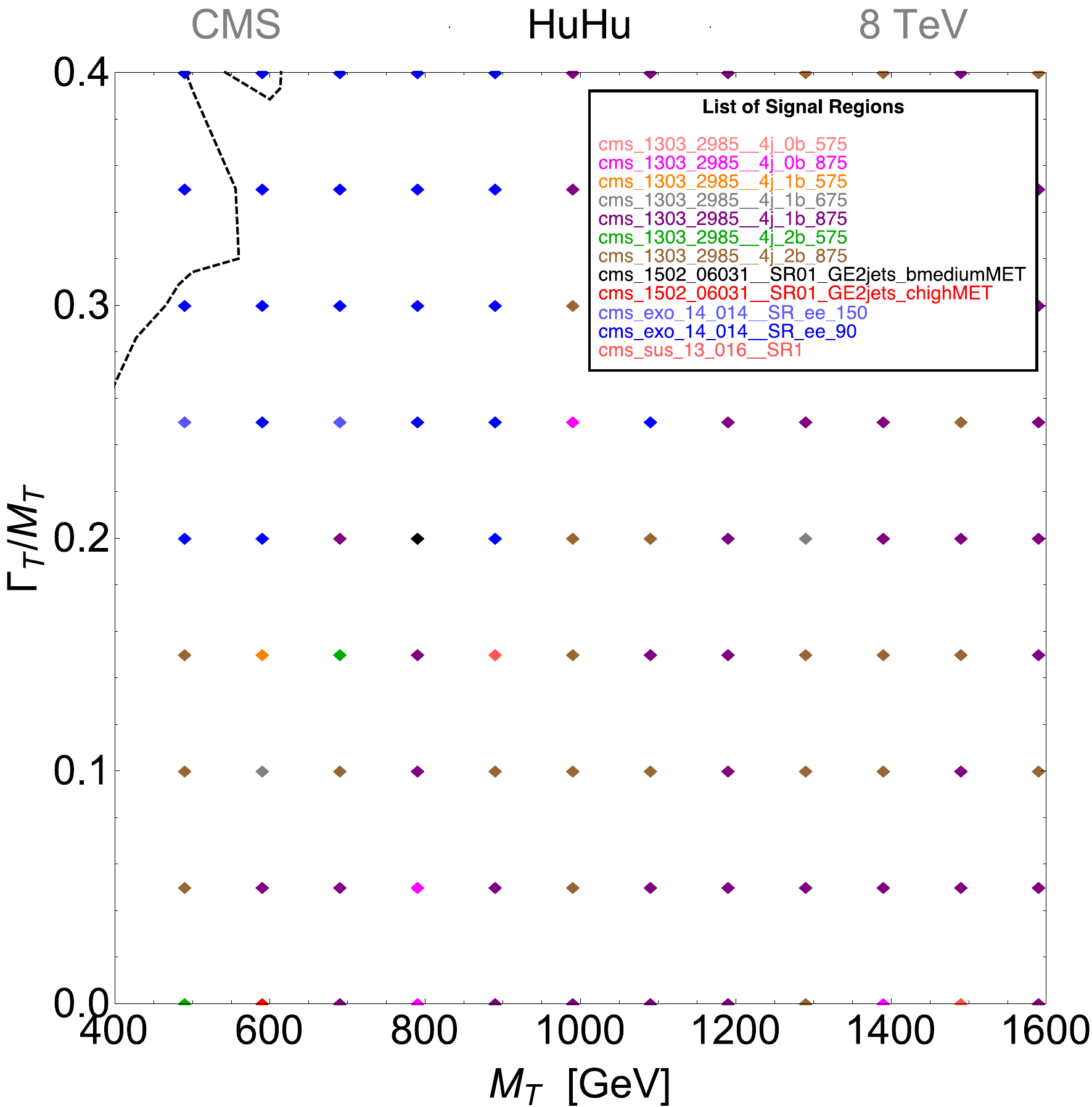}
\caption{\label{fig:Detector8TeV1stgen} Same as Fig.~\ref{fig:Detector8TeV3rdgen} for $T$ mixing with first generation.}
\end{figure}

\begin{figure}[H]
\centering
\includegraphics[width=.3\textwidth]{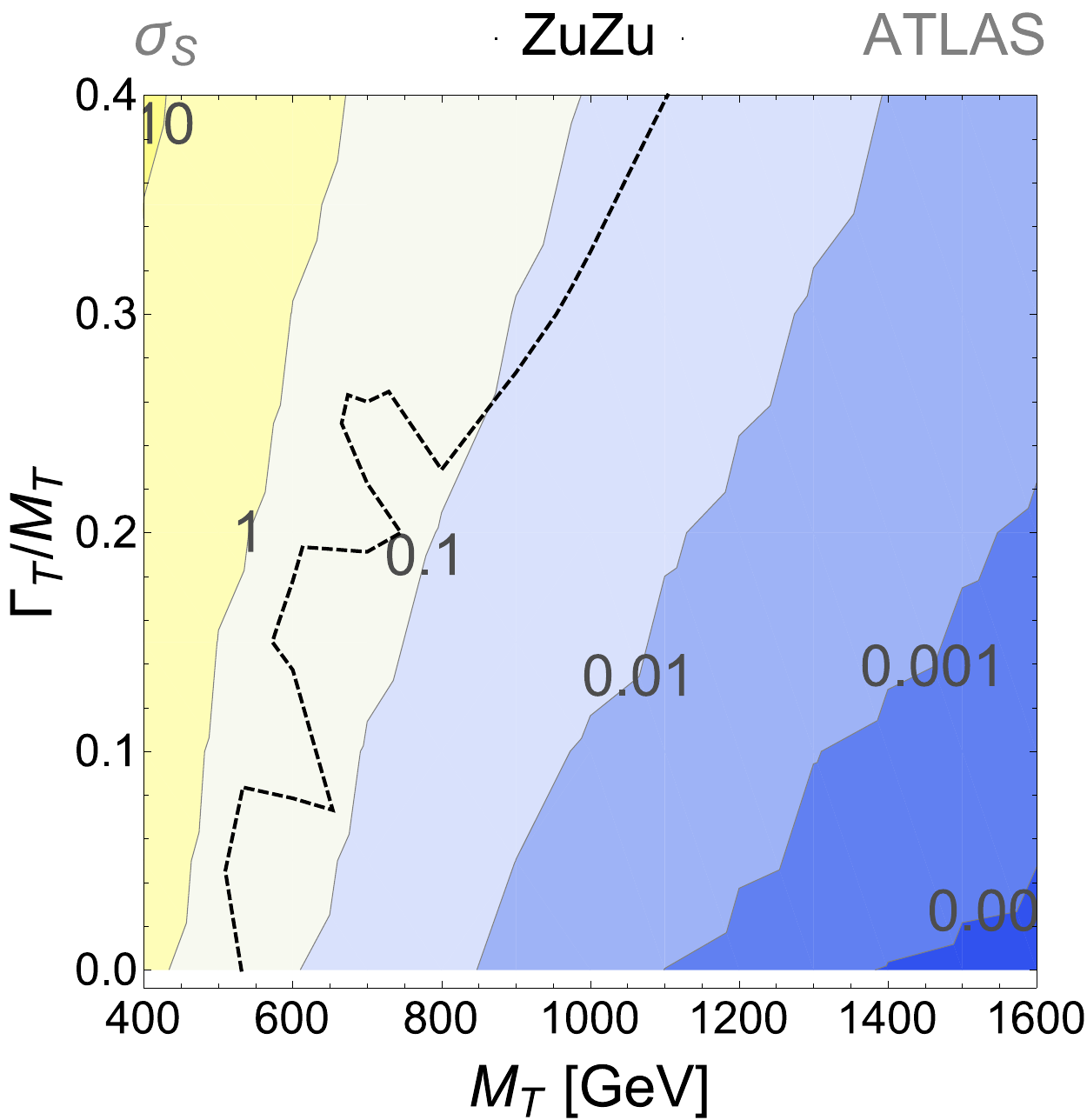}
\includegraphics[width=.3\textwidth]{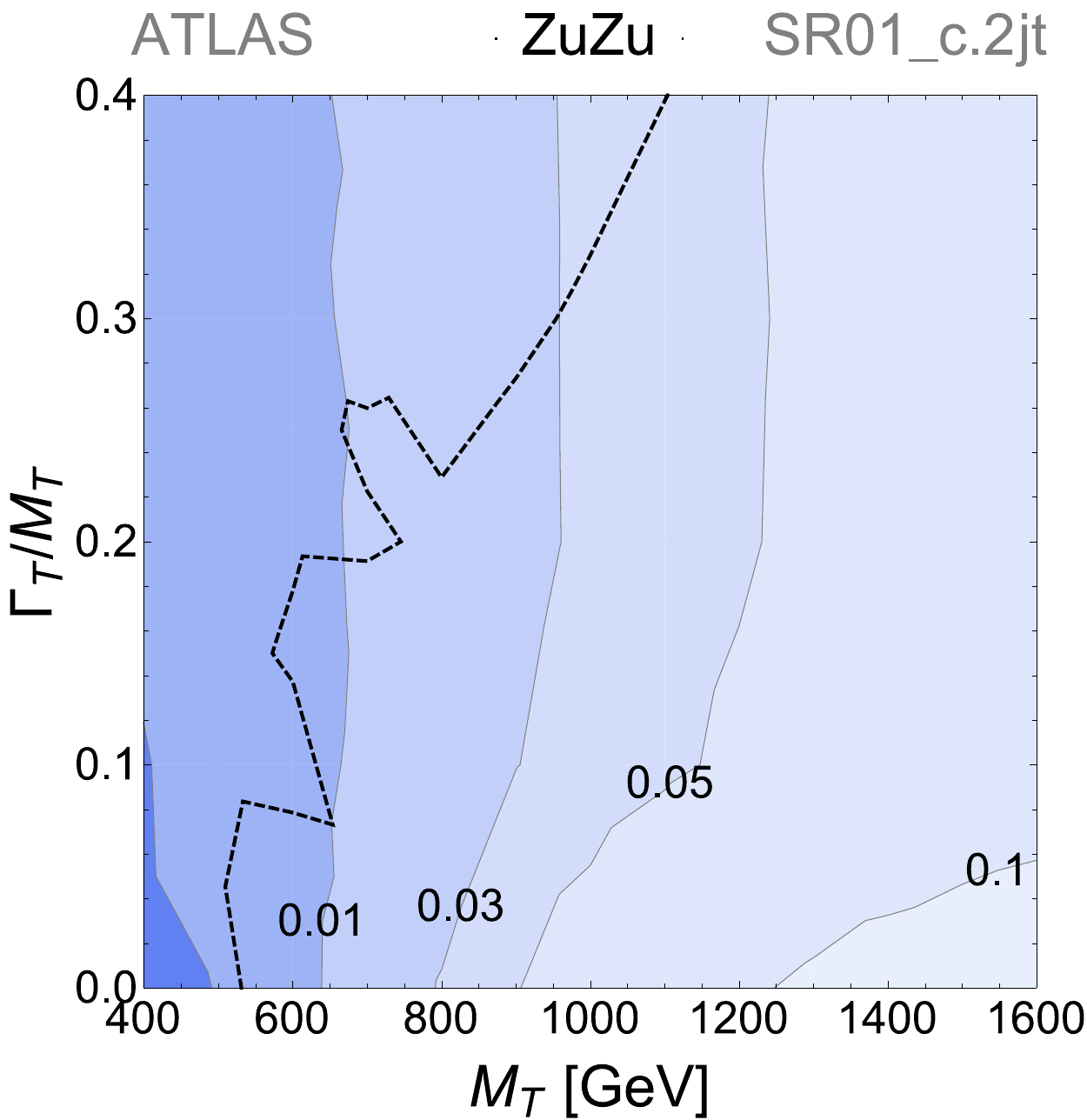}
\caption{\label{fig:CombinedBound8TeVZuZu} Cross-section and efficiency of the best ATLAS SR (SR01\_c.2jt of \cite{Aad:2014wea}) for the $ZuZu$ channel, compared with the bound.}
\end{figure}

Our recast results, obtained considering the same set of ATLAS and CMS searches at 8 TeV as in the case of mixing with third generation, are shown in Fig.~\ref{fig:Detector8TeV1stgen}.
The dependence of the bound on the $T$ width is stronger than in the case of mixing with third generation. For all channels the bound on the $T$ mass becomes stronger as the $T$ width increases. This behaviour has again to be put in relation with the dependence of the signal cross-section, $\sigma_S$, on the $T$ mass and width, shown in the example of Fig.~\ref{fig:CombinedBound8TeVZuZu} for the bound on the $ZuZu$ channel from ATLAS searches. It is possible to see that the bound roughly tracks the cross-section, which unlike in the case of third generation mixing is much more dependent on the width of the $T$, and that the width dependence of the efficiency on the other hand is weakly increasing with both width and mass of $T$ along the bound. 

\section{Conclusions}

We have performed an analysis of off-shell and interference contributions to the process of pair production of heavy quarks at the LHC in the context of minimal scenarios where the SM is extended by adding only a new quark state. As, according to current experimental limits, the latter cannot have the $V-A$ structure of the top quark (unless the Higgs sector is extended, which is not the case in our analysis), we have first assessed how off-shellness impacts on the heavy quark decay signature common to the one of top quark pairs, i.e., $W^-\bar b W^+b$, showing that a $V+A$ chiral structure would be similarly affected over the LHC kinematical regime for pair production of heavy quarks
which can be profiled through a resonance. In this case then, the implementation of finite width effects for heavy quarks can be subsumed under the well established procedures already put in place for the top quark, by simply rescaling the mass of the fermion. Many more decays are however possible for a generic heavy quark pair. Of all the latter, as representative examples, we have chosen to focus on the production and decay of a heavy vector-like top partner $T$ in the singlet representation and considered two scenarios in which it mixes  with either the first or third generation of SM quarks. 

The results of our analysis quantify the relevance of the large width regime in the determination of the cross-section and  the importance of  interference effects between signal and SM background. Clearly, the differences in the cross-section are ultimately reflected in different kinematical distributions, which result in different experimental efficiencies for specific sets of kinematical cuts on the final state. The effect of interference is also found to be generally relevant if the NWA approximation is adopted, while its role is almost negligible if the full signal is considered. Finally, we have evaluated the performance of a set of ATLAS and CMS searches at both 8 and 13 TeV in the determination of the excluded region in the $(M_T,\Gamma_T/M_T)$ plane. We found that the signal regions which are most relevant for the determination of the constraints are weakly sensitive to the $T$ width if the $T$ mixes with the SM top quark, while they can pose higher mass bounds (with respect to the NWA limits) if the $T$ mixes with the up quark. 

To summarise, the conclusion of our analysis is that it is not possible to trivially rescale the mass bounds for VLQs decaying to SM states obtained considering processes of pair production and decay in the NWA to determine constraints for VLQ with large widths. Further, given the weak dependence on the $T$ width of a large set of signal regions of 8 TeV analyses from both ATLAS and CMS, and also of signal regions from a dedicated ATLAS analysis~\cite{TheATLAScollaboration:2016gxs} at 13 TeV looking at pair production of $T$ VLQs, we think that designing different signal regions in experimental analyses to explore the large width regime by taking into account the full kinematical properties of the signal is advisable for a more comprehensive search of heavy quarks at the LHC. A prerequisite to this is to dismiss at MC generation level both the NWA (which leads to severe mis-estimates) and a naive generalisation to a FW approach using the same topologies as in the NWA (which is potentially strongly gauge dependent) in favour of a full determination of every contribution (off-shellness and new topologies) to the signal.

\section*{Acknowledgements}
SM and LP are funded in part through the NExT Institute.

\bibliographystyle{JHEP}
\bibliography{XQCAT}

\noindent

\end{document}